\documentclass{jfm}

\usepackage{caption} 
\usepackage{graphicx}
\usepackage{epstopdf, epsfig}
\usepackage{amsmath}
\usepackage{subfig}
\usepackage{pgfplots}
\usepackage{mathrsfs}
\usepackage{lineno}
\usepackage{setspace}
\usepackage{afterpage}
\usepackage{rotating}
\usepackage{mwe}
\usepackage{xcolor}
\usepackage{verbatim}
\usepackage{array}
\newcolumntype{?}{!{\vrule width 2pt}}

\usepackage{psfrag} 

\pgfplotsset{/pgf/number format/use comma,compat=newest}
\newcommand{\ov}{\overline} 
\newcommand{\wtil}{\widetilde}

\newcommand{\wh}[1]{\widehat{#1}}  
\newcommand{\mc}{\mathcal}
\newcommand{\mb}{\mathbf}
\newcommand{\mr}[1]{\mathrm{#1}}

\newcommand{\maf}[1]{\mathfrak{#1}}

\newcommand{\les}{\left [ }  
\newcommand{\ris}{\right ] }  
\newcommand{\ler}{\left ( }  
\newcommand{\rir}{\right ) }  
\newcommand{\lela}{\left \langle} 
\newcommand{\rira}{\right \rangle} 

\newcommand{\dive}{\nabla \cdot}

\newcommand{\dx}{\mr{d} x}  
\newcommand{\dy}{\mr{d} y}
\newcommand{\dz}{\mr{d} \wh{z}}
 \newcommand{\dt}{\mr{d} \wh{t}} 

\newcommand{\dta}{\mr{d} \tau}

\newcommand{\dl}{\mathrm{d} l}  
\newcommand{\dV}{\mathrm{d} V}  
\newcommand{\dS}{\mathrm{d} S}  

\newcommand{\intdv}[1]{\int_{V} #1 \, \dV } 
\newcommand{\intsg}[2]{\int_{#2} #1 \, \dS^* } 
\newcommand{\intsgn}[2]{\int_{#2} #1 \, \dS } 
\newcommand{\intl}[2]{\int_{#2} #1 \, \dl } 
\newcommand{\flux}[1]{\int_{S} #1 \,\cdot \mathbf{n}\, \dS } 
\newcommand{\avz}[1]{\frac{1}{L_z}\int_0^{L_z} #1 \, \dz } 
\newcommand{\intyz}[5]{\int_{#1}^{#2} \int_{#3}^{#4} #5\, \mr{d} y\, \mr{d} z}       
\newcommand{\avt}[1]{\frac{1}{t_f-t_i}\int_{t_i}^{t_f} #1 \dt}   
\newcommand{\avte}[1]{\frac{1}{N}\sum_{n=0}^{N-1}  #1  \ler x,y,z,\tau+n \, T \rir}   

\newcommand{\avtez}[1]{\lela \wh {#1} \rira}   

\shortauthor{Vinh-Tan Nguyen, Pierre Ricco, Gianluca Pironti}

\title{Separation drag reduction through a spanwise oscillating pressure gradient}

\author{
    Vinh-Tan Nguyen\aff{1},
    Pierre Ricco\aff{2}\corresp{\email{p.ricco@sheffield.ac.uk}} and 
    Gianluca Pironti\aff{1,2}
}

\affiliation{
\aff{1}Institute of High Performance Computing, Agency for Science Technology and Research, 1 Fusionopolis Way, 16-16 Connexis Tower, Singapore 138632
\aff{2}Department of Mechanical Engineering, The University of Sheffield, Mappin Street, S1 3JD Sheffield, United Kingdom
}
\begin{document}

\maketitle

\begin{abstract}
An oscillating spanwise pressure gradient is imposed numerically to control the flow separation and reduce the drag of a turbulent flow in a channel with square bars. The transverse flow produces a maximum drag reduction of 25\%, due to lower pressure and skin-friction drag forces. The pressure drag reduction reaches a maximum of 22\% and is due to a decrease of the positive high pressure in front of the bars and an increase of the low negative pressure behind the bars. The skin-friction drag reduction is caused by a lower wall-shear stress along the cavity between the bars where the flow is fully attached, while the wall-shear stress on the crest of the bars and in the separated region behind the bars is unaffected. The spanwise laminar flow obtained by neglecting the nonlinear terms involving the turbulent velocity fluctuations is used to compute the power spent for oscillating the fluid along the spanwise direction and an excellent agreement is found with the power spent obtained by the averaged turbulent flow. A marginal or negative net power saved is found by subtracting the power employed for controlling the flow from the power saved thanks to the transverse flow. The control reduces the total drag as the integral of the Reynolds stresses along the horizontal line connecting the corners of two consecutive bars is decreased, which in turn impacts on the pressure and wall-shear stress reductions.

\end{abstract}

\newpage
\section{Introduction}

Flow separation occurs when the flow streamlines are not able to follow the contour of a solid surface and depart significantly from it. The problem is of interest as flows around large vehicles and buildings often experience separation \citep{oliveira-younis-2000,choi-etal-2014}.

In many engineering systems, separation causes a large increase in pressure drag that has to be overcome by an undesirable power expenditure, thus contributing negatively to pressing environmental issues, such as excessive fuel consumption, noise and pollutant emissions. The development of flow control techniques that aim to reduce the pressure drag due to flow separation is an extremely active area of academic and industrial research. Control methods can be classified into active and passive, depending on whether energy is supplied or not to the physical system. A clear advantage of active controllers is the large beneficial flow modifications that can be achieved. Active controllers can be deactivated when their action is no longer required, but a weight penalty is introduced due to the actuators. Active methods are also prone to damage and often require expensive maintenance. Passive methods often involve a geometrical modification of the surface, offer the benefit of not requiring external power supply, although the small wall alterations may necessitate regular repairment. The development of systems which, at the same time, guarantee a sizeable drag-reduction margin and low costs for production, maintenance and energy consumption is therefore of paramount importance to achieve economical advantages.           

\subsection{Active methods}

\label{sec:hel}
\subsubsection{Suction and blowing}
The suction of fluid from a slit on the surface was the first active technique used to control flow separation \citep{chang-2014}. The amount of control is typically quantified via a mass flow rate coefficient $C_m$ \citep{nuber-needham-1948} or a momentum coefficient $C_\mu$  \citep{chng-etal-2009}. 

\cite{nuber-needham-1948} found a $25\%$ increase of the maximum lift coefficient for a NACA $64_1A212$ airfoil by applying suction with a coefficient $C_m=1.8\cdot 10^{-3}$ along the upper surface near the leading edge. \cite{seifert-pack-2002} applied steady and oscillating suction and blowing near the separation point of a Glauert-Goldschmied airfoil. Oscillating suction and blowing effectively reduced the pressure drag for $C_\mu<0.1\%$. Separation drag was suppressed by steady suction or blowing using $C_\mu=0.8\%$ and $C_\mu=2\%$ respectively, with suction leading to the best performance. \cite{greenblatt-etal-2006a} removed the fluid around the upper surface of a modified Glauert profile and the form drag was reduced by increasing $C_\mu$, showing a strong dependence on the Reynolds number. In the same study, the separation bubble was suppressed when $C_\mu=2\%$.

Some authors applied blowing over the separation line of supersonic airfoils \citep{bradley-wray-1974,meyer-seginer-1994,wong-kontis-2007-a} or backward facing s.pdf \citep{chun-lee-sung-1999}, effectively reducing the reattachment length. More recent studies, such as \cite{brackston-etal-2016-eif}, have employed extremum-seeking feedback control, implemented via pulsed-jet actuators, to minimize the separation drag given by a bluff-body wake. 

\cite{cho-etal-2016} reduced the extent of the separation bubble in a turbulent boundary layer by means of oscillating suction and blowing. The control, applied near the separation line, gave the best performance for a frequency $f=0.5$, normalized via the free-stream velocity and the uncontrolled bubble length. The creation of intense spanwise vortices energized the fluid near the wall, thereby reducing the bubble.  

Critical aspects of these active methods are the minimization of the pumping power required by the actuation, the skin-friction penalty \citep{gadelhak-pollard-bonnet-1998} due to the enhanced fluid mixing, and the decrease of thrust due to the air drawn from the aircraft engines \citep{chng-etal-2009}.     

\subsubsection{Plasma actuators}

Plasma actuators are devices composed of two metal electrodes separated by a dielectric material and subjected to a high voltage. The fluid is ionized around the actuators, by creating a volume force equivalent to a pressure gradient \citep{corke-etal-2002}.  

\hspace{\parindent}\cite{post-corke-2004} investigated the ability of plasma actuators to reattach the flow over a NACA$66_3-018$ airfoil at the post-stall incidence. The actuators, applied at the leading edge of the airfoil and at the maximum chamber location, created a two-dimensional steady flow field that energized the boundary layer around the body. The lift increased with the forcing magnitude up to a peak value corresponding to the fluid reattachment, while the drag reduced proportionally to the actuation intensity, causing a maximum $400\%$ increase of the aerodynamic efficiency. \cite{post-corke-2006} controlled the dynamic stall and separation on a NACA$0015$ airfoil oscillating around its axis by employing three types of plasma actuators that all reattached the flow. The steady actuator increased the lift, except at high angles of attack, while the unsteady actuator raised the lift only during the pitch-down phases.

\cite{huang-corke-thomas-2006} employed a plasma actuator to control separation on a blade cascade. The actuator was located at two positions upstream of the separation point and led to a reduction of the reattachment length, similar to that obtained with vortex generators in the same configuration. The main drawback of plasma actuators is the energy used for their activation, causing a low efficiency at high Reynolds numbers \citep{neretti-2016}. 

\subsection{Passive methods}

Vortex generators are the most common passive devices for separation control. They are in the form of geometrical modifications, usually placed upstream of the region of separation. They create streamwise vortices that enhance the turbulent mixing and allow the flow to sustain adverse pressure gradients, thereby reducing the tendency of the flow to separate. \cite{calarese-crisler-1985} conducted experiments on a replica of a C-130 aircraft with arrays of vortex generators placed along the fuselage circumference. Vortex generators at the most upstream position were  most effective at an incidence of $4$ degrees, leading a $7\%$ reduction of the total aircraft drag. \cite{bragg-gregorek-1987} applied three kind of vortex generators on a canard wing to alleviate the effect of boundary-layer separation caused by dirt deposition. All vortex generators were able to increase the lift.

\hspace{\parindent}Classical vortex generators create parasite drag even when no separation occurs, and therefore research has focused on developing devices that are small enough to be completely submerged in the boundary layer. \cite{lin-selby-howard-1991} proved that wishbone vortex generators with a height of $20\%$ of the boundary-layer thickness were effective in  reducing the reattachment length in a descending ramp. They created a pair of counter-rotating vortices that enhanced mixing in the area with the lowest momentum. \cite{lin-etal-1994} experimentally studied how delta and trapezoidal submerged vortex generators influenced the flow on a three-section airfoil with a flap. At low incidence ($8$ degrees) vortex generators were placed at a quarter of the flap chord and the separation around the flap was controlled effectively as the wake thickness was reduced by $60\%$. At the same angle of attack they increased the lift by $11\%$ and generated a drag reduction of $38\%$ with an improvement of $80\%$ on the aerodynamic efficiency.

The main shortcoming of submerged vortex generators is their tendency to break because of their small size.  

\subsection{Objectives and structure of the paper}

The aim of this numerical work is to study the effectiveness of a spanwise oscillating pressure gradient in controlling flow separation. The objectives are to reduce the pressure drag caused by the flow separation around square bars located on the opposite walls of a channel and to understand the physics behind the flow alterations induced by the spanwise oscillating pressure gradient.

Spanwise oscillating walls and pressure gradients have been used extensively for turbulent skin-friction drag reduction \citep{jung-mangiavacchi-akhavan-1992,trujillo-bogard-ball-1997,quadrio-ricco-2004}, but only a very few studies have focused on the effect of spanwise forcing on large flow separation. \cite{jukes-choi-2012} were able to reattach the flow on a $20$-degree inclined ramp by using spanwise jets, while \cite{brackston-etal-2016} and \cite{brackston-etal-2018} used oscillating flaps positioned at the back sides of a bluff body in a wind tunnel to control the separation wake via reduced-order feedback control. The recent numerical study of \cite{banchetti-etal-2020} showed that streamwise travelling waves of spanwise wall velocity can reduce the pressure drag of flows separating marginally behind smooth obstacles.

In the present study, square bars are chosen because the pressure drag and the skin-friction drag are completely distinct around these obstacles. The pressure drag is produced by the difference in the surface integrated pressures in front and behind the bars and the skin-friction drag is the result of the surface integrated wall-shear stress on the bar crests and over the channel walls between bars. The contribution of the oscillating pressure gradient to the global drag reduction can therefore be precisely quantified by the separate reductions of the pressure drag and the skin-friction drag.
For the chosen geometry, the pressure drag contributes significantly to the total drag as the ratio between the pressure drag and the skin-friction drag is 4.3, that is, the pressure drag is 81\% of the total drag. The selection of an uncontrolled flow with such a feature is central to our study because we are particularly interested in the effect of spanwise forcing on the separated flow behind the square bars.

Section \ref{sec:numerics} illustrates the flow configuration and discusses the numerical method and the parameters used to evaluate the control performance. Section \ref{sec:laminar} describes the derivation and the numerical solution of the laminar auxiliary problem employed to compute the power spent for the actuation. In \S\ref{sec:uncon} the mean pressure and the mean streamlines of the uncontrolled flow are first discussed and the drag-reduction margin, the power spent for the control actuation, and the net power are quantified. In \S\ref{sec:physics}, the effect of the actuation on the flow statistics is further investigated and the role of the Reynolds stresses between the bars is outlined. Section \ref{sec:conclusions} presents the conclusions of our work.

\section{Numerical procedures}
\label{sec:numerics}

\begin{figure}
\centering
\includegraphics[width=1.0\textwidth]{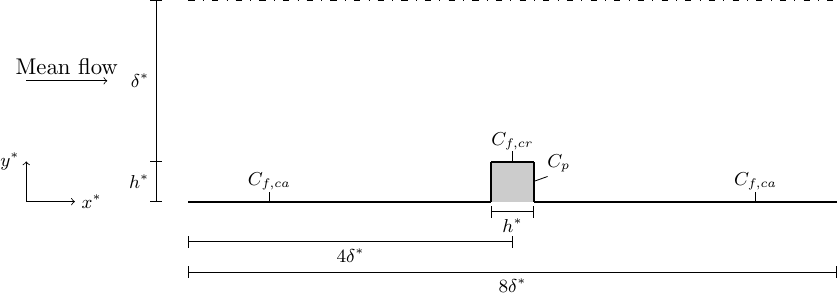}
\caption{Sketch of the system configuration. The bar is indicated by the gray square and the drag coefficients, assigned to each surface, are defined in \S\ref{sec:perfor}. The pressure coefficient $C_p$ is computed by considering both sides of the bar. Only the bottom half of the channel is shown. The dash-dotted line represents the symmetry plane at $y^*=\delta^*+h^*$ and the height of the bar is $h^*=0.2\delta^*$. The distance between the bar and the domain sides is not drawn to scale.}
\label{fig:fig1}
\end{figure}

We study the turbulent channel flow depicted in figure \ref{fig:fig1} where, within the computational domain, two identical square bars are located on the opposite walls and oriented perpendicularly to the mean streamwise direction. 

The boundary conditions are periodic over the streamwise and spanwise directions, while the no-slip condition is imposed on the solid walls. Dimensional quantities are indicated with an asterisk. The Cartesian coordinates are $x^*$, $y^*$ and $z^*$ along the the streamwise, vertical and spanwise directions respectively, where $x^*$ is defined from the upstream start of the computational domain and $y^*$ is defined from the bottom channel wall. The side of the square section of a bar is of length $h^*$ and $\delta^*=5 h^*$ is half of the distance between the bar crests. The centre of each bar is located at $x^*=4\delta^*$. 

The flow used to initiate the computations is of zero velocity in the two strips between the bars, $0<y^*<h^*$ and $2\delta^*+h^*<y^*<2(\delta^*+h^*)$, and a laminar Poiseuille flow plus a random three-dimensional perturbation velocity field of zero mean in the range $h^*<y^*<2\delta^*+h^*$. The centerline velocity of this Poiseuille flow is $U_p^*$. The channel studied in this work differs from the one of \cite{leonardi-etal-2003}, where multiple bars were used, because in our case only one bar is placed on each wall within the computational domain and symmetry between top and bottom wall applies. The distance between bars on the same wall is $39 h^*$, therefore higher than those used by \cite{leonardi-etal-2003} ($0.33h^*-19 h^*$). We have chosen this large distance because \cite{leonardi-etal-2003} showed that there is no influence between the recirculation regions for distances larger than $7 h^*$. All the variables are scaled using $\delta^*$ and $U_p^*$ and the Reynolds number is $Re_p=U_p^* \delta^*/\nu^*=4200$, where $\nu^*$ is the kinematic viscosity of the fluid. The simulations are performed at a constant mass flow rate. The components of the velocity vector along $x^*$, $y^*$ and $z^*$ are indicated by $u^*$, $v^*$ and $w^*$, respectively. The dimensions of the computational domain are $L_x=8$, $L_y=2.4$ and $L_z=\pi$, where the subscripts denote the Cartesian coordinates.
The pressure $p^*$, scaled using the density of the fluid $\rho^*$ as $p=p^*/(\rho^*{U_p^*}^2)$, is defined as $p(x,y,z,t)=\phi(x,y,z,t)+\Pi_x(t)\, (x-L_x)+\Pi_z(t)\, z$, where $\phi$ is a periodic function of $x$ and $z$, $\Pi_x <0$ is the spanwise-averaged streamwise pressure gradient, and $\Pi_z$ is the spanwise pressure gradient used for controlling the flow. The latter has the form:
\begin{equation}
\Pi^*_z=A^* \cos\left(\frac{2\pi t^*}{T^*}\right),
\end{equation}
where $A^*$ and $T^*$ are the amplitude and the period, scaled as $A=A^*\,\delta^*/ \ler\rho^* {U^*_p}^2 \rir$ and $T=T^* U^*_p/ \delta^*$. Analogous to the studies on the wall-oscillation technique where the peak-to-peak displacement is used as a third forcing parameter \citep{ricco-quadrio-2008}, we define the peak-to-peak impulse $J=J^*/\rho^* U_p^*=2\max_t\left(\int_0^t \Pi_z \mathrm{d}{\hat t}\right)=AT/\pi$ as an additional parameter related to $A$ and $T$. The impulse $J^*$ has units kg/m$^2$s and therefore it represents the change of momentum along $z$ per unit volume imposed by the pressure gradient. 
 
\subsection{Direct numerical simulations}
\label{sec:numpro}
 
\begin{figure}
\centering
\includegraphics[width=0.7\textwidth]{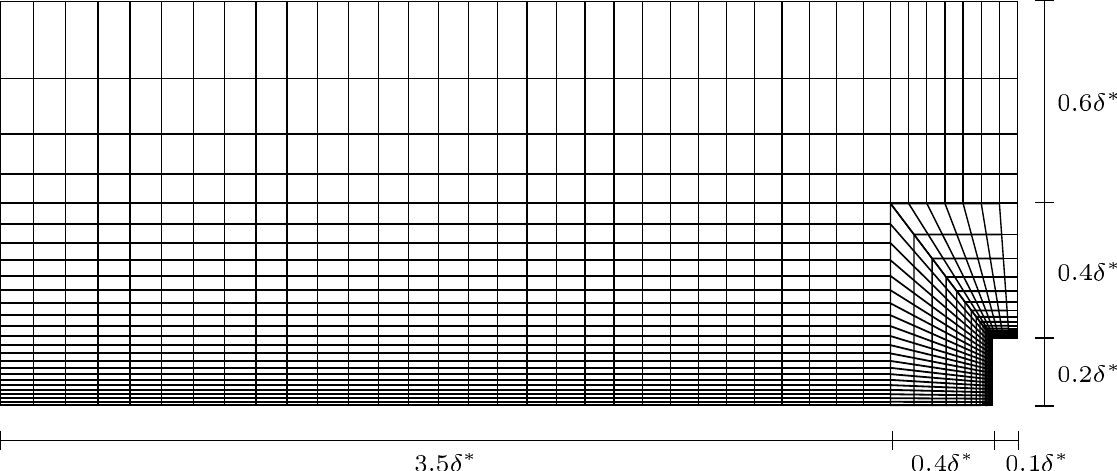}
\caption{Mesh for the channel with bars. The grid is symmetrical around the axes $x^*=4\delta^*$ and $y^*=1.2\delta^*$ and only the quarter of the computational domain in the range $\les 0,4\delta^* \ris \times \les 0,1.2\delta^* \ris$ is shown. Only the hexahedral elements are shown.}
\label{fig:fig2}
\end{figure}

The fluid flow is simulated by the spectral element code Nek5000 \citep{fischer-2017}. The code employs the spectral finite element method (SEM), based on the Galerkin discretization of the weak form of the Navier-Stokes equations. The fluid domain is split into hexahedral elements where each fluid quantity $q$ is expressed in terms of Lagrange polynomials with nodes following the distribution of Gauss-Lobatto-Legendre (GLL) points. Each element is discretized with $l_x l_y l_z=l_x^3$ GLL points, where $l_x=l_y=l_z$ are the number of points used for each edge of the element.          
The Navier-Stokes equations are discretized in time using an operator integration factor splitting scheme \citep{maday-patera-ronquist-1990} that permits large time s.pdf $\Delta t$ due to its semi-implicit nature. The time step is set to the constant value $\Delta t= 1.69\cdot 10^{-4}$ and the related Courant-Friedrichs-Lewy number is in the range 1.5-2.0. Figure \ref{fig:fig2} depicts the mesh around the lower bar. The mesh radiates from the bar to guarantee a high clustering of points especially around the left and the right top corners, where singular behaviours of the local pressure and the local wall-shear stress occur. A total number of $76496$ hexahedral elements are employed. The mesh is identical on each $x$-$y$ plane and 14 identical elements are used along $z$. A number of $6$ GLL points is adequate to compute the total drag accurately, as shown by the convergence study in Appendix \ref{sec:dragconv}.
The first point above the cavity is at a distance $\Delta y=1.1\cdot 10^{-3}$ from the wall, while the last point along $y$ is at a distance $\Delta y=1.3\cdot 10^{-2}$ from the centerline. The maximum streamwise resolution is $\Delta x=1.5\cdot 10^{-2}$. The minimal $\Delta x=\Delta y =5.7\cdot 10^{-5}$ refer to the grid elements at the side edges and the crests of the bars, respectively. Even though the spectral elements along $z$ are identical, the spanwise resolution varies from $\Delta z_{min}=2.6\cdot 10^{-2}$ to $\Delta z_{max}=6.4\cdot 10^{-2}$.       
All the simulations are performed on the high performance computing facility of the A*STAR Computational Resource Centre in Singapore, employing 2000-8000 processors. The flow statistics and the drag coefficients are computed after the end of an initial transient, evaluated by the inspection of the time history of total drag. This is crucial to assess that the flow has reached a statistically steady state.  

\subsection{Averaging operators and flow decomposition} 
\label{sec:averdecomp}

The average of a quantity $q$ along the spanwise direction is

\begin{equation}
\lela q \rira(x,y,t)=\avz{q(x,y,\wh{z},t)},
\end{equation}   

where the integral is approximated via the Gauss-Legendre quadrature that takes into account the distribution of the GLL points. The time-ensemble average is

\begin{equation}
\wh{q}\ (x,y,z,\tau)=\avte{q}, \hspace{1cm} 0\leq \tau<  T,
\end{equation}

where $N$ is the number of periods, and the time average between times $t_i$ and $t_f$ is

\begin{equation}
\ov{q}(x,y,z)=\avt{q\left(x,y,z,\wh{t}\ \right)}.
\end{equation} 
A quantity $q$ of the flow field is decomposed as

\begin{equation}
q=q_m\ler x,y \rir+q_{osc}\ler x,y,\tau \rir+q_t\ler x,y,z,t \rir,
\end{equation}

where $q_m\ler x,y \rir=\lela \ov{q} \rira$ is the mean flow, $q_{osc}\ler x,y,\tau \rir=\lela \wh{q} \rira-q_m$ is the time-periodic flow, and $q_t=q-\lela \wh{q} \rira$ is the turbulent fluctuation.

\subsection{Parameters defining the actuation performance}
\label{sec:perfor}

For a separated flow the total drag force is composed of a pressure drag force $F_p^*$ and a skin-friction drag force $F_f^*$, defined as

\begin{equation}
F_p^*=\intsg{p^*\, \mb{i} \cdot \mb{n}}{S^*}, \hspace{1cm} F_f^*=-\intsg{\mu^* \frac{\p u^*}{\p \mb{n}}\bigg \vert_w}{S^*}, 
\label{eq:dragdef}
\end{equation}

where $\mb{n}$ is the normal unit vector pointing out of the fluid domain, $\mb{i}$ is the unit vector along the $x$ direction, $S^*$ is the solid surface, and the $w$ denotes the wall. 
For the vertical surfaces of the bar $\mb{n} \cdot \mb{i}\neq 0$, and therefore the pressure drag force is

\begin{equation}
F_p^*=\intsg{\ov{p^*}}{S^*_l}-\intsg{\ov{p^*}}{S^*_r},
\label{eq:pd}
\end{equation}

where $S^*_l$ and $S^*_r$ are the left and right vertical surfaces of the bar of area $h^*L_z^*$.
The skin-friction drag is

\begin{equation}
F_f^*=F^*_{ca}+F^*_{cr}=\intsg{\mu^* \frac{\p \ov{u^*}}{\p y^*}\bigg |_w }{S^*_{ca}}+\intsg{\mu^* \frac{\p \ov{u^*}}{\p y^*}\bigg |_w }{S^*_{cr}}, 
\label{eq:fricdef}
\end{equation}
 
where $S^*_{ca}$ and $S^*_{cr}$ are the cavity and crest surfaces, respectively. Note that the minus sign in \eqref{eq:dragdef} is absent in \eqref{eq:fricdef} because, for the bottom horizontal wall, $\mb{n}=-\mb{j}$, where $\mb{j}$ is the unit vector along the $y$ axis. The forces $F^*_{ca}$, $F^*_{cr}$, and $F_p^*$ include contributions from the bottom and top walls. The drag coefficients of the skin-friction along the cavity ($C_{ca}$), the skin-friction along the crest ($C_{cr}$), the pressure ($C_p$) are defined by dividing $F^*_{ca}$, $F^*_{cr}$, and $F_p^*$  by $\rho^*{U^*_p}^2 L^*_xL^*_z$, respectively. The skin-friction coefficient is composed of the crest and the cavity drag coefficients, i.e., $C_f=C_{cr}+C_{ca}$. 
We do not define the drag coefficients on the left and the right bar sides and their reductions because, as remarked by \cite{banchetti-etal-2020}, it is not meaningful to define a local pressure coefficient for an incompressible flow as the pressure values can be shifted by the any constant. 

We define a percentage reduction for each drag coefficient with respect to its value in the uncontrolled case, denoting these reductions with the letter $\mathcal{R}$ and the same subscript of the drag coefficients. For instance, the percentage reduction of the skin-friction coefficient along the cavity is 

\begin{equation}
\mc{R}_{ca}(\%)=100 \ler  1-\frac{{C_{ca}}_o}{C_{ca}} \rir,
\label{eq:0}
\end{equation}

where $C_{ca}$ and ${C_{ca}}_o$ are the uncontrolled and controlled values, respectively (the subscript $o$ stands for oscillating). The total drag reduction $\mc{R}(\%)$ is defined as

\begin{equation}
\mc{R}(\%)
=
100 \ler  1-  \frac{{F_p}_{o}^*+{F_f}_{o}^*}{F_p^*+F_f^*} \rir
=
100 \ler  1-  \frac{{C_f}_{o}+{C_p}_{o}}{{C_f}+{C_p}} \rir,
\label{eq:1}
\end{equation}
where the pressure drag force $F_p^*$ is defined in \eqref{eq:pd} and the skin-friction drag force $F_f^*$ is defined in \eqref{eq:fricdef}. \cite{banchetti-etal-2020} fixed the total pressure at the inflow location of their computational domain in order to quantify the effect of the pressure drop given by their control method. In all our cases, we instead maintain the pressure constant at the further end of our computational domain.

As we study an active technique, it is crucial to account for the power $\mc P_z(\%)$ supplied to the fluid for controlling the flow, defined as 

\begin{equation}
\mc P_z(\%)=\frac{100}{\ov{\Pi}_x U_b (t_f-t_i)} \int_{t_i}^{t_f} \Pi_z\left( \wh{t} \right)\; W_b\left( \wh{t} \right) \; {\dt}, 
\label{eq:2}
\end{equation}

expressed as a percentage of the power used for driving the fluid along the streamwise direction. In \eqref{eq:2}
$\ov{\Pi}_x$ is the mean pressure gradient driving the flow along $x$ in the uncontrolled case, $U_b$ is the constant volume-averaged velocity, $U_b=V^{-1} \intdv{\ov{u}}$, $W_b(t)$ is the spanwise volume-averaged velocity due to the control, $W_b=V^{-1} \intdv{w(x,y,z,t)}$, and $V$ is the fluid volume. Appendix \ref{sec:dragbalpow} presents the derivation for the power used to drive the fluid along $x$, $\mr{P}_x=-\ov{\Pi}_x U_b>0$. 

It is verified numerically that the force exerted by the pressure gradient to drive the fluid along the streamwise direction matches the sum of the drag forces, i.e., $-\ov{\Pi}_x L_y=C_f+C_p$ and $-{\ov{\Pi}_x}_o L_y={C_f}_o+{C_p}_o$.
By using this force balance, the percentage of power saved is
\begin{equation}
\mc{P}_{sav}(\%)
=
100 \ler 1- \frac{{\Pi_x}_o U_b}{{\ov{\Pi}_x} U_b}\rir
=
100 \ler 1- \frac{{\ov{\Pi}_x}_o}{{\ov{\Pi}_x} }\rir
=
100 \ler  1-  \frac{{C_f}_{o}+{C_p}_{o}}{{C_f}+{C_p}} \rir=\mc{R}(\%),
\end{equation}
because the volume-averaged velocity $U_b$ is constant.

The difference between the power $\mc{P}_{sav}(\%)$ saved through the oscillating pressure gradient and the power $\mc{P}_z(\%)$ spent for controlling the flow is the net power saved:

\begin{equation}
\mc{P}_{net}(\%)=\mc{R}(\%)-\mc{P}_z(\%).
\label{eq:3}
\end{equation}

\section{Laminar-flow results}
\label{sec:laminar}
Previous studies \citep{ricco-quadrio-2008,quadrio-ricco-2011,ricco-hahn-2013} showed that useful information about control in the turbulent case can be extracted from the corresponding laminar flow solutions. Here, for the first time, we apply the same strategy to the study of a separated flow. We consider the time-ensemble and $z$-averaged spanwise momentum equation

\begin{equation}
\frac{\p \avtez{w}}{\p \tau}+\frac{\p \avtez{u w}}{\p x}+ \frac{\p \avtez{v w}}{\p y}=-\frac{1}{L_z}\les \wh{p}(x,y,L_z,\tau)-\wh{p}(x,y,0,\tau)\ris+\frac{1}{Re_p}\ler \frac{\p^2 \avtez{w}}{\p x^2}+\frac{\p^2 \avtez{w}}{\p y^2} \rir,
\label{eq:eq4}
\end{equation}

where 

\begin{equation}
-\frac{1}{L_z}\les \wh{p}(x,y,L_z,\tau)-\wh{p}(x,y,0,\tau)\ris=
-\biggl\langle \;\frac{\p \wh{p}}{\p z} \;\biggr\rangle.
\end{equation}

Substitution of the pressure $p=\phi+\ov{\Pi}_x\, (L_x - x) + \Pi_z\, z$, the Reynolds decomposition $w=\avtez{w} +w_t$,  and the $z$-averaged continuity equation

\begin{equation} 
\frac{\p \avtez{u}}{\p x}+\frac{\p \avtez{v}}{\p y}=0
\end{equation}

into \eqref{eq:eq4} leads to

\begin{equation} 
\frac{\p \avtez{w}}{\p \tau}=\frac{1}{Re_p}\ler \frac{\p^2 \avtez{w}}{\p x^2}+\frac{\p^2 \avtez{w}}{\p y^2} \rir-\Pi_z-\avtez{u}\frac{\p \avtez{w}}{\p x} -\avtez{v}\frac{\p \avtez{w}}{\p y}  -\frac{\p \avtez{u_t w_t}}{\p x}- \frac{\p \avtez{v_t w_t}}{\p y}.
\label{eq:5}
\end{equation}

\cite{ricco-quadrio-2008} showed that, for a smooth channel with oscillating walls, there is a very good agreement between the $x$- and $z$-averaged spanwise velocity profile and the laminar Stokes layer profile because the term $\p \avtez{v_t w_t}/\p y$ is negligible in the spanwise momentum equation. Following \cite{ricco-quadrio-2008}, we assume that $\p \avtez{v_t w_t}/ \p y$ and $\p \avtez{u_t w_t}/\p x$ are negligible and obtain 

\begin{equation}
\frac{\p w_l}{\p \tau}=\frac{1}{Re_p} \ler \frac{\p^2 w_l}{\p x^2}+\frac{\p^2 w_l}{\p y^2} \rir -\Pi_z- \ler \avtez{u}  \frac{\p w_l}{\p x} +\avtez{v} \frac{\p w_l}{\p y} \rir,
\label{eq:stokes}
\end{equation}

where $w_l$ is a laminar approximation to $\avtez{w}$.
We further note that the velocities $\avtez{u}$ and $\avtez{v}$ are independent of $w_l$ as shown by the time-ensemble and $z$-averaged momentum equations, 

\begin{subequations}
\begin{equation} 
\frac{\p \avtez{u}}{\p \tau}=\frac{1}{Re_p} \ler \frac{\p^2 \avtez{u}}{\p x^2}+\frac{\p^2 \avtez{u}}{\p y^2} \rir-\ler \frac{\p \avtez{uu}}{\p x}+\frac{\p \avtez{uv}}{\p y} \rir-\frac{\p \avtez{p}}{\p x},          
\label{eq:xmom}
\end{equation} 
\begin{equation} 
\frac{\p \avtez{v}}{\p \tau}=\frac{1}{Re_p} \ler \frac{\p^2 \avtez{v}}{\p x^2}+\frac{\p^2 \avtez{v}}{\p y^2} \rir-\ler \frac{\p \avtez{uv}}{\p x}+\frac{\p \avtez{vv}}{\p y} \rir-\frac{\p \avtez{p}}{\p y}.     
\label{eq:ymom}
\end{equation} 
\end{subequations}
We assume that $w_l$ is a periodic function of $\tau$ and thus Fourier series expansions can be employed. The quantities $\Pi_z$ and $w_l$ are expressed as 

\begin{equation}
w_l=\sum_{n=-\infty}^{+\infty}\frac{\wh{w}_n}{2} e^{i n  \omega \tau}, \hspace{1cm} \Pi_z=A \cos\ler \omega  \tau \rir =\frac{A}{2} \ler  e^{i\omega \tau}+e^{-i\omega \tau} \rir,
\label{eq:fourier1}
\end{equation}

where $\omega=2\pi/T$.
Assuming that the mean velocities $\avtez{u}$ and $\avtez{v}$ do not depend on $\tau$, substitution of \eqref{eq:fourier1} into \eqref{eq:stokes} leads to

\begin{equation}
\begin{aligned}
\sum_{n=-\infty}^{+\infty}
&
\les\frac{i\;n\;\omega}{2} \wh{w}_n -\frac{1}{Re_p}\ler \frac{\p^2 \wh{w}_n}{\p x^2}+\frac{\p^2 \wh{w}_n}{\p y^2} \rir+\avtez{u} \frac{\p \wh{w}_n}{\p x}+\avtez{v}\frac{\p  \wh{w}_n}{\p y} \ris e^{i\,n\,\omega\,\tau}
=\\
&-\frac{A}{2} \ler e^{i\,\omega\,\tau}+e^{-i\,\omega\,\tau}  \rir.
\label{eq:fourier2}
\end{aligned}
\end{equation}
The assumptions of the neglect of the Reynolds stresses, the steadiness of $\avtez{u}$ and $\avtez{v}$, and the periodicity of $w_l$ on $\tau$ are confirmed by the turbulent-flow numerical calculations, discussed in \S\ref{sec:flowstatistics}. 
The modes $\wh{w}_n$ satisfy:

\begin{subequations}
\begin{equation} 
i\,n\,\omega  \wh{w}_n =\frac{1}{Re_p}\ler  \frac{\p^2 \wh{w}_n}{\p x^2} + \frac{\p^2 \wh{w}_n}{\p y^2}\rir -\avtez{u}\frac{\p \wh{w}_n}{\p x}-\avtez{v}\frac{\p \wh{w}_n}{\p y}, \hspace{1cm} n\neq \pm 1,
\end{equation}
and
\begin{equation}
\ler \pm  i\,\omega \rir \wh{w}_{\pm 1} =\frac{1}{Re_p}\ler  \frac{\p^2 \wh{w}_{\pm 1}}{\p x^2}+ \frac{\p^2 \wh{w}_{\pm 1}}{\p y^2} \rir-\avtez{u}\frac{\p \wh{w}_{\pm 1}}{\p x}-\avtez{v}\frac{\p \wh{w}_{\pm 1}}{\p y} -A.
\label{eq:helmolts1}
\end{equation}
\end{subequations}

The boundary conditions for \eqref{eq:helmolts1} are 

\begin{subequations}
\begin{equation}
\wh{w}_{\pm 1}\ler 0,y \rir =\wh{w}_{\pm 1}\ler h,y \rir=0,  \hspace{1cm}   0\leq y\leq h,
\label{eq:bca}
\end{equation}
\begin{equation}
\wh{w}_{\pm 1}\ler x,h \rir=\wh{w}_{\pm 1}\ler x,0\rir=0,   \hspace{1cm}   h\leq x\leq L_x,
\label{eq:bcb}
\end{equation}
\begin{equation}
\wh{w}_{\pm 1}\ler h,y\rir=0,  \hspace{1cm}   0\leq y\leq h,
\label{eq:bcc}
\end{equation}
\begin{equation}
\frac{\p \wh{w}_{\pm 1}}{\p y}\ler x,1+h \rir =0,  \hspace{1cm} 0\leq x\leq L_x,
\label{eq:bcd}
\end{equation}
\begin{equation}
\wh{w}_{\pm 1} \ler 0,y \rir=\wh{w}_{\pm 1} \ler L_x,y \rir, \hspace{1cm} 0\leq y\leq 1+h.
\label{eq:bce}
\end{equation}
\end{subequations}

Conditions \eqref{eq:bca}, \eqref{eq:bcb}, \eqref{eq:bcc} are the no-slip conditions at the walls, condition \eqref{eq:bcd} expresses the symmetry along the centerline, and \eqref{eq:bce} denotes the periodicity along $x$. The only non-null modes are those for which $n=\pm 1$ and from \eqref{eq:fourier1} it follows that 

\begin{equation}
w_l\ler x,y,\tau \rir=\frac{1}{2}\ler \wh{w}_{1}\,e^{i\,\omega\,\tau}+\wh{w}_{-1}\,e^{-i\,\omega\,\tau}  \rir.
\label{eq:hermite}
\end{equation}

The spanwise velocity $w_l\ler x,y,\tau \rir $ is real and thus $\wh{w}_{1}=\wh{w}_{-1}^{cc}$ because of the Hermitian property (the superscript $cc$ indicates the complex conjugate).

In order to eliminate the non-homogeneous term $A$ from \eqref{eq:helmolts1}, we introduce $\wtil{w}_1\ler x,y \rir=\wh{w}_1\ler x,y \rir- A i/ \omega $, which transforms \eqref{eq:helmolts1} into 

\begin{equation}
\frac{\p^2 \wtil{w}}{\p x^2}+ \frac{\p^2 \wtil{w}}{\p y^2}-i\,\omega\,Re_p \wtil{w}-Re_p\ler \avtez{u}\frac{\p \wtil{w}}{\p x}+\avtez{v}\frac{\p \wtil{w}}{\p y} \rir=0,
\label{eq:helmolts2}
\end{equation}

where the subscript of $\wtil{w}_1$ is removed for clarity.
The boundary conditions for \eqref{eq:helmolts2} are 

\begin{subequations}
\begin{equation}
\wtil{w}\ler 0,y\rir=\wtil{w}\ler h,y\rir =-\frac{A}{\omega}\, i,  \hspace{1cm}   0\leq y\leq h,
\label{eq:bc_stokes_1}
\end{equation}
\begin{equation}
\wtil{w}\ler x,h \rir =\wtil{w}\ler x,0 \rir=-\frac{A}{\omega}\, i,  \hspace{1cm}   h\leq x\leq L_x,
\end{equation}
\begin{equation}
\wtil{w}\ler h,y \rir=-\frac{A}{\omega}\, i,  \hspace{1cm}   0\leq y\leq h,
\end{equation}
\begin{equation}
\frac{\p \wtil{w}}{\p y}\ler x,1+h\rir=0,  \hspace{1cm} 0\leq x\leq L_x,
\label{eq:bc_stokes_4}
\end{equation}
\begin{equation}
\wtil{w} \ler 0,y\rir=\wtil{w} \ler L_x,y\rir, \hspace{1cm} 0\leq y\leq 1+h.
\label{eq:bc_stokes_5}
\end{equation}
\end{subequations}

The laminar power spent is   

\begin{equation}
\mc{P}_{z,l}=-\frac{1}{T}\int_0^T \Pi_z(\tau)\; {W_b}_l(\tau) \; {\dta},
\label{eq:eq14}
\end{equation}

where the laminar volume-averaged velocity ${W_b}_l=V^{-1} \intdv{w_l(x,y,\tau)}$ can be simplified by using \eqref{eq:hermite},

\begin{equation}
{W_b}_l=\frac{1}{2 V}\intdv{\ler \wh{w}_{1}\,e^{i\,\omega\,\tau}+\wh{w}^{cc}_{1}\,e^{-i\,\omega\,\tau} \rir}
=
\ler \wh{w}_b e^{i\,\omega\,\tau}+\wh{w}^{cc}_b e^{-i\,\omega\,\tau}  \rir/2,
\label{eq:eq15}
\end{equation}  

where $\wh{w}_b=V^{-1} \intdv{\wh{w}_1}$. Substitution of \eqref{eq:eq15} and the second of \eqref{eq:fourier1} into \eqref{eq:eq14} leads to

\begin{equation}
\mc{P}_{z,l}=-\frac{A}{4}\left( \wh{w}_b+\wh{w}^{cc}_b \right). 
\label{eq:eq16}
\end{equation}

By introducing $\wh{w}_b=A\wh{w}_{u,b}$, where $\wh{w}_{u,b}$ is the value of $\wh{w}_b$ obtained when $A=1$, expression \eqref{eq:eq16} can be written in terms of $\wh{w}_u$ as

\begin{equation}
\mc{P}_{z,l}=-\frac{A^2}{4} \ler \wh{w}_{u,b} +\wh{w}^{cc}_{u,b} \rir=A^2 P_{z,lu},
\label{eq:eq18}
\end{equation}

where $P_{z,lu}$ is the laminar power obtained for $A=1$.
The laminar power \eqref{eq:eq18} can also be expressed as the percentage of the power employed for driving the uncontrolled turbulent flow along $x$,

\begin{equation}
\mc{P}_{z,l}(\%)=- \frac{100 \mc{P}_{z,l}}{\ov{\Pi}_x U_b}=\frac{25 A^2}{\ov{\Pi}_x U_b} \ler \wh{w}_{u,b} +\wh{w}^{cc}_{u,b} \rir.
\label{eq:eq19}
\end{equation}  

Equation \eqref{eq:helmolts2} is solved numerically via a finite-difference method, as described in Appendix \ref{sec:numhelm}. The laminar and the averaged turbulent flow solutions are compared in \S\ref{sec:flowstatistics}.
We also solve equation \eqref{eq:helmolts2} by neglecting the last two terms in parenthesis on the left hand side. We compare the solutions in these two cases to quantify the effect of the advective terms on $w_l$ to find the locations where these terms are important and to determine whether they have an impact on the laminar power spent $\mc{P}_{z,l}$. 

\begin{figure}
\includegraphics[width=0.45\textwidth]{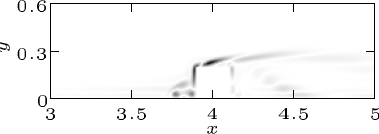}
\includegraphics[width=0.485\textwidth]{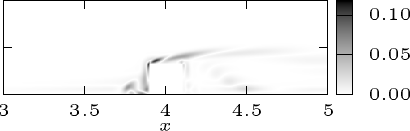}
\caption{Absolute values of the differences between the solution $\wh{w}_{\pm 1}$ of \eqref{eq:helmolts1} including or neglecting the advective terms for $A=1$ and $T=2.3$. The contour of the real parts is depicted on the left, while the contour of the imaginary parts is shown on the right.}  
\label{fig:lamsol}
\end{figure}

Figure \ref{fig:lamsol} shows the absolute values of the differences between the real parts (left) and the imaginary parts (right) of the solutions obtained including and excluding the advective terms. The contours are more intense in front of the cavity and on the crest, while a mild disagreement occurs around the separated region.

For a smooth channel, it is possible to find an analytical expression relating the power $\mc{P}_{z,l}$ to the forcing parameters $A$ and $T$ \citep{ricco-quadrio-2008}. We herein use the analytical expression for $\mc{P}_{z,l}$ in the smooth channel case without bars to obtain an empirical relationship linking the power spent $\mc{P}_{z,l}$ to the period $T$. In the smooth channel case, the streamwise viscous term and the spatial advection terms are null. The streamwise viscous term is zero because the boundary conditions are independent of the $x$ direction. The laminar flow thus satisfies

\begin{equation}
\begin{cases}
\frac{\mr{d}^2 \wtil{w}}{\dy^2}-i\omega Re_p\wtil{w}=0, \\
\wtil{w}(0)=\wtil{w}(2)=-\frac{A\,i}{\omega}.
\end{cases}
\label{eq:smoothlam1}
\end{equation}

By imposing the boundary conditions on the general solution $\wtil{w}=C_1e^{\xi y}+C_2e^{-\xi y}$ and using $\wh{w}_1=\wtil{w}+A i/\omega$, one finds

\begin{equation}
\wh{w}=\frac{A i}{\omega} \les 1-\frac{e^{\xi (y-2)}+e^{-\xi y}}{1+e^{-2\xi}} \ris,
\label{eq:smoothlam2}
\end{equation}

where $\xi=\sqrt{i\omega Re_p}$.
The integrated velocity $\wh{w}_b$ for the smooth channel is therefore

\begin{equation}
\wh{w}_b=\frac{1}{2} \int_0^{2} \wh{w} \dy=\int_0^{1} \wh{w} \dy=\frac{A i}{\omega} \les 1- \frac{1-e^{-2 \xi}}{\xi(1+e^{-2 \xi})} \ris.
\label{eq:smoothlam3}
\end{equation}

By noting that $\wh{w}_b+\wh{w}^{cc}_b$ is twice the real part of $\wh{w}_b$, we obtain 

\begin{equation}
\wh{w}_b+\wh{w}^{cc}_b=-\frac{2A}{\omega} \maf{Re} 
\left[ \frac{i\left(1-e^{-2 \xi}\right)}{\xi \left(1+e^{-2 \xi}\right)}\right]
=-\frac{\sqrt{2}A}{\omega^{3/2} \sqrt{Re_p}}\maf{Re}\les (i+1)  \frac{1-e^{-2 \xi}}{1+e^{-2 \xi}}\ris,
\label{eq:smoothlam4}
\end{equation}

where $\maf{Re}$ indicates the real part. When $2\sqrt{\omega Re_p}\gg 1$, which is always satisfied in the studied cases, equation \eqref{eq:smoothlam4} can be approximated by 

\begin{equation}
\wh{w}_b+\wh{w}^{cc}_b\approx-\frac{\sqrt{2}A}{\omega^{3/2} \sqrt{Re_p}}.
\label{eq:smoothlam5}
\end{equation}

\begin{figure}
\centering
\includegraphics[width=0.5\textwidth]{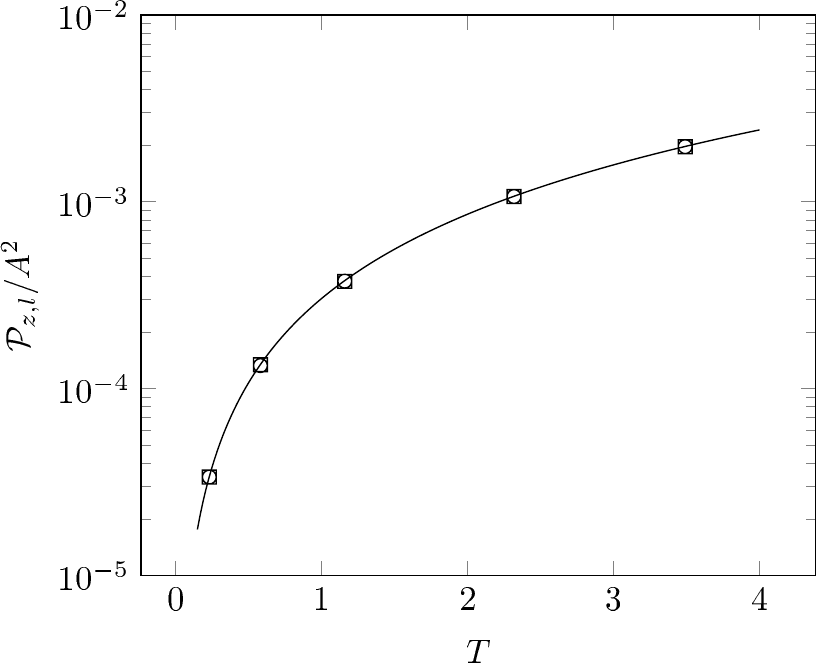}
\caption{Power spent $\mc{P}_{l}/A^2$ as a function of $T$ for \eqref{eq:helmolts2} with (squares) and without (circles) the mean advective terms. The solid line represents the function $\mc{P}_{l}/A^2 =3.03 \cdot 10^{-4} T^{1.5}$.}  
\label{fig:lamenergy}
\end{figure}

The laminar power spent for a smooth channel can be obtained by substituting \eqref{eq:smoothlam5} into \eqref{eq:eq16} to find

\begin{equation}
\mc P_{sc}=\frac{\sqrt{2} A^2\,\omega^{-3/2}}{4 \sqrt{Re_p}}=\frac{A^2 \, T^{3/2}}{8 \pi^{3/2}\sqrt{Re_p}},
\label{eq:smoothlam6}
\end{equation}

where $sc$ stands for smooth channel. Equation \eqref{eq:smoothlam6} suggests that an expression proportional to $A^2 T^\gamma$ could be an approximation to $\mc{P}_{l}$. Best-line fitting of the laminar data points gives

\begin{equation}
\mc{P}_{z,l}(\%) \approx -\frac{3.03\cdot 10^{-2} A^2 T^{3/2}}{\ov{\Pi}_x U_b}.
\label{eq:lampower}
\end{equation}

The same exponent $\gamma=3/2$ as in the smooth channel case is found. Expression \eqref{eq:lampower} is plotted in figure \ref{fig:lamenergy} together with the values obtained via the laminar solution. The agreement is excellent, proving that the mean advection and the mean viscous diffusion along $x$ have a small impact on $\mc{P}_{z,l}$, as assumed when deriving \eqref{eq:smoothlam6}. 
\cite{ricco-quadrio-2008} found that the laminar power spent to oscillating the wall is proportional to $T^{-1/2}$, while we find that $\mc{P}_{z,l}$ is proportional to $T^{3/2}$ in our case. The reason of this apparent discrepancy is explained by considering a transformation relating the two laminar problems, as discussed in Appendix \ref{app:powoscwall}.

\section{Turbulent-flow performance quantities}
\label{sec:uncon}

In this section, we present results on the uncontrolled turbulent flow and the performance parameters in the controlled cases.

\subsection{Uncontrolled flow}
\begin{figure}
\hspace{3mm}
\includegraphics[width=0.83\textwidth,height=0.2344\textwidth]{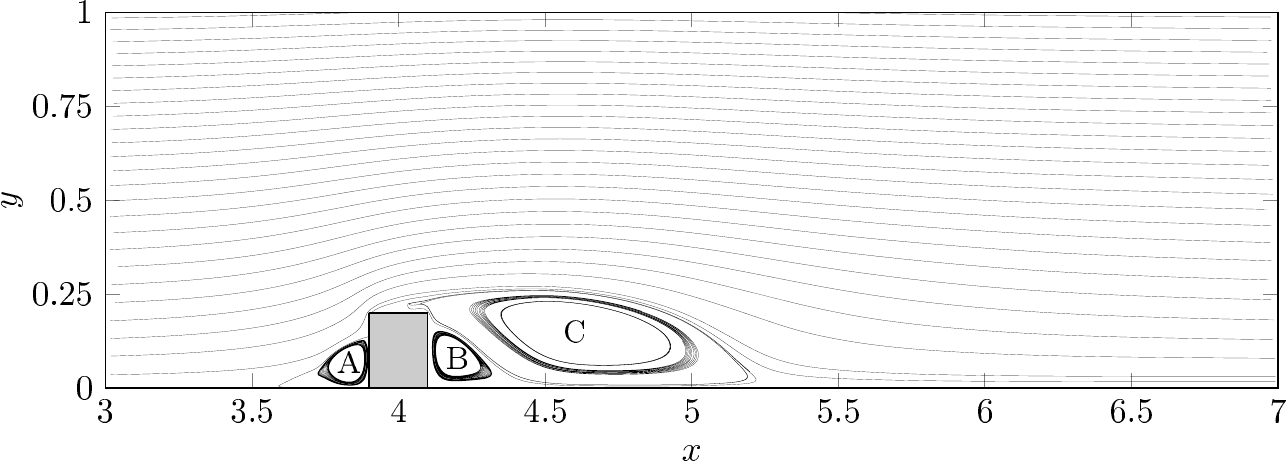}
\caption{Spanwise- and time-averaged streamlines for the flow without control. The recirculation vortices are indicated by the letters A, B, and C.}
\label{fig:streamlines}
\end{figure}

The mean-flow streamlines in the region surrounding the bar in the uncontrolled flow case are depicted in figure \ref{fig:streamlines}. The separation zone is occupied by three recirculating vortices: two of them, A and B, are adjacent to the vertical surfaces of the bar and have a size comparable with the bar height. Vortex C is located downstream and on top of vortex B, extending for a length of more than five bar heights along the streamwise direction and covering the crest where a mild separation occurs. Vortices A and C rotate clockwise, while vortex B rotates anticlockwise. Sufficiently far from the bar ($x<3$ and $x>6.5$) the flow is not influenced by the obstacle and the mean flow streamlines become straight and aligned with the streamwise direction. We define the reattachment points as the positions where the time and spanwise-averaged wall-shear stress is zero. They separate the forward and backward mean flows near the wall and are found at $x=3.56, 4.39, 5.38$. The mean streamlines agree visually with those of \cite{leonardi-etal-2003} and \cite{ikeda-durbin-2007}, although their Reynolds number is slightly larger.

\begin{figure}
\centering
\includegraphics[width=0.95\textwidth]{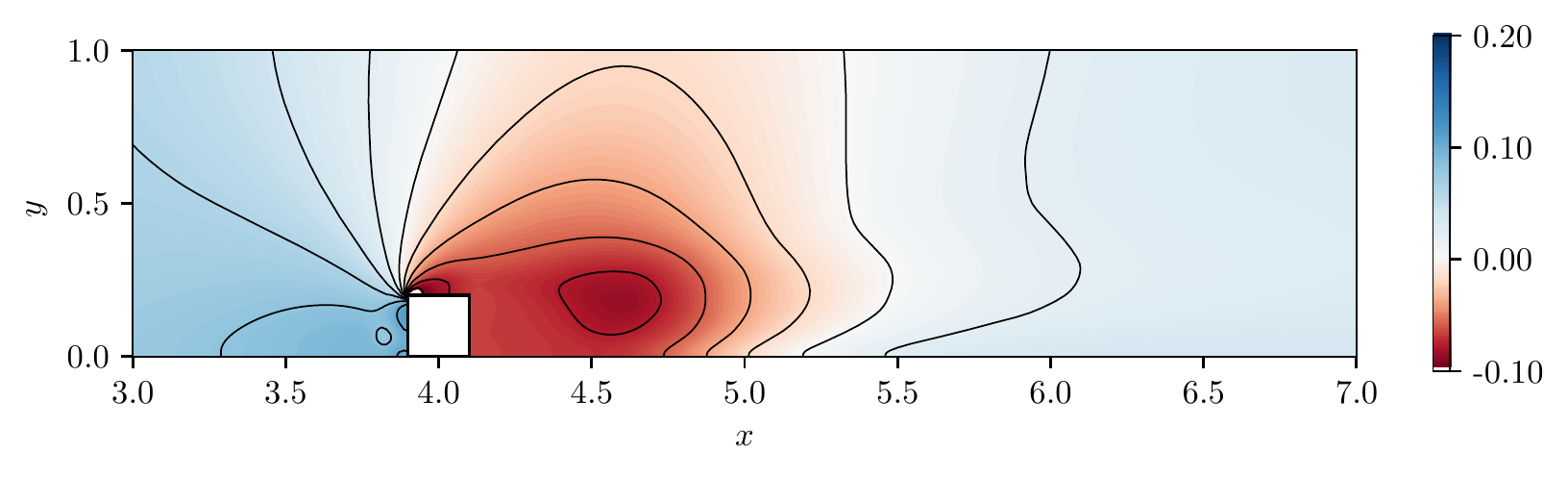}
\caption{Colour map and contour lines of the mean pressure in the uncontrolled case. The isoline values are equispaced at intervals of $0.02$ in the range $[-0.25,0.1]$.}
\label{fig:fig4}
\end{figure}

\begin{figure}
\centering
\includegraphics[width=0.85\textwidth]{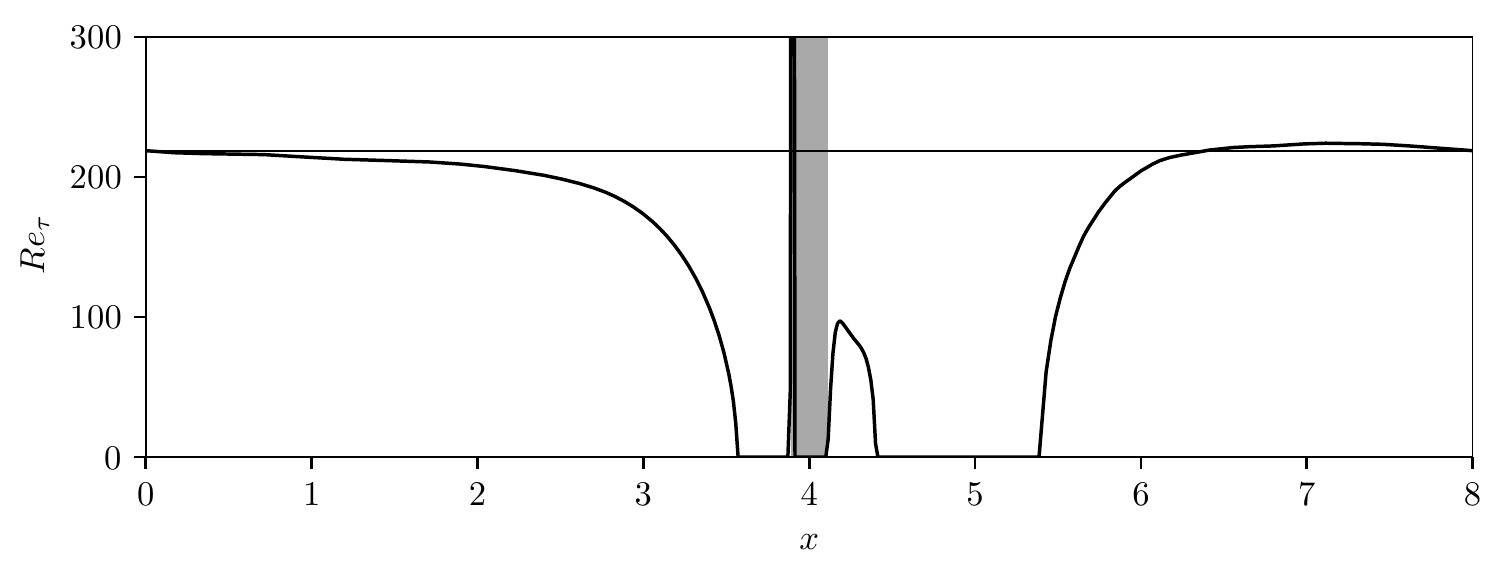}
\caption{Local friction Reynolds number along the channel in the uncontrolled case. The location of the bar is marked by the shaded region. The thin solid line denotes the friction Reynolds number $Re_{\tau}=218$, averaged in the constant-friction region, $x<2$ and $x>6.5$.}
\label{fig:channelRetau}
\end{figure}

Figure \ref{fig:fig4} shows the mean pressure of the uncontrolled flow. The pressure reaches high values right upstream of the bar where vortex A occurs, with the maximum pressure occurring along the left vertical surface of the bar. The core of vortex C, located approximately at $x=4.6$, is characterized by a low pressure.
The effect of the bar on the wall-shear stress is clearly shown in figure~\ref{fig:channelRetau} by the distribution of the local friction Reynolds number $Re_\tau=u_\tau^* (h^*+\delta^*)/\nu^*$ as a function of the streamwise direction, where $u_\tau^*=\sqrt{\tau_w^*/\rho^*}$ is the local wall-friction velocity and $\tau_w^*$ is the time- and spanwise-averaged wall-shear stress. The Reynolds number $Re_\tau$ is only computed where the wall-shear stress is positive. 
It varies significantly in the proximity of the bar, while it is almost constant at $Re_{\tau}=218$ away from the bar, for $x<2$ and $x>6$. It must be remarked that scaling quantities with the local wall-friction velocity is only meaningful in the region where the flow is fully attached and relatively undisturbed from the separation regions so that the skin-friction approaches a positive constant. It is instead not appropriate to employ wall-friction scaling in the recirculating regions where the skin-friction is either vanishingly small or negative. The constant wall-friction velocity in the fully attached region, $u_\tau=0.044$, is used to scale quantities in wall units, as follows

\begin{align}
A^+ &= \frac{A^* \nu^*}{\rho^* u_{\tau}^{*3}} = \frac{A(h+1)^3 Re_p^2}{h Re_\tau^3}, \\
T^+ &= \frac{T^* u_{\tau}^{*2}}{\nu^*} = \frac{T Re_\tau^2}{(h+1)^2 Re_p}, \\
J^+ &= \frac{J^*}{\rho^* u^*_{\tau}} = \frac{J (h+1) Re_p}{Re_\tau}.
\end{align}
Scaling the forcing parameters in wall units is useful to compare the performance of the actuation in reducing the skin-friction drag of the attached flow along the cavity with the widely studied spanwise wall oscillation technique \citep{quadrio-ricco-2004}. We also find that the uncontrolled scaled wall-shear stress of the attached flow away from the bars, $Re_p^{-1}\mathrm{d}U/\mathrm{d}y\vert_{y=0}=0.0034$, is less that 1\% different from the coefficient found from the empirical correlation $Re_p^{-1}\mathrm{d}U/\mathrm{d}y\vert_{y=0}=0.0336 U_b^2 Re_\tau^{-0.273}$ \citep{pope-2000}, valid for fully-developed turbulent channel flows over flat walls.


\subsection{Skin-friction and pressure drag coefficients} 
\label{sec:data}

As the current control strategy aims at reducing both the viscous drag along the streamwise-parallel surfaces and the significant form drag experienced by the bar because of the recirculation regions, it is central to precisely quantify the drag coefficients and their changes in the controlled cases. 

\begin{figure}
\begin{tabular}{cc}
\includegraphics[width=0.5\textwidth]{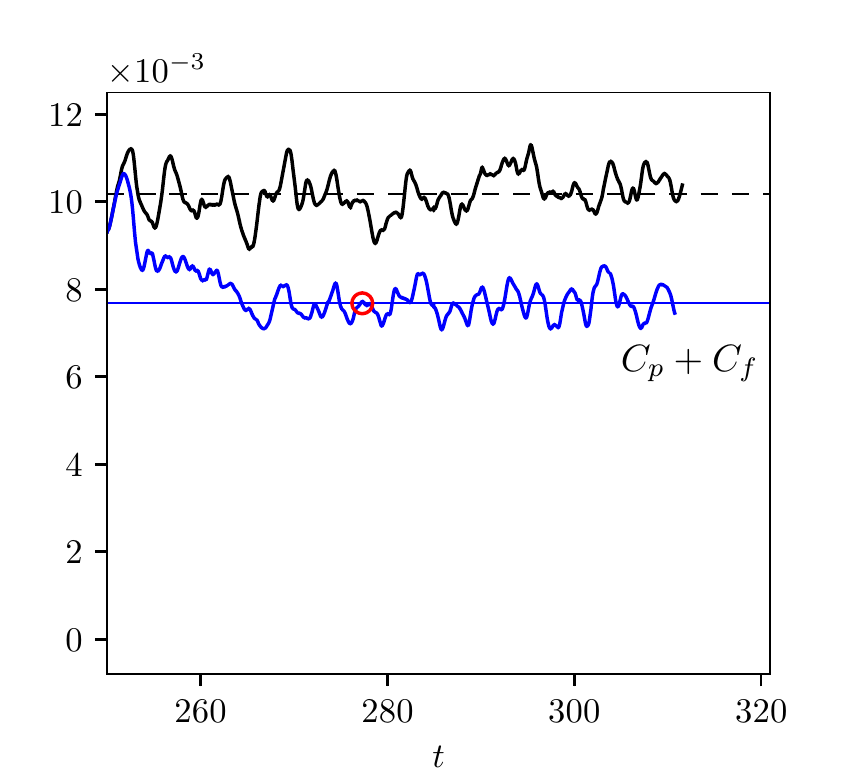} &
\includegraphics[width=0.5\textwidth]{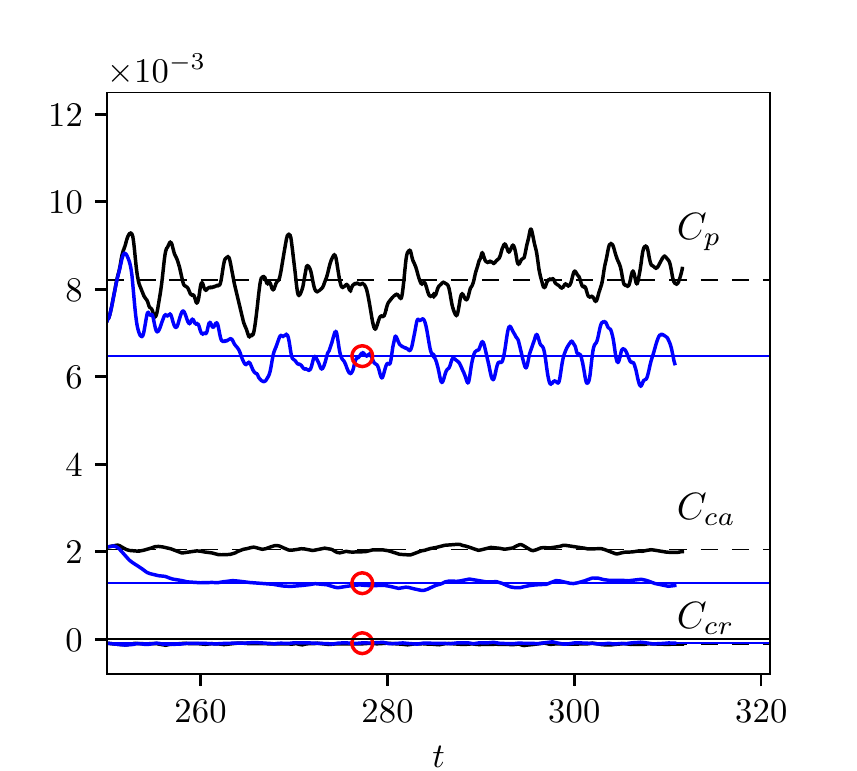}\\
\end{tabular}
\caption{Left: time evolutions of the total drag coefficient $C_p+C_f$. Right: time evolutions of the pressure-drag coefficients $C_{p}$ and the skin-friction coefficients $C_{cr}$ and $C_{ca}$. The control parameters are $T^+=50$ and $J^+=50$. The control is activated at $t=250$ and the circles denote the start of averaging. The horizontal dashed and solid lines indicate the mean uncontrolled and controlled coefficients, respectively.}
\label{fig:fig5}
\end{figure}

The effect of the control on the drag coefficients for a large drag-reduction case with $T=6.33$ and $A=1.08$ ($J^+=50$, $T^+=50$) is presented in figure \ref{fig:fig5}. The flow responds quickly to the control activation at $t=250$ and the coefficients adjust to the new value monotonically. This transient response is shorter and smoother than in the case of spanwise wall oscillations, where the wall-shear stress may evolve non-monotonically and reach a very low instantaneous value before adjusting to the new fully developed value \citep{quadrio-ricco-2003}. The time-averaged coefficients are computed after the end of the transient phase, estimated at around $t=270$, i.e., when all the coefficients reach a new statistically steady state. The control guarantees a substantial reduction of total drag $C_p+C_f$ due to a reduction of both $C_p$ and $C_f$. This result proves that a spanwise oscillating pressure gradient can reduce the skin-friction drag of a cavity flow and the pressure drag given by the difference on the bar sides in the integrated pressures of a fully separated flow. 
The drop of total pressure drag may not be achieved by the traditional separation control methods, such as vortex generators and jets, which may introduce a viscous drag penalty due to the enhanced flow mixing. The time histories of the pressure coefficient is much more oscillatory than that of the skin-friction coefficient, which is expected because the pressure drag is affected by the intense fluctuations in the separated region downstream of the bar.
The skin-friction reduction is mostly due to the decrease of the cavity drag. The friction drag on the crest is very small and negative because of the reversed flow in the small recirculation region occurring due to separation. The spanwise forcing has a marginal effect on the crest friction.

\begin{table}
\begin{center}
\begin{tabular}{ccc|cc|cc|c|c|c}
\hline
 $J^+$ &  $T^+$ &  $A^+$ &   $T$  &   $A$ &  $C_{ca}$  &  $C_{cr}$ &  $C_f$  &  $C_p$ &  $\ov{\Pi}_x L_y$ \\
       &        &        &        &       &  $\times 10^{-3}$ & $\times 10^{-4}$ & $\times 10^{-3}$ & $\times 10^{-3}$ &   $\times 10^{-3}$ \\
 \hline
  0.00 &   $\infty$ &   0.00 &   $\infty$ &  0.00 & 2.01 & -1.10 & 1.89 &   8.11 &  10.00 \\
  0.14 &   1.57 &   0.29 &  0.20 &  0.10 & 2.04 & -1.08 & 1.93 &   8.06 &   9.99 \\
  0.73 &   1.57 &   1.46 &  0.20 &  0.50 & 1.94 & -1.08 & 1.84 &   8.06 &   9.90 \\
  1.46 &   1.57 &   2.92 &  0.20 &  1.00 & 2.02 & -1.09 & 1.91 &   8.05 &   9.96 \\
  3.21 &   1.57 &   6.43 &  0.20 &  2.20 & 2.03 & -1.08 & 1.92 &   8.18 &  10.10 \\
 14.60 &   1.57 &  29.21 &  0.20 & 10.00 & 2.04 & -1.09 & 1.93 &   8.01 &   9.94 \\
  4.40 &   4.73 &   2.92 &  0.60 &  1.00 & 2.03 & -1.10 & 1.92 &   8.18 &  10.10 \\
  4.40 &   9.47 &   1.46 &  1.20 &  0.50 & 1.95 & -1.08 & 1.84 &   7.90 &   9.74 \\
  8.80 &   9.47 &   2.92 &  1.20 &  1.00 & 1.91 & -1.09 & 1.80 &   8.01 &   9.81 \\
 16.88 &  18.16 &   2.92 &  2.30 &  1.00 & 1.84 & -1.10 & 1.73 &   7.83 &   9.56 \\
 50.64 &  18.16 &   8.76 &  2.30 &  3.00 & 1.55 & -1.08 & 1.44 &   7.35 &   8.79 \\
 15.00 &  15.00 &   3.14 &  1.90 &  1.08 & 1.92 & -1.09 & 1.81 &   7.93 &   9.74 \\
 15.00 &  30.00 &   1.57 &  3.80 &  0.54 & 1.76 & -1.08 & 1.65 &   7.66 &   9.31 \\
 15.00 &  50.00 &   0.94 &  6.33 &  0.32 & 1.73 & -1.10 & 1.62 &   7.54 &   9.16 \\
 15.00 &  80.00 &   0.59 & 10.13 &  0.20 & 1.75 & -1.05 & 1.65 &   7.18 &   8.83 \\
 15.00 & 100.00 &   0.47 & 12.67 &  0.16 & 1.77 & -1.07 & 1.66 &   7.41 &   9.07 \\
 15.00 & 150.00 &   0.31 & 19.00 &  0.11 & 1.81 & -1.05 & 1.71 &   7.40 &   9.11 \\
 50.00 &  18.00 &   8.73 &  2.28 &  2.99 & 1.49 & -1.09 & 1.39 &   7.35 &   8.73 \\
 50.00 &  30.00 &   5.24 &  3.80 &  1.79 & 1.34 & -1.04 & 1.24 &   6.93 &   8.17 \\
 50.00 &  40.00 &   3.93 &  5.07 &  1.34 & 1.32 & -0.98 & 1.22 &   6.74 &   7.96 \\
 \bf{50.00} &  \bf{50.00} &   \bf{3.14} &  \bf{6.33} &  \bf{1.08} & \bf{1.28} & \bf{-0.92} & \bf{1.19} &   \bf{6.47} &   \bf{7.66} \\
 50.00 &  80.00 &   1.96 & 10.13 &  0.67 & 1.27 & -0.85 & 1.18 &   6.30 &   7.49 \\
 50.00 & 100.00 &   1.57 & 12.67 &  0.54 & 1.30 & -0.85 & 1.22 &   6.34 &   7.56 \\
 50.00 & 150.00 &   1.05 & 19.00 &  0.36 & 1.43 & -0.90 & 1.34 &   6.66 &   8.00 \\
\hline
\end{tabular}
\end{center}
\caption{Control parameters and mean drag coefficients. In this table and in tables \ref{tab:tab2} and \ref{tab:lampower}, the case highlighted in bold is studied in \S\ref{sec:physics}.}
\label{tab:tab1}
\end{table}
\begin{table}
\begin{center}
\begin{tabular}{ccc|cc|cc|c|c|c}
\hline
$J^+$  & $T^+$  & $A^+$ & $T$ & $A$ & $\mc{R}_{ca}(\%)$ & $\mc{R}_{cr}(\%)$ & $\mc{R}_f(\%)$ & $\mc{R}_{p}(\%)$ &  $\mc{R}(\%)$ \\
\hline
  0.14 &    1.57 &   0.29 &   0.20 &   0.10 & -1.5 &  1.8 & -2.1 &  0.6 &  0.1 \\
  0.73 &    1.57 &   1.46 &   0.20 &   0.50 &  3.5 &  1.8 &  2.6 &  0.6 &  1.0 \\
  1.46 &    1.57 &   2.92 &   0.20 &   1.00 & -0.5 &  0.9 & -1.1 &  0.7 &  0.4 \\
  3.21 &    1.57 &   6.43 &   0.20 &   2.20 & -1.0 &  1.8 & -1.6 & -0.9 & -1.0 \\
 14.60 &    1.57 &  29.21 &   0.20 &  10.00 & -1.5 &  0.9 & -2.1 &  1.2 &  0.6 \\
  4.40 &    4.73 &   2.92 &   0.60 &   1.00 & -1.0 & -0.0 & -1.6 & -0.9 & -1.0 \\
  4.40 &    9.47 &   1.46 &   1.20 &   0.50 &  3.0 &  1.8 &  2.6 &  2.6 &  2.6 \\
  8.80 &    9.47 &   2.92 &   1.20 &   1.00 &  5.0 &  0.9 &  4.8 &  1.2 &  1.9 \\
 16.88 &   18.16 &   2.92 &   2.30 &   1.00 &  8.5 & -0.0 &  8.5 &  3.5 &  4.4 \\
 50.64 &   18.16 &   8.76 &   2.30 &   3.00 & 22.9 &  1.8 & 23.8 &  9.4 & 12.1 \\
 15.00 &   15.00 &   3.14 &   1.90 &   1.08 &  4.4 &  0.6 &  4.1 &  2.2 &  2.6 \\
 15.00 &   30.00 &   1.57 &   3.80 &   0.54 & 12.6 &  2.2 & 12.7 &  5.6 &  6.9 \\
 15.00 &   50.00 &   0.94 &   6.33 &   0.32 & 13.8 &  0.3 & 14.2 &  7.1 &  8.4 \\
 15.00 &   80.00 &   0.59 &  10.13 &   0.20 & 12.8 &  4.2 & 12.9 & 11.4 & 11.7 \\
 15.00 &  100.00 &   0.47 &  12.67 &   0.16 & 12.1 &  2.9 & 12.1 &  8.7 &  9.3 \\
 15.00 &  150.00 &   0.31 &  19.00 &   0.11 &  9.8 &  4.4 &  9.6 &  8.8 &  8.9 \\
 50.00 &   18.00 &   8.73 &   2.28 &   2.99 & 25.6 &  0.8 & 26.7 &  9.4 & 12.7 \\
 50.00 &   30.00 &   5.24 &   3.80 &   1.79 & 33.1 &  5.4 & 34.4 & 14.5 & 18.3 \\
 50.00 &   40.00 &   3.93 &   5.07 &   1.34 & 34.3 & 10.6 & 35.4 & 16.9 & 20.4 \\
 \bf{50.00} &   \bf{50.00} &   \bf{3.14} &   \bf{6.33} &   \bf{1.08} & \bf{36.3} & \bf{16.1} & \bf{37.2} & \bf{20.2} & \bf{23.4}\\
 50.00 &   80.00 &   1.96 &  10.13 &   0.67 & 36.8 & 22.4 & 37.3 & 22.3 & 25.1 \\
 50.00 &  100.00 &   1.57 &  12.67 &   0.54 & 35.3 & 22.8 & 35.6 & 21.8 & 24.4 \\
 50.00 &  150.00 &   1.05 &  19.00 &   0.36 & 28.9 & 18.1 & 29.2 & 17.9 & 20.0 \\
\hline
\end{tabular}
\end{center}
\caption{Percentage variations of the drag coefficients. Positive values indicate drag reduction and negative values denote drag increase.}
\label{tab:tab2}
\end{table}

Table \ref{tab:tab1} lists the values of the drag coefficients for different amplitudes $A^+$, periods of oscillations $T^+$, and impulses in the range $J^+=1-50$. In uncontrolled conditions, the contribution of the pressure drag to the total drag is more than four times the contribution of the skin-friction drag. 

Table \ref{tab:tab2} presents the computed percentage variations of the drag coefficients. Note that, while the partial skin-friction and pressure drag coefficients in table \ref{tab:tab1} add to give the corresponding total coefficients, the partial reductions of the coefficients in table \ref{tab:tab2} do not because each reduction is defined with respect to the corresponding value in the uncontrolled case, as clear from the definitions \eqref{eq:0} and \eqref{eq:1}. 

\begin{figure}
\centering
\includegraphics[width=0.7\textwidth]{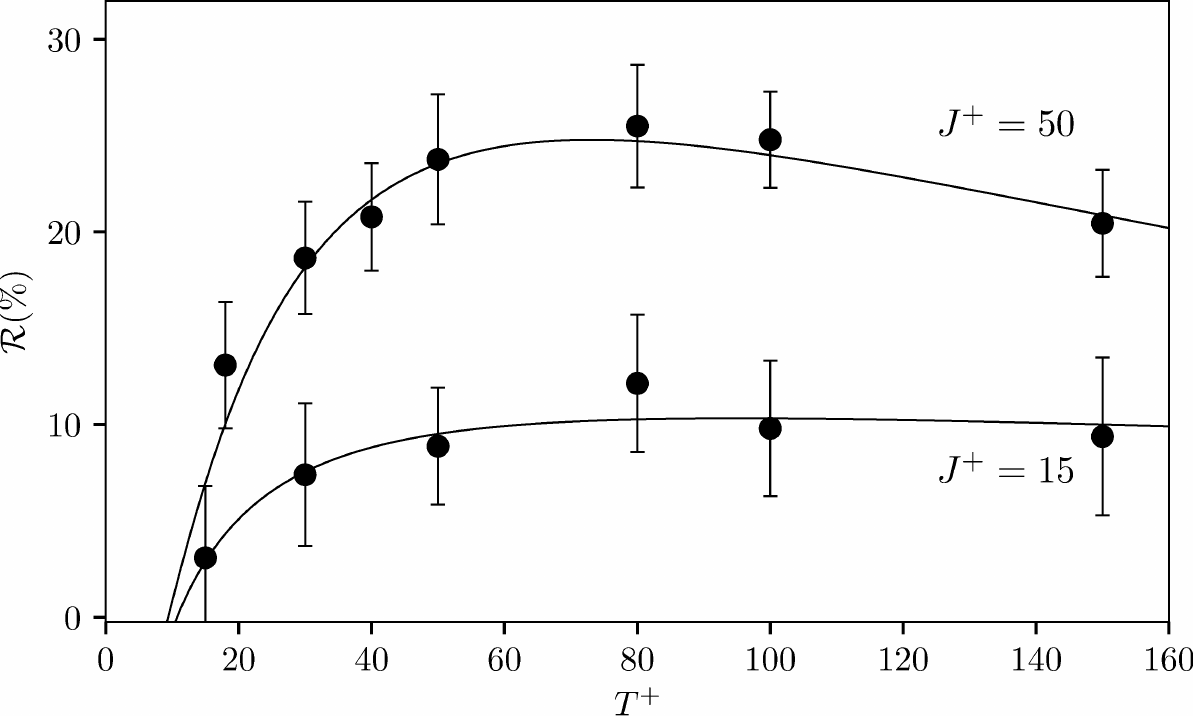}
\caption{Drag-reduction margin as a function of the oscillating period. The circles indicate the numerical results and the solid lines are the best fits at constant $J^+=A^+ T^+/\pi$.}
\label{fig:DR}
\end{figure} 
\begin{figure}
\centering
\begin{tabular}{cc}
\includegraphics[width=0.66\textwidth]{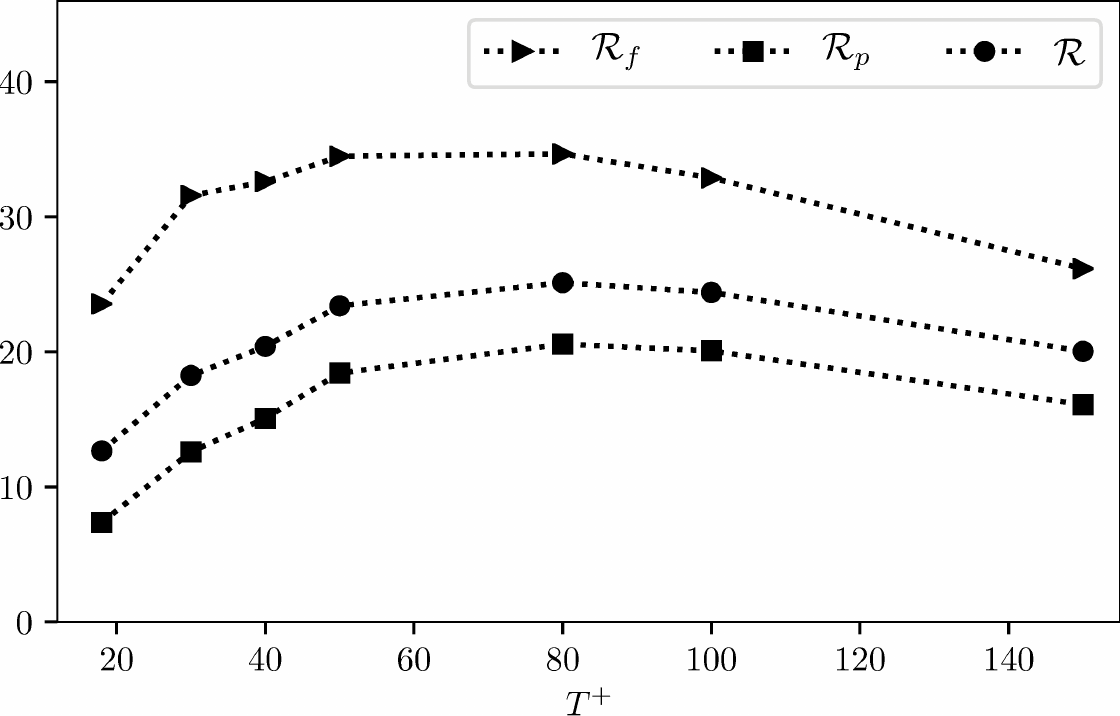}
\end{tabular}
\caption{Drag-reduction margins as functions of the oscillating period for an impulse $J^+=50$.}
\label{fig:DR-components}
\end{figure}

The control with small period $T=0.2$ ($T^+=1.57$) reduces the total drag marginally as the impulse $J$ is too small. The small drag reduction is due to the small reduction of $C_p$, while both the cavity and the total skin-friction increase slightly. For such a low period, a variation of $A$ of two orders of magnitude in the range $0.1-10$ ($A^+=0.29-29$) does not cause any improvement. The drag increases slightly in some cases for $J^+=3.2$ and $4.4$. Both the skin-friction drag and the pressure drag decrease at higher impulse values, $J^+=15, 50$. 
The case with $T=10.1$ and $A=0.67$ ($T^+=80$ and $J^+=50$) shows the maximum total drag-reduction margin, $\mathcal{R}=25\%$, the maximum skin-friction drag-reduction margin, $\mathcal{R}_f=37\%$, and the maximum pressure drag-reduction margin, $\mathcal{R}_p=22\%$. 

Figure~\ref{fig:DR} shows the total drag-reduction margin as a function of the actuation period at the impulse values $J^+=15$ and 50. For a constant $T^+$, the drag-reduction margin is larger as the impulse increases, while, for a constant $J^+$, it increases with the period up to a maximum at $T^+=80$. The existence of an optimum period of oscillation is consistent with similar control strategy for skin-friction drag reduction, such as the oscillating-wall technique in smooth channels \citep{ricco-quadrio-2008}. The optimal period of wall oscillation in the flat-wall case is $T^+=100$ \citep{touber-leschziner-2012}, which gives maximum drag-reduction margins of $35-40\%$, thus in good agreement with the maximum found in the channel with bars altered by a spanwise pressure gradient.

Figure \ref{fig:DR-components} reveals that the reduction in viscous drag along the cavity contributes to most of the total viscous reduction because the skin-friction on the bar crest is negligible. It is also shown that the total friction drag reduction $\mathcal{R}_f$ is almost double the total pressure drag reduction $\mathcal{R}_p$. The viscous drag reduction reaches a maximum of about $37\%$ for $T^+=50-100$, while the pressure drag reduction reaches its maximum value of about $22\%$ for $T^+=80-100$. The amount of skin-friction drag reduction is comparable with the maximum reduction brought about by spanwise wall oscillations in a fully-developed channel flow for $T^+=100$, and similar wall-friction Reynolds numbers and maximum velocity amplitude of oscillation \citep{quadrio-ricco-2004}. While the scaled forces are additive, i.e., $-\ov{\Pi}_x L_y=C_f+C_p$, the reductions of viscous drag, $\mathcal{R}_f$, and of the pressure drag, $\mathcal{R}_p$, lead to $\mathcal{R}$ via:
\begin{equation}
\mathcal{R} = 100 - \frac{1}{\overline{C}+1}\left[ 100 - \mathcal{R}_f + \overline{C} \left(100-\mathcal{R}_p \right) \right],
\end{equation}
where $\overline{C}=C_p/C_f=4.3$.

\subsection{Power spent and net power saved}
\label{sec:spent-net}

Table~\ref{tab:lampower} gives the values of the drag-reduction margin, $\mc{R}(\%)$ (which coincides with the power saved), the power spent to activate the control, $\mc{P}_z(\%)$, and the laminar power $\mc{P}_{z,l}(\%)$ for different parameters $T$ and $A$. Very marginal savings in terms of $\mc{P}_{net}(\%)$ are found when the drag-reduction effect is small and these values are within the uncertainty of the computed $\mc{R}(\%)$. For the cases with large drag reduction, the decrease in pressure gradient along the streamwise direction comes at the expense of a high consumption of energy, which leads to no savings in terms of $\mc{P}_{net}(\%)$. This finding opens new avenues of research to improve the net energy gain with a view to render this technique competitive with vortex generators which, at this stage, are more effective and guarantee estimated net drag reductions of order of $10\%$ \citep{calarese-crisler-1985}.

\begin{figure}
\centering
\subfloat{\includegraphics[width=0.5\textwidth]{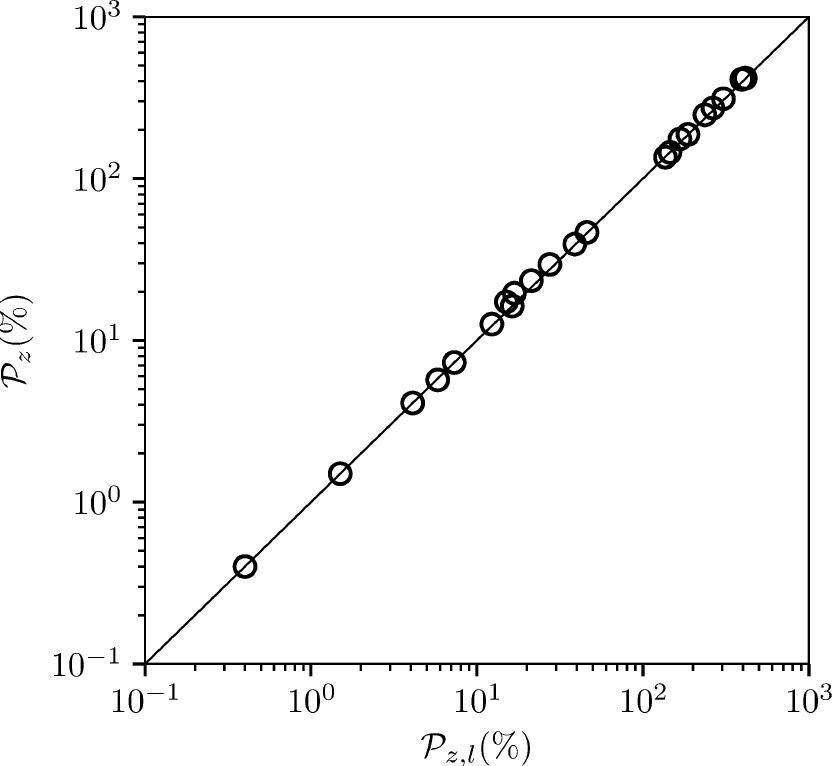}}
\caption{Turbulent power spent $\mc{P}_z(\%)$ versus its laminar prediction $\mc{P}_{z,l}(\%)$. The solid line indicates the line $\mc{P}_z(\%)=\mc{P}_{z,l}(\%)$.}
\label{fig:turbpower}
\end{figure}

Figure \ref{fig:turbpower} shows the values of $\mc{P}_z$ obtained via the the direct numerical simulations versus the laminar prediction $\mc{P}_{z,l}$. The laminar power $\mc{P}_{z,l}$ gives an excellent approximation of $\mc{P}_z$ for all the values of $A$ and $T$ considered. The laminar solution found in \S\ref{sec:laminar} is therefore able to predict the turbulent power spent very well.          

\begin{table}
\begin{center}
\begin{tabular}{ccc|cc |cc |c |c}
 $J^+$ & $T^+$ & $A^+$ & $T$ & $A$ & $\mc{R}(\%)$ & $\mc{P}_z(\%)$ & $\mc{P}_{z,l}(\%)$ & $\mc{P}_{net}(\%)$ \\
\hline
  0.14 &    1.57 &   0.29 &   0.20 &   0.10 &  0.3 &   0.0 &   0.0 &    0.3 \\
  0.73 &    1.57 &   1.46 &   0.20 &   0.50 &  1.1 &   0.4 &   0.4 &    0.8 \\
  1.46 &    1.57 &   2.92 &   0.20 &   1.00 &  0.5 &   1.5 &   1.5 &   -1.0 \\
  3.21 &    1.57 &   6.43 &   0.20 &   2.20 & -1.3 &   7.3 &   7.3 &   -8.6 \\
 14.60 &    1.57 &  29.21 &   0.20 &  10.00 &  0.8 & 145.3 & 145.5 & -144.5 \\
  4.40 &    4.73 &   2.92 &   0.60 &   1.00 & -0.5 &   5.7 &   5.8 &   -6.2 \\
  4.40 &    9.47 &   1.46 &   1.20 &   0.50 &  2.8 &   4.1 &   4.1 &   -1.3 \\
  8.80 &    9.47 &   2.92 &   1.20 &   1.00 &  2.1 &  16.3 &  16.3 &  -14.2 \\
 16.88 &   18.16 &   2.92 &   2.30 &   1.00 &  4.6 &  46.5 &  46.0 &  -41.9 \\
 50.64 &   18.16 &   8.76 &   2.30 &   3.00 & 12.3 & 418.0 & 414.2 & -405.8 \\
 15.00 &   15.00 &   3.14 &   1.90 &   1.08 &  2.6 &  39.4 &  38.8 &  -36.8 \\
 15.00 &   30.00 &   1.57 &   3.80 &   0.54 &  6.9 &  29.5 &  27.5 &  -22.5 \\
 15.00 &   50.00 &   0.94 &   6.33 &   0.32 &  8.4 &  23.4 &  21.3 &  -15.0 \\
 15.00 &   80.00 &   0.59 &  10.13 &   0.20 & 11.7 &  19.6 &  16.8 &   -7.9 \\
 15.00 &  100.00 &   0.47 &  12.67 &   0.16 &  9.3 &  17.3 &  15.0 &   -8.0 \\
 15.00 &  150.00 &   0.31 &  19.00 &   0.11 &  8.9 &  12.6 &  12.3 &   -3.7 \\
 50.00 &   18.00 &   8.73 &   2.28 &   2.99 & 12.7 & 410.9 & 394.0 & -398.2 \\
 50.00 &   30.00 &   5.24 &   3.80 &   1.79 & 18.3 & 311.9 & 305.2 & -293.7 \\
 50.00 &   40.00 &   3.93 &   5.07 &   1.34 & 20.4 & 273.5 & 264.3 & -253.1 \\
 \bf{50.00} &   \bf{50.00} &   \bf{3.14} &   \bf{6.33} &   \bf{1.08} & \bf{23.4} & \bf{248.1} & \bf{236.4} & \bf{-224.7} \\
 50.00 &   80.00 &   1.96 &  10.13 &   0.67 & 25.1 & 187.8 & 186.9 & -162.6 \\
 50.00 &  100.00 &   1.57 &  12.67 &   0.54 & 24.4 & 176.0 & 167.2 & -151.6 \\
 50.00 &  150.00 &   1.05 &  19.00 &   0.36 & 20.0 & 135.7 & 136.5 &  -81.5 \\
\end{tabular}
\end{center}
\caption{Power saved $\mc{P}_z(\%)$, net power $\mc{P}_{net}(\%)$ and  laminar power $\mc{P}_{z,l}(\%)$ as functions of the control parameters $A$ and $T$.}
\label{tab:lampower}
\end{table}

\section{Turbulent-flow physics}
\label{sec:physics}

In this section, instantaneous flow fields and flow statistics are presented for the controlled case with $A=1.08$ and $T=6.33$ ($J^+=50$ and $T^+=50$), which features a large total drag-reduction margin, $\mathcal{R}=23\%$.

\subsection{Controlled instantaneous fields and average flow}
\label{sec:meancon}

\begin{figure}
\centering
\includegraphics[width=0.7\textwidth]{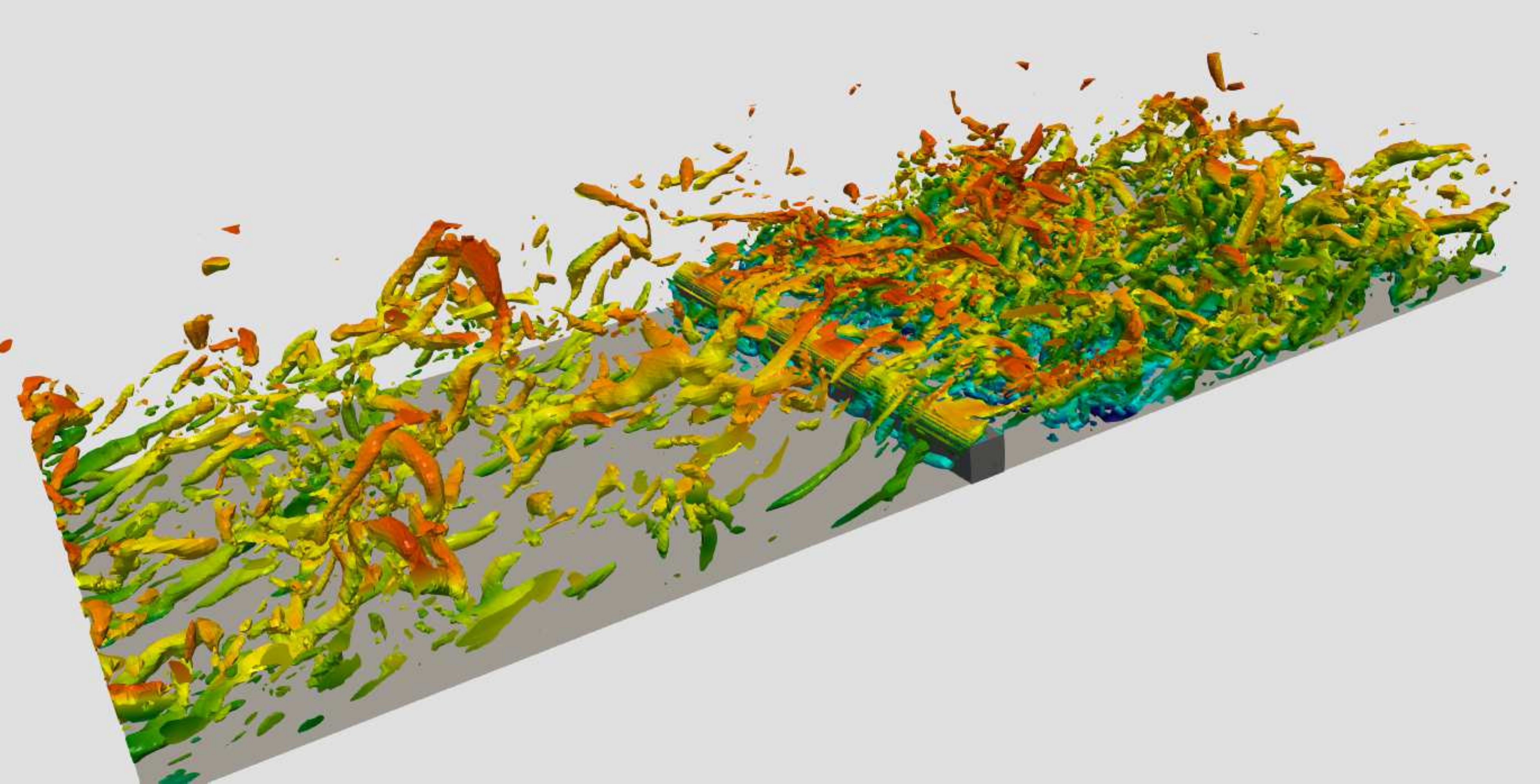}
\includegraphics[width=0.7\textwidth]{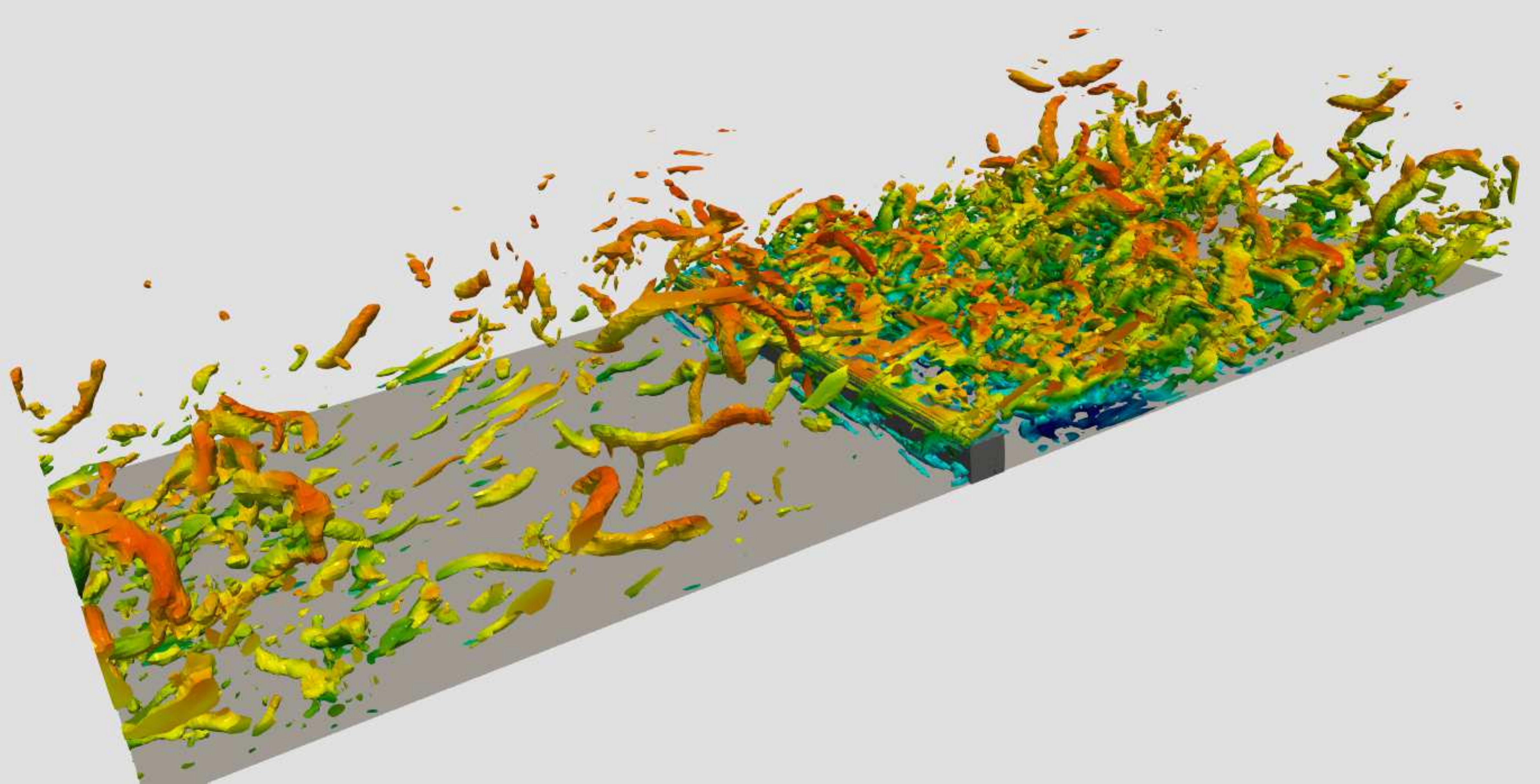}
\caption{Isosurfaces of $\lambda_2=-1.0$ in the channel for the uncontrolled case (top) and the controlled case (bottom). The isosurface contours are coloured by the streamwise velocity.}
\label{fig:lambda2}
\end{figure}

\begin{figure}
\centering
\hspace{2mm}
\includegraphics[width=0.9\textwidth]{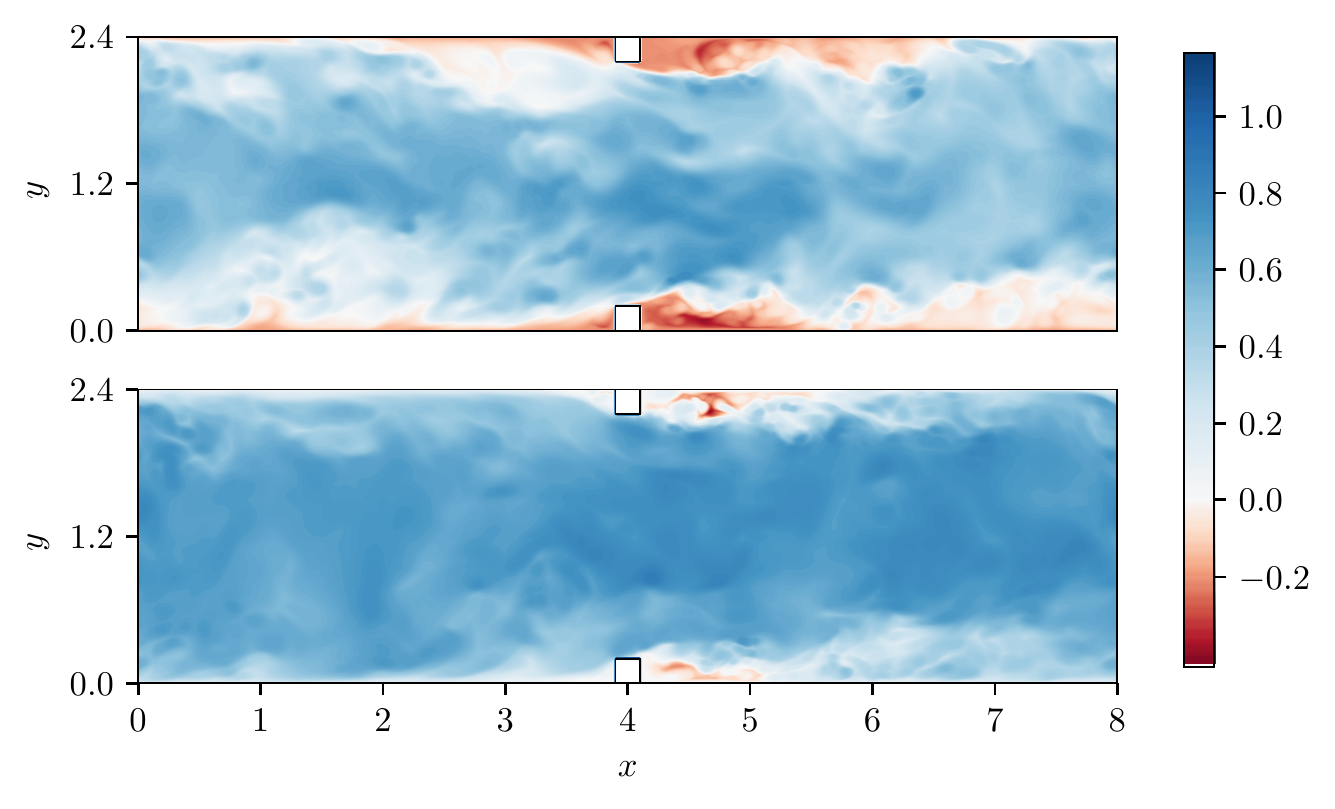}
\caption{Contour plots of instantaneous streamline velocity $u$ in the channel for the uncontrolled case (top) and the controlled case (bottom).}
\label{fig:uvel}
\end{figure}

\begin{figure}
\centering
\hspace{2mm}
\includegraphics[width=0.9\textwidth]{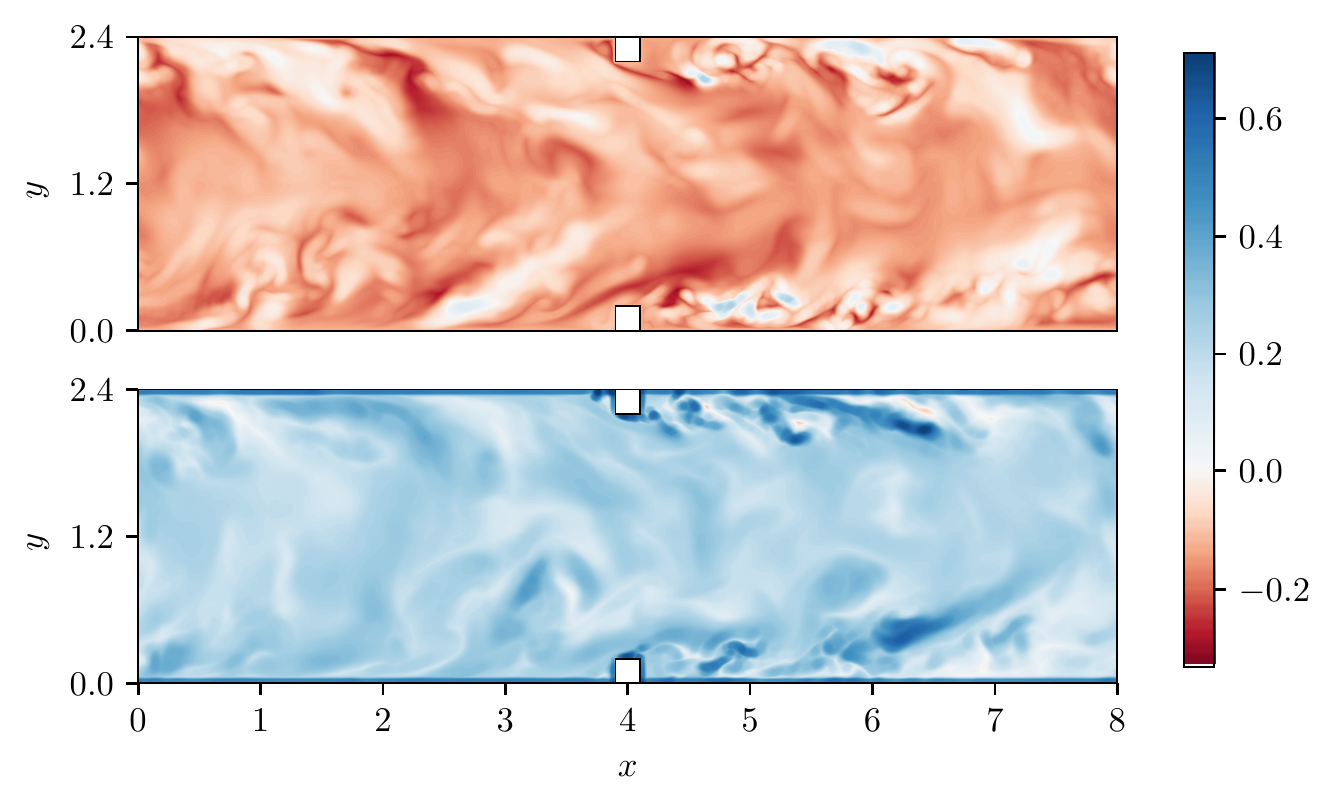}
\caption{Contour plots of instantaneous spanwise velocity, $w$ in the channel for uncontrolled (top) and controlled case (bottom).}
\label{fig:wvel}
\end{figure}

Instantaneous flow fields are shown in figures \ref{fig:lambda2}, \ref{fig:uvel}, and \ref{fig:wvel} to illustrate the effects of the square bar and of the spanwise-forcing control on the flow dynamics. 
Figure~\ref{fig:lambda2} shows isosurfaces of the invariant $\lambda_2$, proposed by \cite{jeong-hussain-1995} to visualize vortical structures. Vortices are formed on top and downstream of the bar, especially where the flow fully separates, and are mostly attenuated upstream of the bar.
Figures~\ref{fig:uvel} and \ref{fig:wvel} depict instantaneous streamwise and spanwise velocity fields in the central $x-y$ plane at $z=\pi/2$. The controlled streamwise velocity is much more uniform than in the reference case, both upstream and downstream of the bar. In the controlled case, the spanwise velocity is positive everywhere in the visualization plane and much larger than in the reference case because of the large spanwise pressure gradient at this phase of the oscillation.  


\begin{figure}
\centering
\includegraphics[width=0.87\textwidth,height=0.51\textwidth]{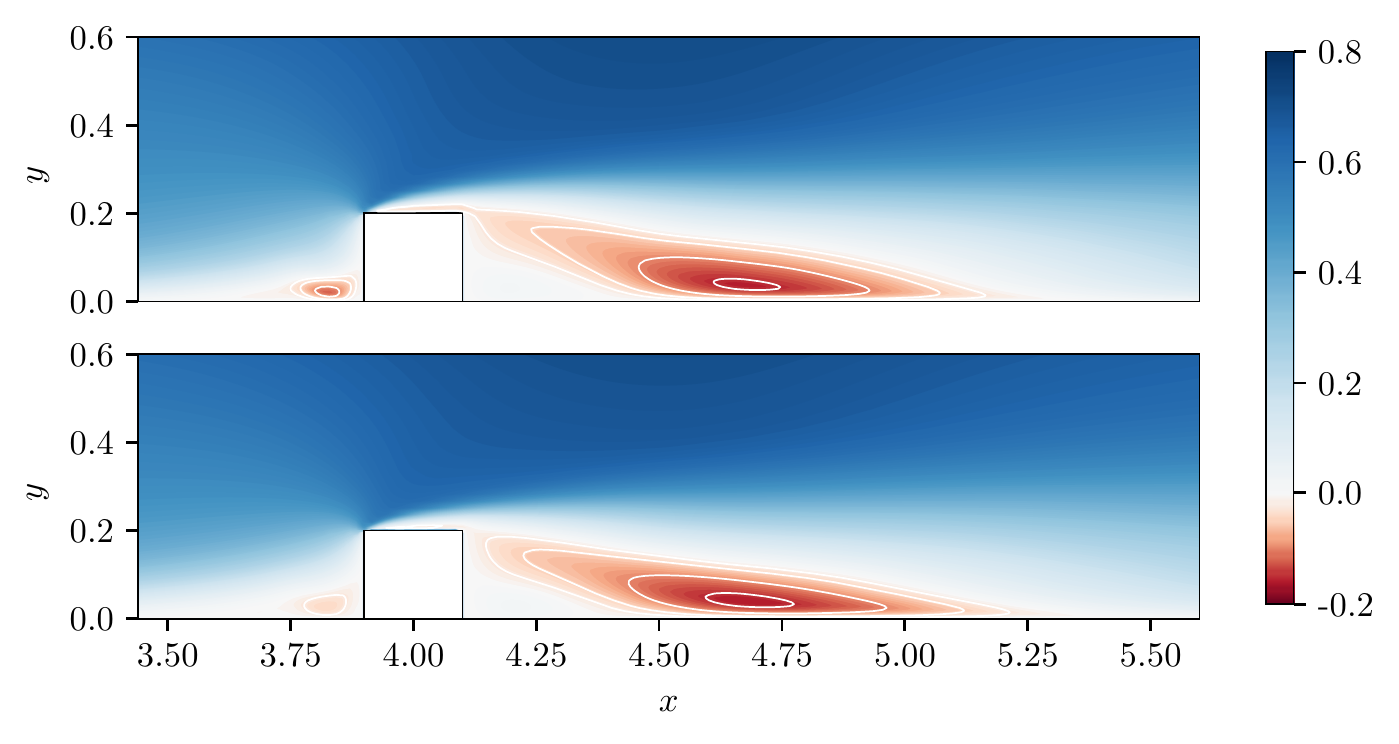}
\caption{Contour plots of time- and spanwise-averaged velocity $\langle \ov{u} \rangle$ in the channel for uncontrolled (top frame) and controlled case (bottom frame). The contour lines are for the negative velocity range of $\langle \ov{u} \rangle$ in  $(-0.2,-0.15,-0.1,-0.05,-0.025)$.}
\label{fig:umean}
\end{figure}

\begin{figure}
\centering
\subfloat{\hspace{-1.5cm}
\includegraphics[width=0.8\textwidth]{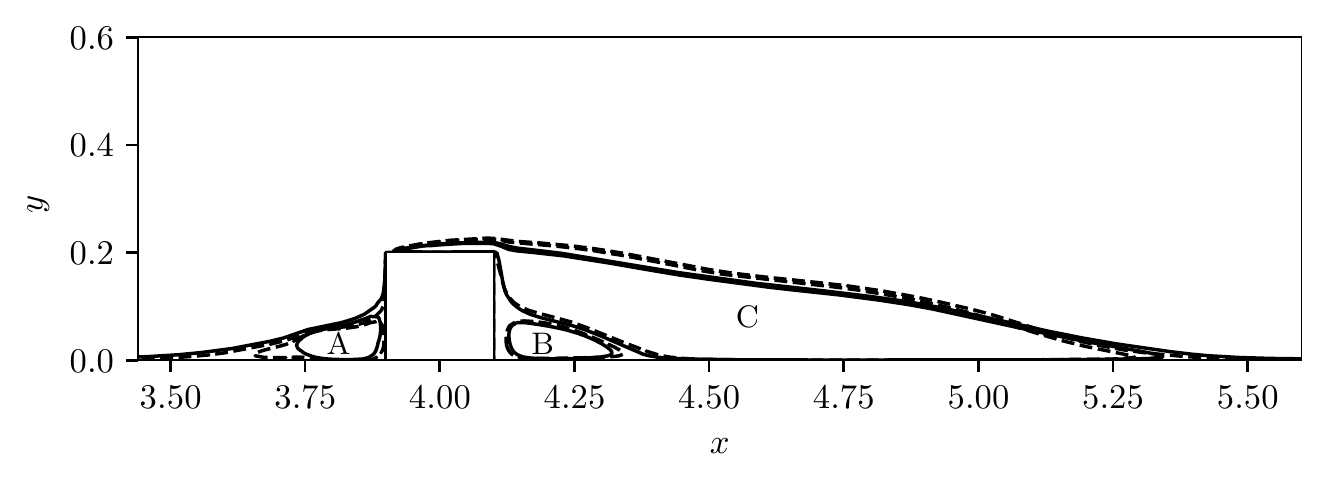}}
\caption{Effect of the control on the separation areas. Dashed and solid lines indicate the time- and spanwise-averaged streamlines for the uncontrolled and controlled flow, respectively.}
\label{fig:fig7}
\end{figure}

The contours of the mean streamwise velocity, shown in figure~\ref{fig:umean}, confirm that the main flow features around the bar are qualitatively unvaried in the controlled case. The three recirculation regions are still present; the large recirculation area behind the bar is slightly compressed towards the wall and the one upstream of the bar is weakened.
Figure \ref{fig:fig7} compares the mean streamlines in the uncontrolled and controlled cases. The forcing reduces the size of vortices B and C and brings them closer to the wall, thus rendering the flow more attached downstream of the bar. The reattachment point upstream of the vortex A displaces slightly and a small vortex forms upstream of vortex A with its core at $x=3.72$. After the bar, although the control shifts the second reattachment point downstream, this change only marginally affects the pressure-drag balance, which is instead mainly influenced by the dynamics in the whole separated region and by the flow modifications upstream of the bar. Similar changes of the separation area as an effect of the control were also found in previous separation-drag studies \citep{banchetti-etal-2020,minelli-etal-2019}.    

\begin{figure}
\hspace*{0.6cm}
\includegraphics[width=0.9\textwidth]{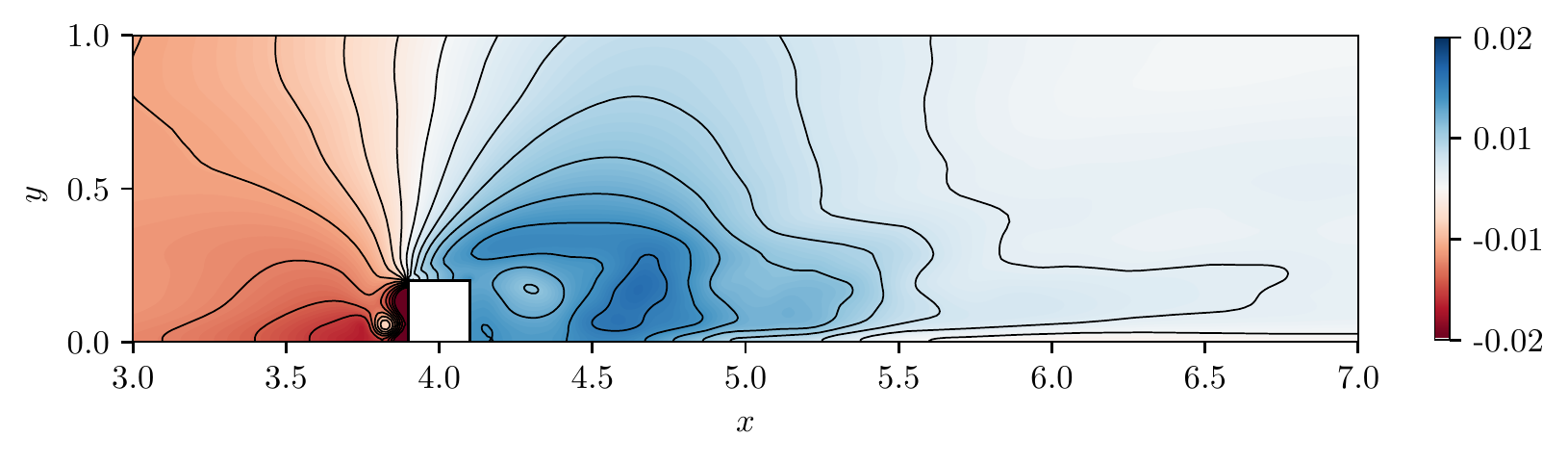}
\caption{Colour map and contours of the difference of the mean pressure between the uncontrolled case and the controlled case. The positive values indicate a pressure increase and the negative values denote a pressure decrease. The isoline values are equispaced at intervals of $0.002$ in the range $[-0.02,0.02]$.}
\label{fig:difpre}
\end{figure}

\subsection{Distribution of pressure and skin-friction drag coefficients}
\label{sec:presfriction}

The difference of the time- and spanwise-averaged pressure between the controlled case and the uncontrolled case is shown in the contour plot of figure \ref{fig:difpre}. The pressure is unvaried in a region that extends from the bar upward toward the channel centerline. The pressure upstream of the bar decreases, while the pressure instead increases in the region downstream the bar with the maximum increase inside the area delimited by the isoline around the point $(x,y)=(4.75,0.1)$. This dual effect on the pressure distribution is the main cause of the reduced pressure difference on the two sides of the bar, which gives a lower overall pressure drag. 

\begin{figure}
\centering
\hspace{-4mm}
\includegraphics[width=0.8\textwidth]{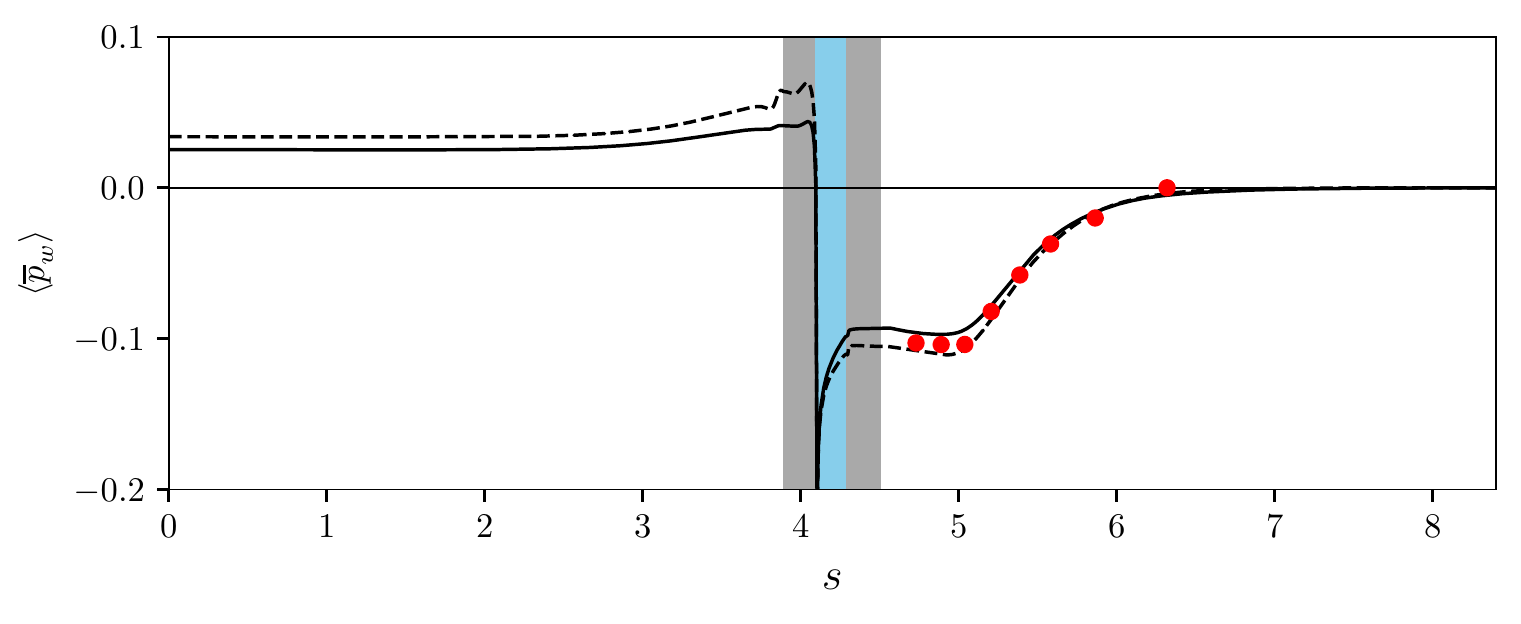} \\
\hspace{-5mm}
\includegraphics[width=0.815\textwidth]{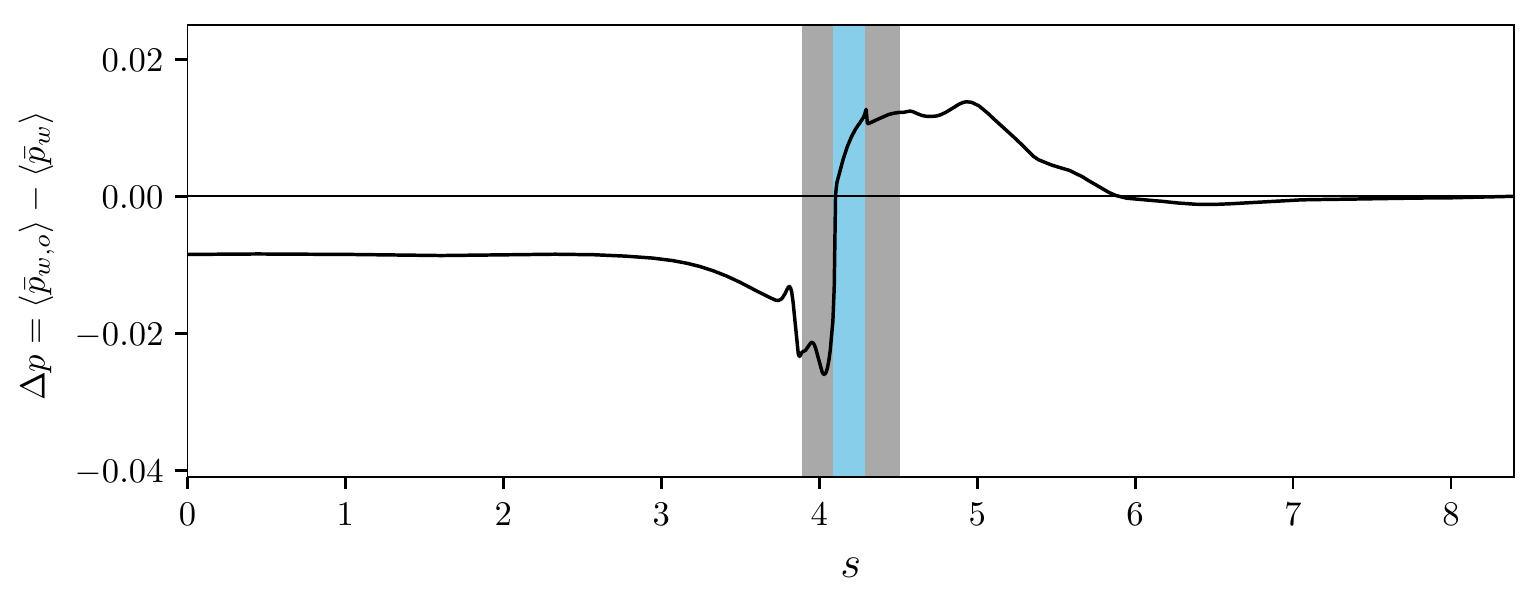}
\caption{Top: time- and spanwise-averaged total pressure in the uncontrolled case (dashed line) and the controlled case (solid line) as a function of the coordinate $s$ running along the wall surface. The sides of the bar are shown in grey and the crest of the bar is marked in light blue. The red circles are data from the direct numerical simulations of \cite{leonardi-etal-2003}. Bottom: difference between the time- and spanwise-averaged pressure in the controlled and uncontrolled cases.}
\label{fig:presprofile}
\end{figure}

Figure \ref{fig:presprofile} depicts the mean pressures in the controlled and uncontrolled cases (top) and their difference (bottom). The change of pressure is much smaller away from the bar, especially downstream of the large separation region, in the fully attached region for $s>6.5$, where $s$ is the streamwise coordinate running along the solid wall. 
The pressures are kept equal to zero at the further end of the domain, as discussed in \S\ref{sec:perfor}.
Excellent agreement is found between our data and the data behind the bar from the direct numerical simulations of \cite{leonardi-etal-2003} (red circles), obtained for a flow over a squared bar with the same height as ours, a slightly different Reynolds number, and a distance between bars which is about half of ours ($h/L_x=19$). This match confirms that the recirculation area is largely unaffected once the bars are sufficiently spaced apart. 
The distribution of the pressure difference in the separated region is qualitatively similar to that found by \cite{banchetti-etal-2020} for a channel flow modified by a smooth bump and altered by wall travelling waves.
In their flow, the total pressure sufficiently downstream of the obstacle recovers the negative constant slope behaviour proper of channel flows over smooth walls. In our case, figure \ref{fig:presprofile} (top) instead shows that the pressure further downstream of the bar tends to a constant value because the effect of the intense separation is still present. The same trend was found by \cite{leonardi-etal-2003}. This finding is expected because \cite{banchetti-etal-2020}'s flow is characterized by a mild separation being their obstacles smoother and thinner than our bars and more spaced apart along the streamwise direction.

\begin{figure}
\centering
\includegraphics[width=0.85\textwidth]{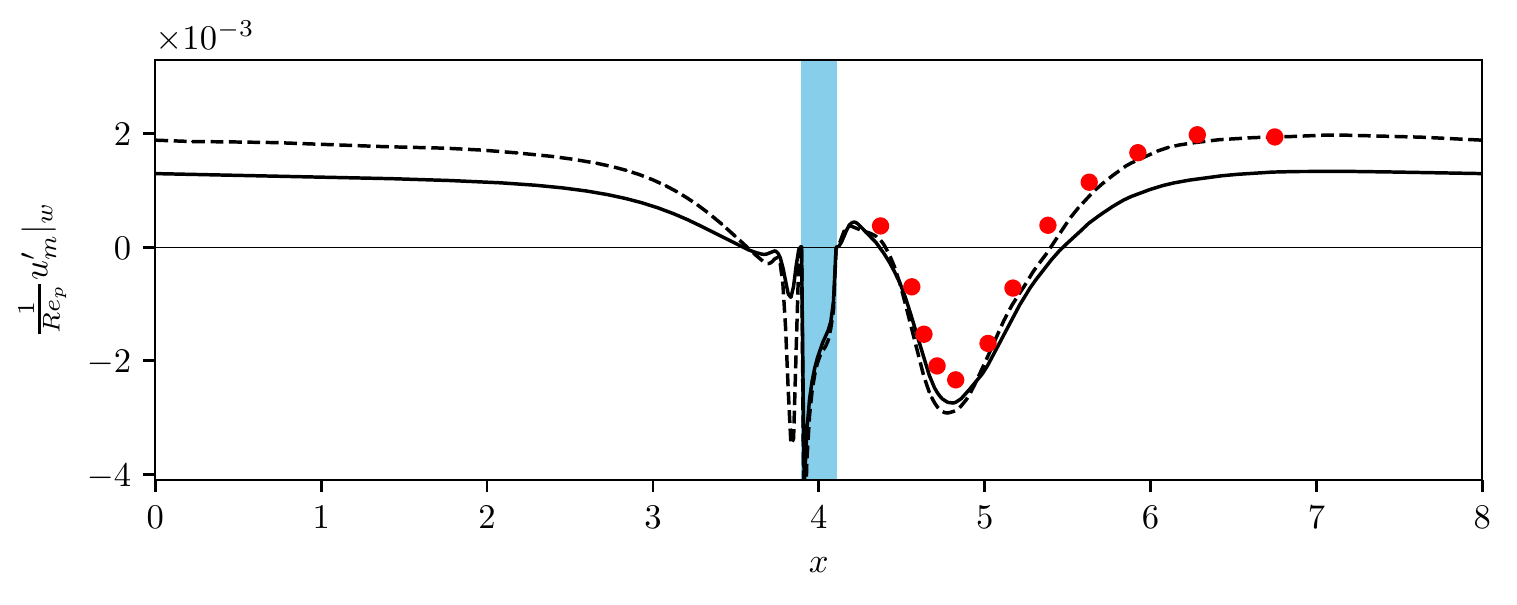} \\
\includegraphics[width=0.87\textwidth]{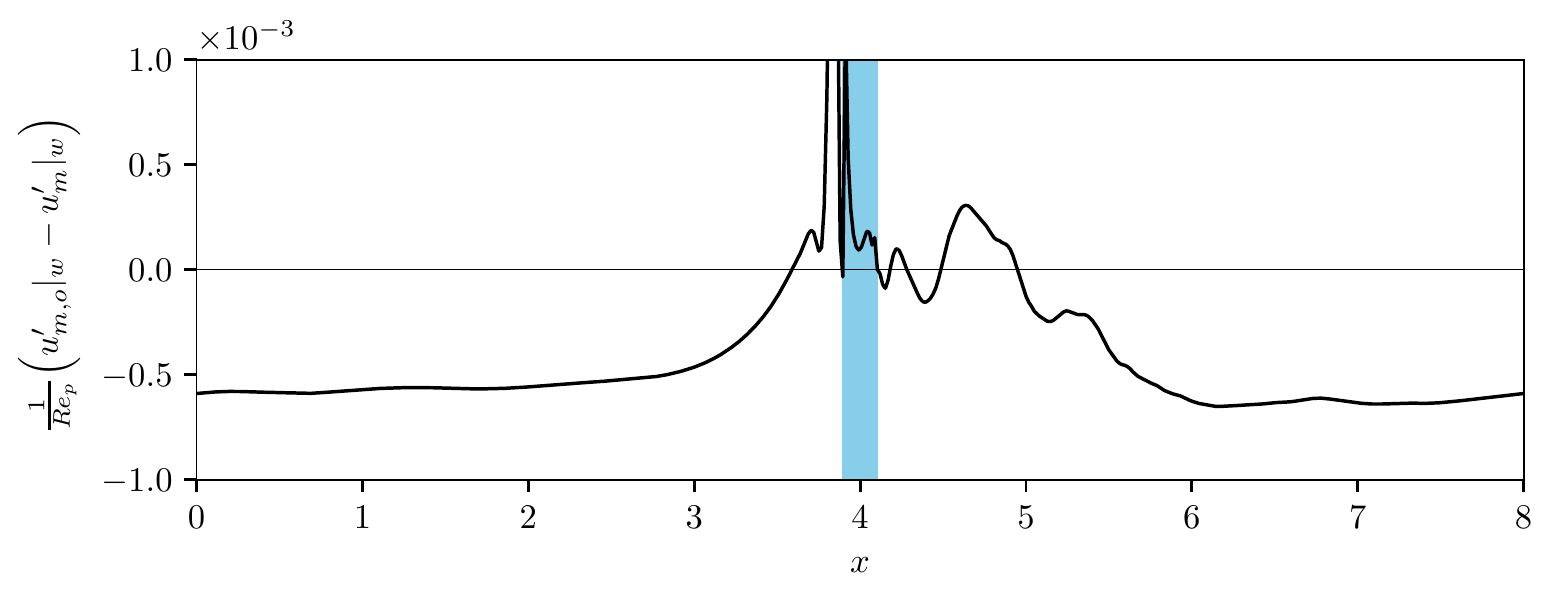}
\caption{Top: time- and spanwise-averaged wall-shear stress in the uncontrolled case (dashed line) and the controlled case (solid line). The red circles are data from the direct numerical simulations of \cite{leonardi-etal-2003}. Bottom: difference between the time- and spanwise-averaged wall-shear stress caused by the control (bottom). The prime indicates derivation with respect to $y$.}
\label{fig:skinfriction}
\end{figure}

The graphs of figure~\ref{fig:skinfriction} show that the wall-shear stress decreases along the major part of the cavity area where the flow is attached ($x<3.5$ and $x>5.4$), while there is no significant change of skin-friction upstream the bar ($3.5<x<3.9$). The wall-shear stress is mostly affected by the forcing when it is positive, i.e., where the flow is attached along the cavity, reaching a reduction of $31\%$. This value is consistent with the drag-reduction margin of $\mathcal{R}=35\%$ obtained by forcing a flat-wall turbulent channel flow by spanwise wall oscillations at a similar friction Reynolds number, period $T^+=50$, and maximum spanwise velocity of about $27u_\tau^*$ \citep{quadrio-ricco-2004}. Like for the pressure distribution, the red circles show that the skin-friction distribution in the recirculation area agrees very well with that of \cite{leonardi-etal-2003}.

The negative skin-friction in the region behind the bar is instead largely unperturbed. The wall-shear stress in the separation-area B between $x=4.1$ and $x=4.4$, defined in figure \ref{fig:fig7}, is unaffected by the forcing, although the near-wall flow is along the streamwise direction. This finding is likely to be due to the stress being small and the flow not being in the fully-developed turbulent regime there. This result is consistent with the spanwise forcing not having an effect on a streamwise laminar flow \citep{ricco-hicks-2018}. Figure~\ref{fig:skinfriction} (top) also shows that the reattachment points upstream of the bar and immediately downstream of the bar are unaffected by the control, while the reattachment point further downstream is shifted only marginally in the streamwise direction.

\subsection{Turbulent-flow statistics}
\label{sec:flowstatistics}

\begin{figure}
\centering
\includegraphics[width=0.9\textwidth]{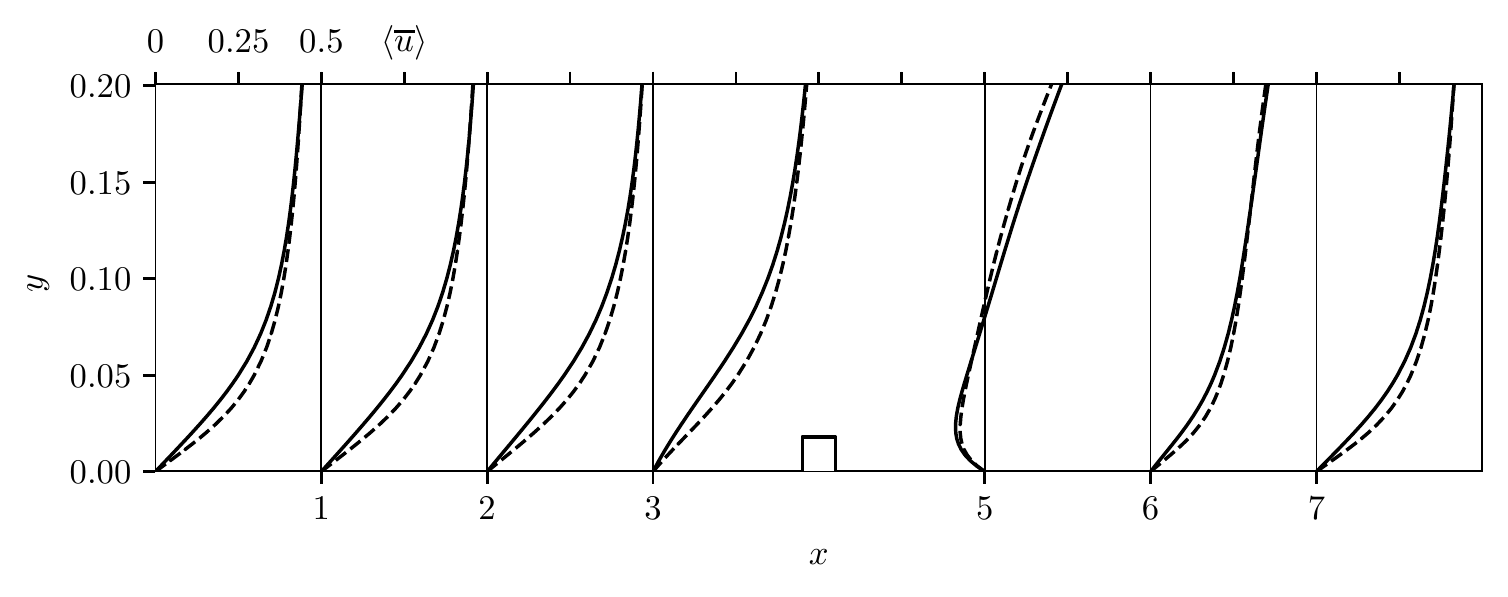}\\
\includegraphics[width=0.9\textwidth]{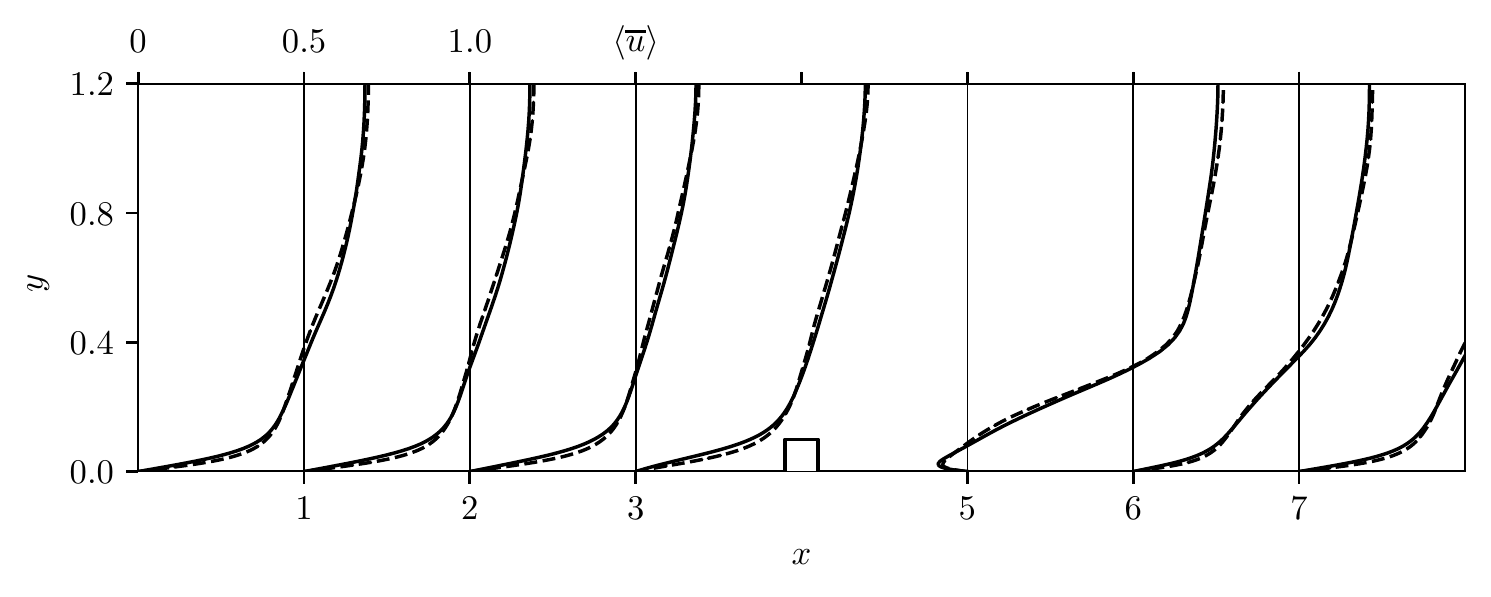}
\caption{Profiles of streamwise mean velocity $\langle \ov{u} \rangle$  at different $x$ locations along the channel. The velocity profiles are shown up to the height of the bar (top) and to the channel centerline (bottom). In this figure and in figures \ref{fig:vmprofile} and \ref{fig:uplus}, solid and dashed lines show profiles in the controlled and uncontrolled cases, respectively. The top horizontal axis denotes the velocity magnitude. The bars are not drawn to scale.}
\label{fig:umprofile}
\end{figure}

\begin{figure}
\centering
\includegraphics[width=0.9\textwidth]{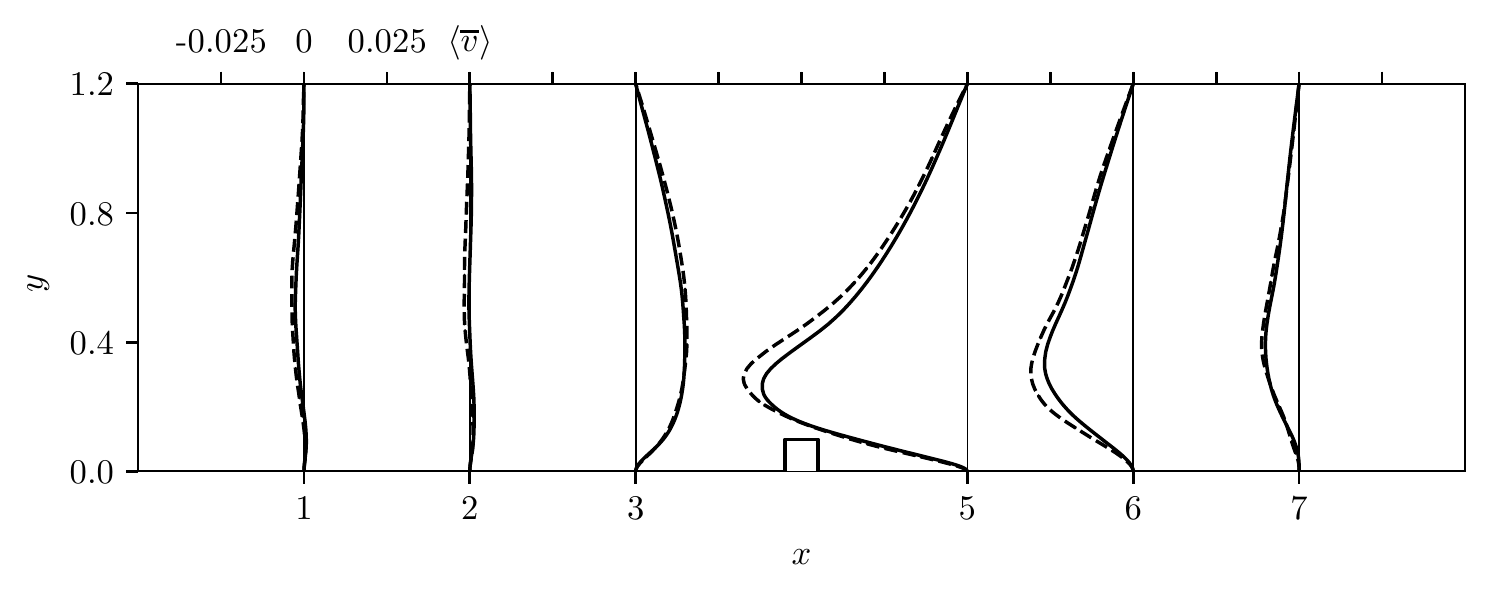}
\caption{Profiles of the wall-normal mean velocity $\langle \ov{v} \rangle$ at different $x$ locations along the channel. The bar is not drawn to scale.}
\label{fig:vmprofile}
\end{figure}

Figures~\ref{fig:umprofile} and \ref{fig:vmprofile} show the time- and spanwise-averaged profiles of the streamwise and wall-normal mean velocities along the channel for the uncontrolled and controlled cases. Close to the wall, the streamwise velocity is smaller in the controlled case resulting in the lower skin-friction. The conservation of mass flow rate manifests itself in the wall-normal location increasing from the wall. Away from the wall, the streamwise velocity in the controlled case becomes first larger and then smaller than in the uncontrolled case as the channel centerline is approached when the flow is streamwise only at every wall-normal location. The opposite occurs when the wall-shear stress is instead negative. The decrease of pressure on the left side of the bar could be due to the reduced dynamic pressure of the flow encountering the bar, due to the lower mean velocity, shown in the first three profiles of figure \ref{fig:umprofile} (top). The wall-normal velocity is unaffected by the control when the flow is fully attached, while it slightly increases in the recirculation region.

\begin{figure}
\centering
\includegraphics[width=0.485\textwidth]{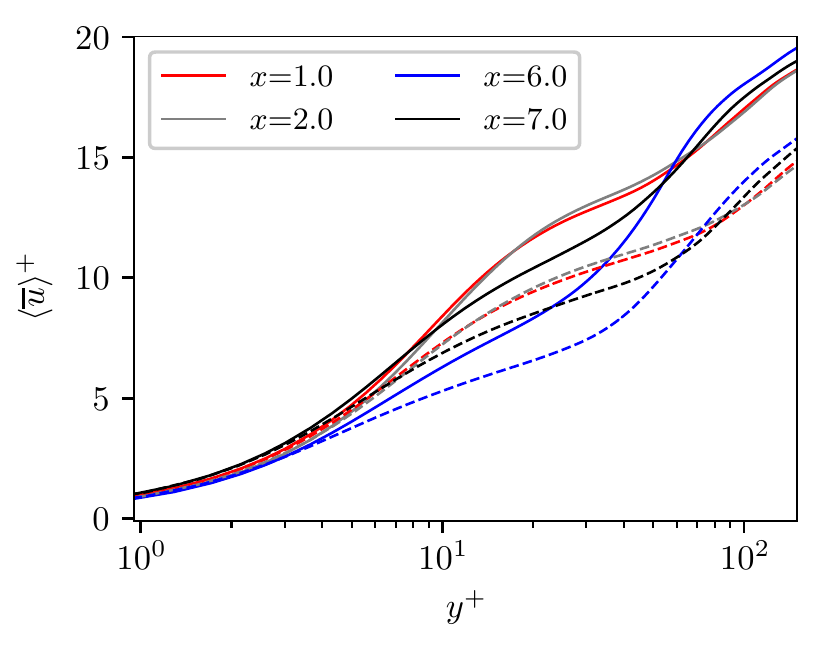}
\includegraphics[width=0.495\textwidth]{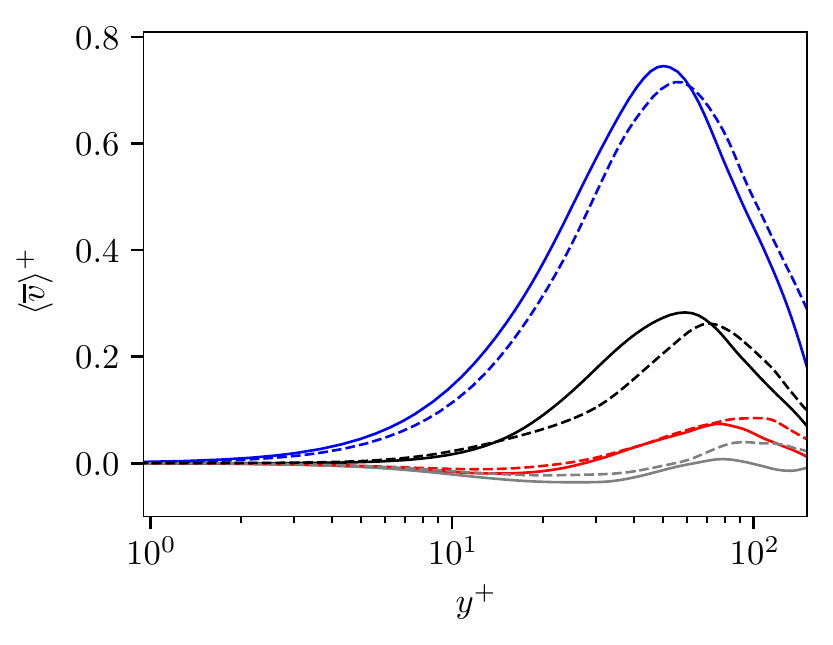}
\caption{Mean streamwise velocity (left) and wall-normal velocity (right) profiles scale in viscous units at different locations along the channel. The wall-friction velocity for each case, defined in \S\ref{sec:uncon}, is used for scaling.}
\label{fig:uplus}
\end{figure}

Figure~\ref{fig:uplus} shows the velocity profiles scaled in viscous units and at locations where the streamwise flow is attached. The wall-friction velocity of each case is used. The streamwise velocity profile shifts upwards in the controlled case, thus showing a thicker viscous layer and buffer region, a well-known feature of drag-reduced attached flows. The wall-normal velocity is instead only marginally affected. 

\begin{figure}
\centering
\includegraphics[width=0.9\textwidth]{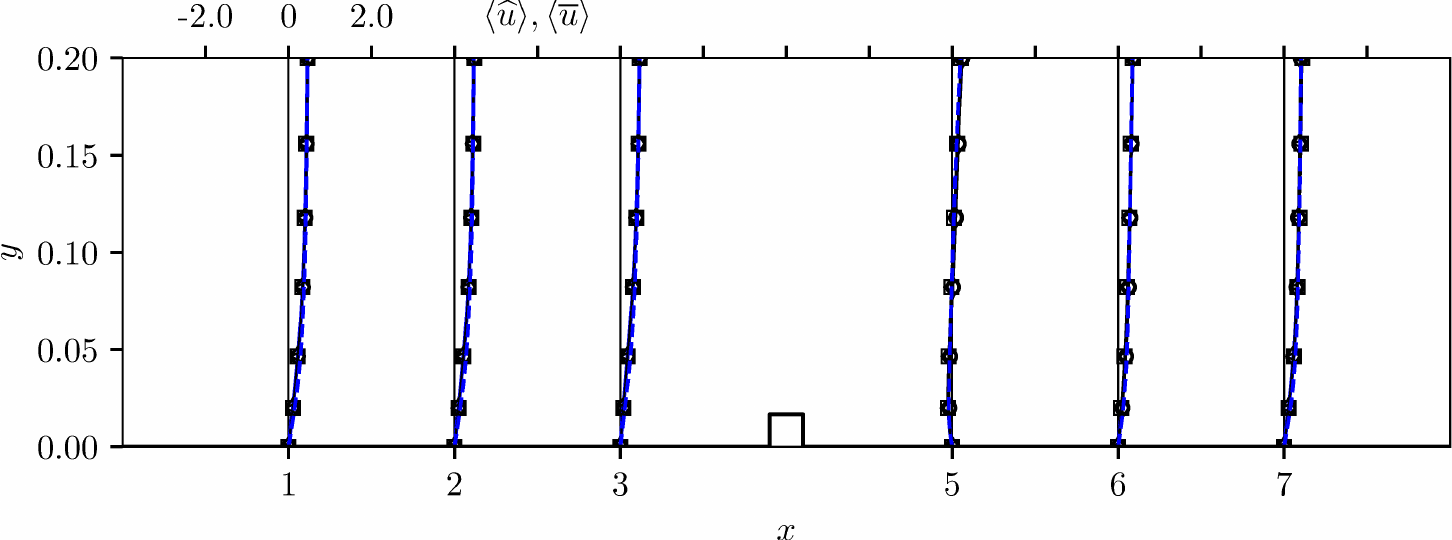}\\
\includegraphics[width=0.9\textwidth]{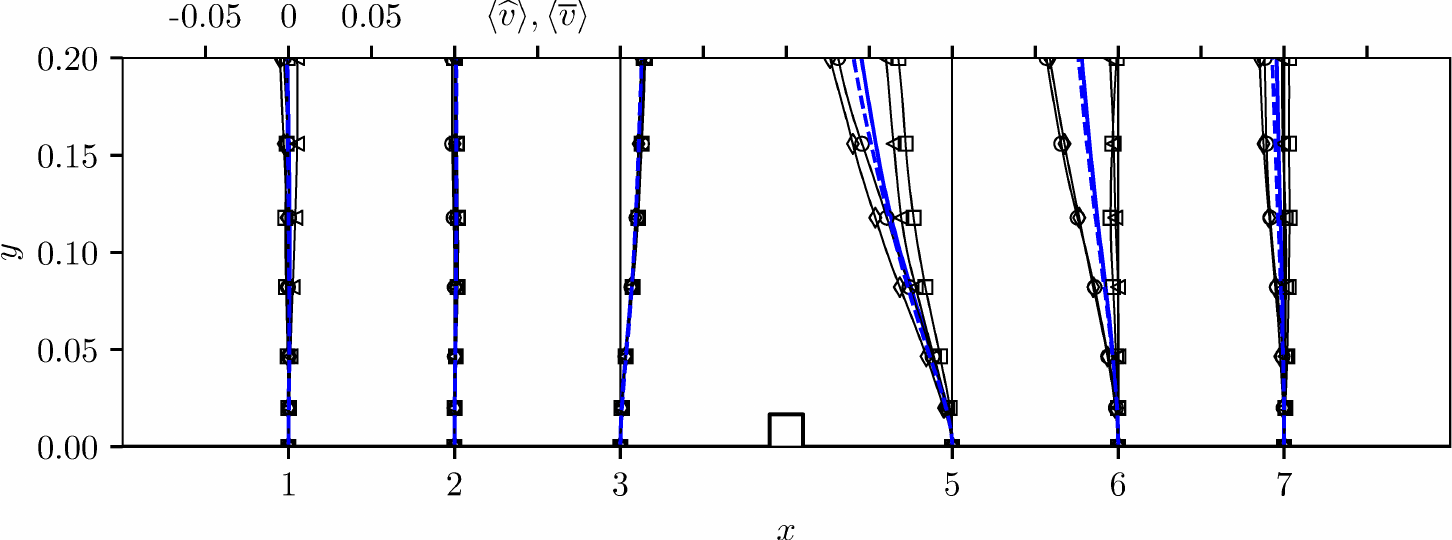}\\
\includegraphics[width=0.9\textwidth]{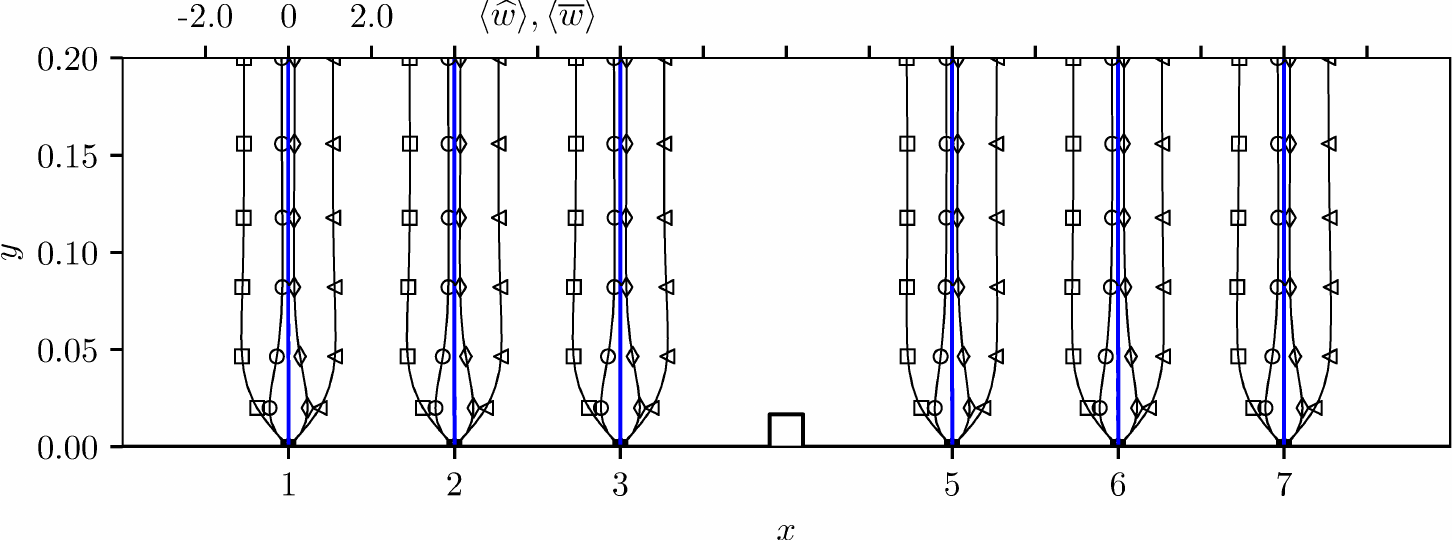}\\
\caption{Profiles of spanwise- and phase-averaged velocities $\lela \wh{u}\rira$, $\lela \wh{v}\rira$ and $\lela \wh{w}\rira$ at different $x$ locations. Thin solid lines with symbols are spanwise- and phase-averaged profiles at different phases (circle: $\tau=0$; square: $\tau=T/4$; diamond: $\tau = T/2$ and triangle: $\tau=3T/4$). The lines indicate the time- and spanwise-averaged velocities $\lela \overline{u}\rira$, $\lela \overline{v}\rira$ and $\lela \overline{w}\rira$  for the controlled case (solid line) and the uncontrolled case (dashed line). The bars are not drawn to scale.}
\label{fig:uphasemean}
\end{figure}

Figure \ref{fig:uphasemean} depicts the spanwise- and phase-averaged velocity components at different streamwise locations and phase of oscillation $\tau$. The velocities $\lela \wh{u} \rira$ and $\lela \wh{v} \rira$ do not depend on $\tau$ and overlap with the time-averaged velocities $\lela \ov{u} \rira$ and $\lela \ov{v} \rira$ when the flow is fully attached and show only a marginal modulation with the phase in the recirculation region downstream of the bar. This result is in accordance with the behaviour of turbulent channel flows with spanwise oscillating walls \citep{jung-mangiavacchi-akhavan-1992} and validates the assumption of the steadiness of $\lela \wh{u} \rira$ and $\lela \wh{v} \rira$, adopted for the laminar-flow analysis of \S\ref{sec:laminar}.
The velocity $\lela \wh{v} \rira$ is smaller than $\lela \wh{u} \rira$ for every $x$ and $\tau$, and is negligible outside the recirculation areas ($x=7$). The spanwise profiles $\lela \wh{w} \rira$ at $\tau=0$ and $\tau=T/4$ are $90$ degrees out of phase with the $\lela \wh{w} \rira$ profiles at $\tau=T/2$ and $\tau=3/4 T$. We also note that the spanwise velocity $\lela \wh{w} \rira$ is comparable with $\lela \ov{u} \rira$ in the bulk of the channel. 

\begin{figure}
\centering
\begin{tabular}{cc}
\includegraphics[width=0.47\textwidth]{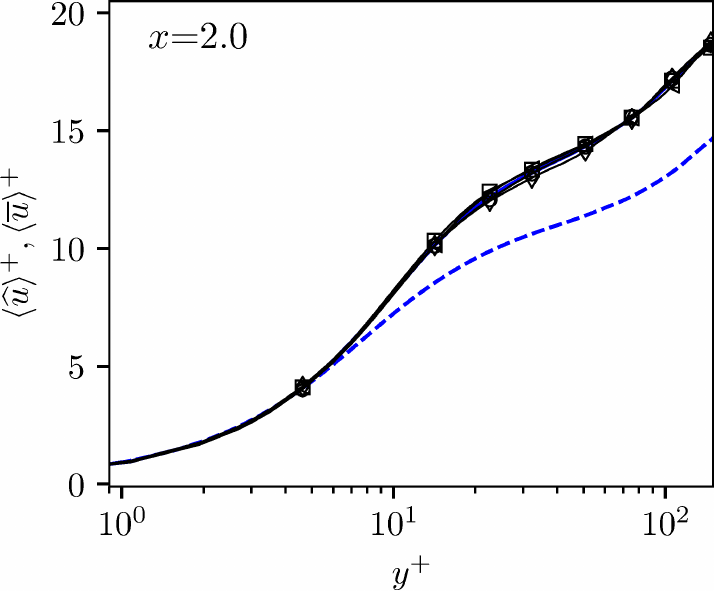}&
\includegraphics[width=0.47\textwidth]{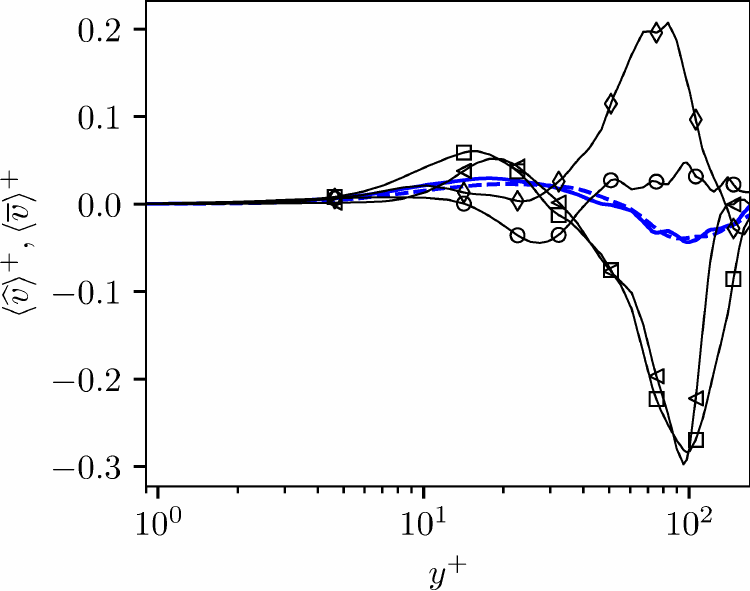}\\
\includegraphics[width=0.47\textwidth]{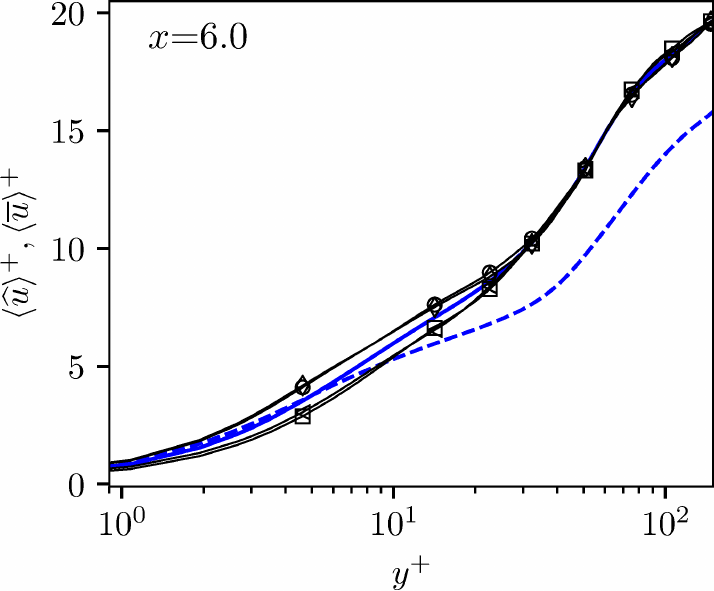}&
\includegraphics[width=0.47\textwidth]{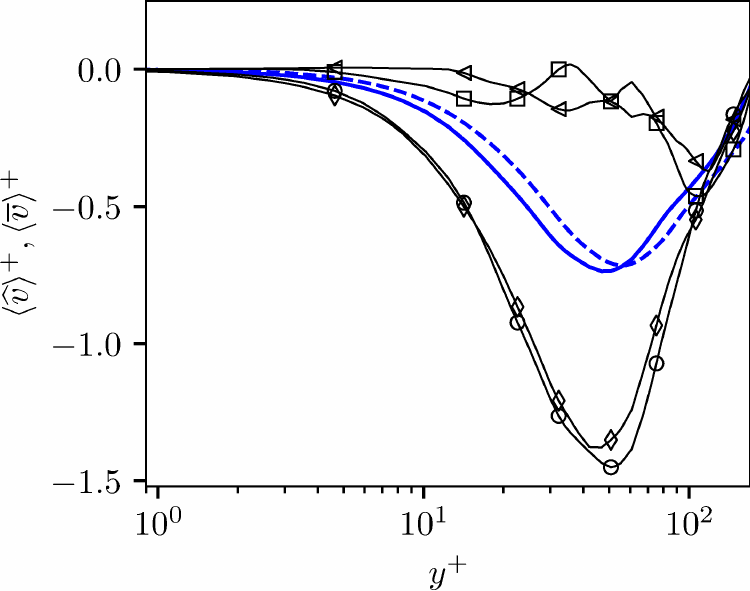}
\end{tabular}
\caption{Profiles of spanwise- and phase-averaged velocities $\lela \wh{u}\rira$, $\lela \wh{v}\rira$ scaled in viscous units at two $x$ locations. Symbols and lines are the same in figure~\ref{fig:uphasemean}.}
\label{fig:uvplus}
\end{figure}

Figure~\ref{fig:uvplus} displays the profiles of the viscous-scaled phase-averaged velocities $\lela \wh{u} \rira^+$ and $\lela \wh{v} \rira^+$ at two locations where the flow is attached. Both velocity components fluctuate more downstream than upstream of the bar. The velocity $\lela \wh{v} \rira$ increases up to an absolute maximum and then decreases monotonically to zero as the channel centerline is approached. 
\begin{figure}
\begin{tabular}{cc}
\includegraphics[width=0.47\textwidth]{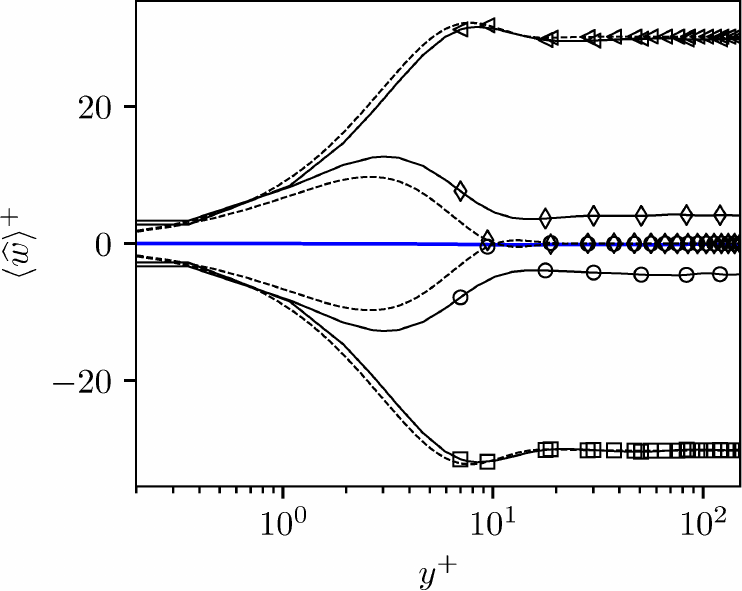}&
\includegraphics[width=0.47\textwidth]{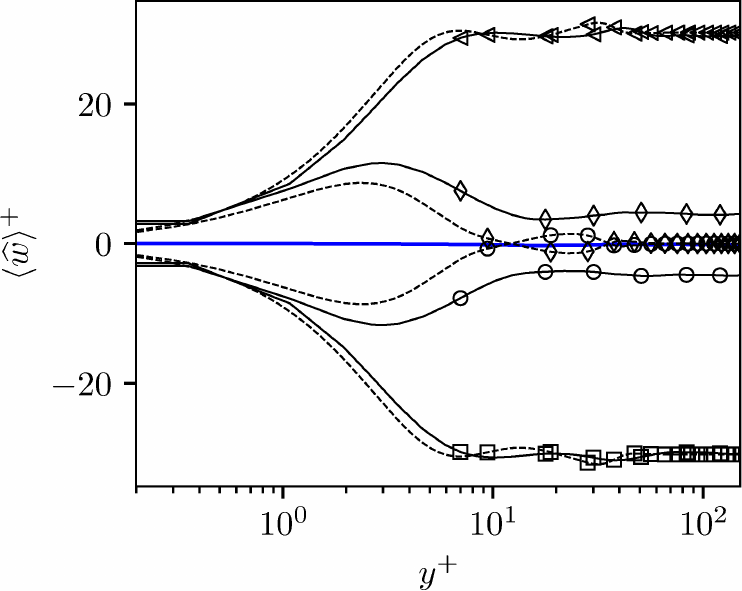}\\
(a) $x=2$ & (b) $x=5$\\
\includegraphics[width=0.47\textwidth]{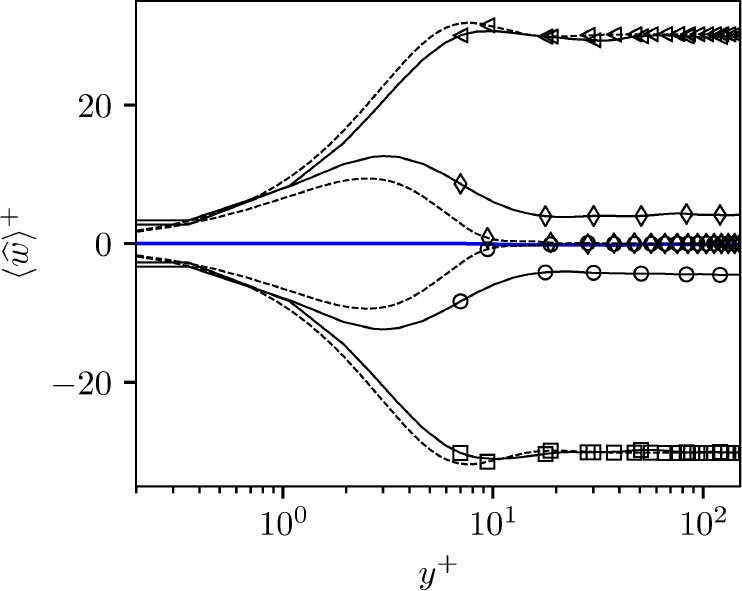}&
\includegraphics[width=0.47\textwidth]{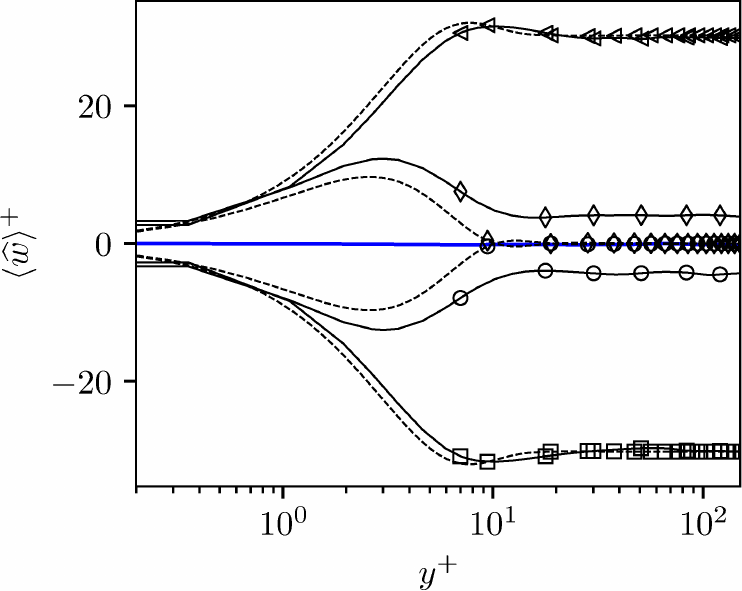}\\
(c) $x=6$ & (d) $x=7$
\end{tabular}
\caption{Comparisons between the profiles of $\lela \wh{w} \rira$ (solid lines) and the laminar $w_l$ obtained via \eqref{eq:stokes} the last term on the right hand side (dashed lines). Profiles are at $\tau=0$ (circles), $\tau=T/4$ (squares), $\tau = T/2$ (diamonds) and $\tau=3T/4$ (triangles).}
\label{fig:wplus}
\end{figure}
\begin{figure}
\centering
\includegraphics[width=0.9\textwidth]{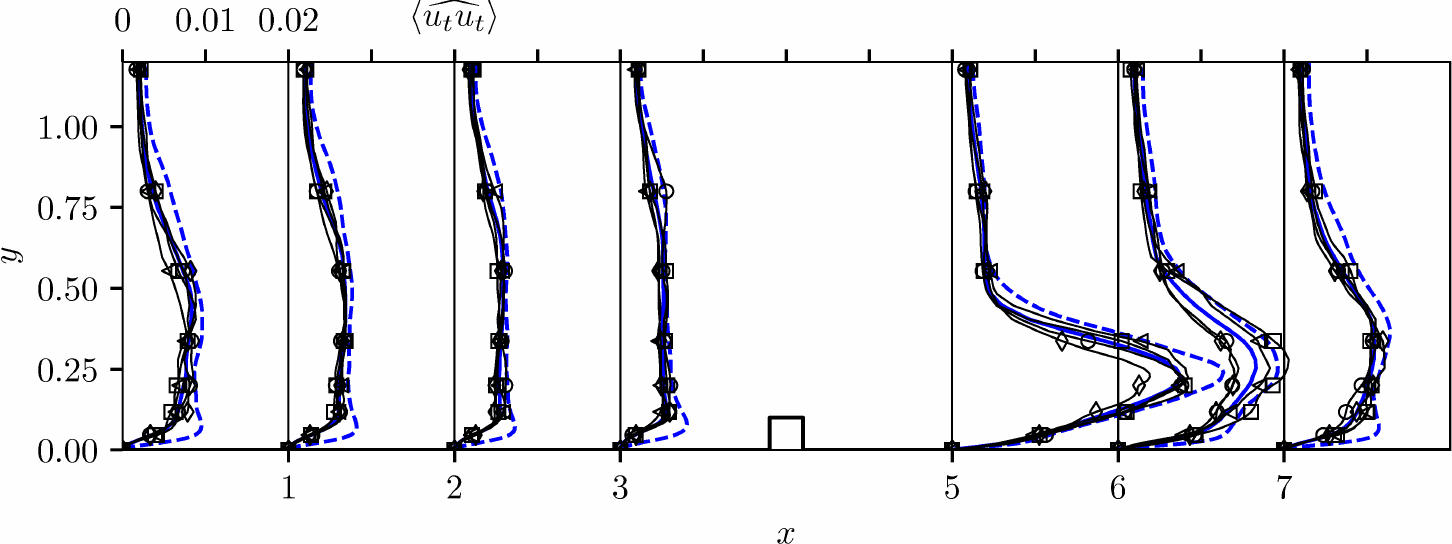}\\
\includegraphics[width=0.9\textwidth]{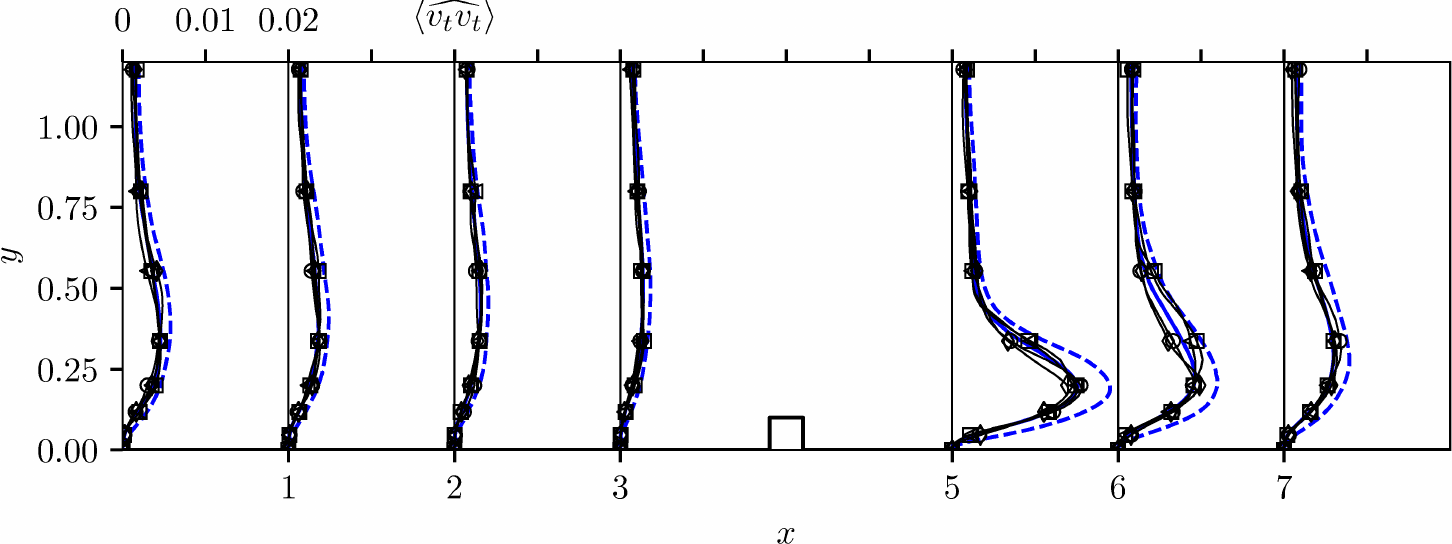}\\
\includegraphics[width=0.9\textwidth]{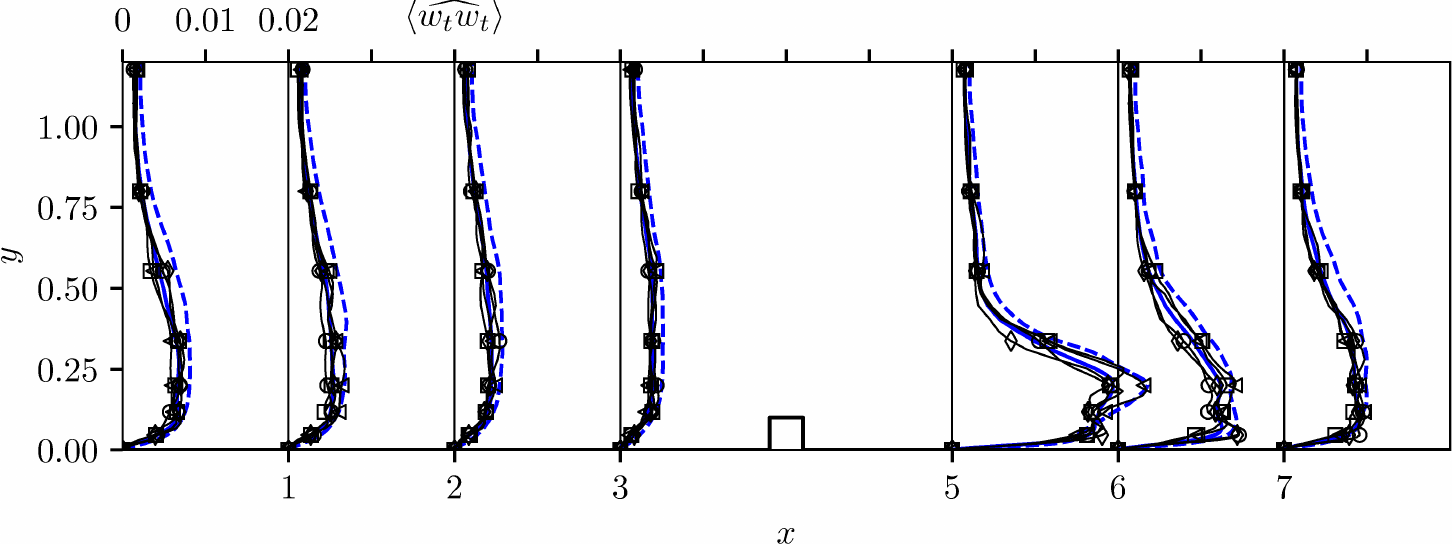}\\
\caption{Profiles of $\lela \wh{u_t u_t}\rira$, $\lela \wh{v_t v_t}\rira$, $\lela \wh{w_t w_t}\rira$ at different $x$ stations and $\tau$ (thin solid lines with symbols). Profiles are at $\tau=0$ (circle), $\tau=T/4$ (square), $\tau = T/2$ (diamond) and $\tau=3T/4$ (triangle). The phase-averaged quantities are compared with time-averaged flow statistics $\lela \ov{u_t u_t}\rira$, $\lela \ov{v_t v_t}\rira$, $\lela \ov{w_t w_t}\rira$ for the control (solid lines) and uncontrolled (dashed lines) case. The bars are not drawn to scale.}
\label{fig:uuphase}
\end{figure}
\begin{figure}
\centering
\includegraphics[width=.9\textwidth]{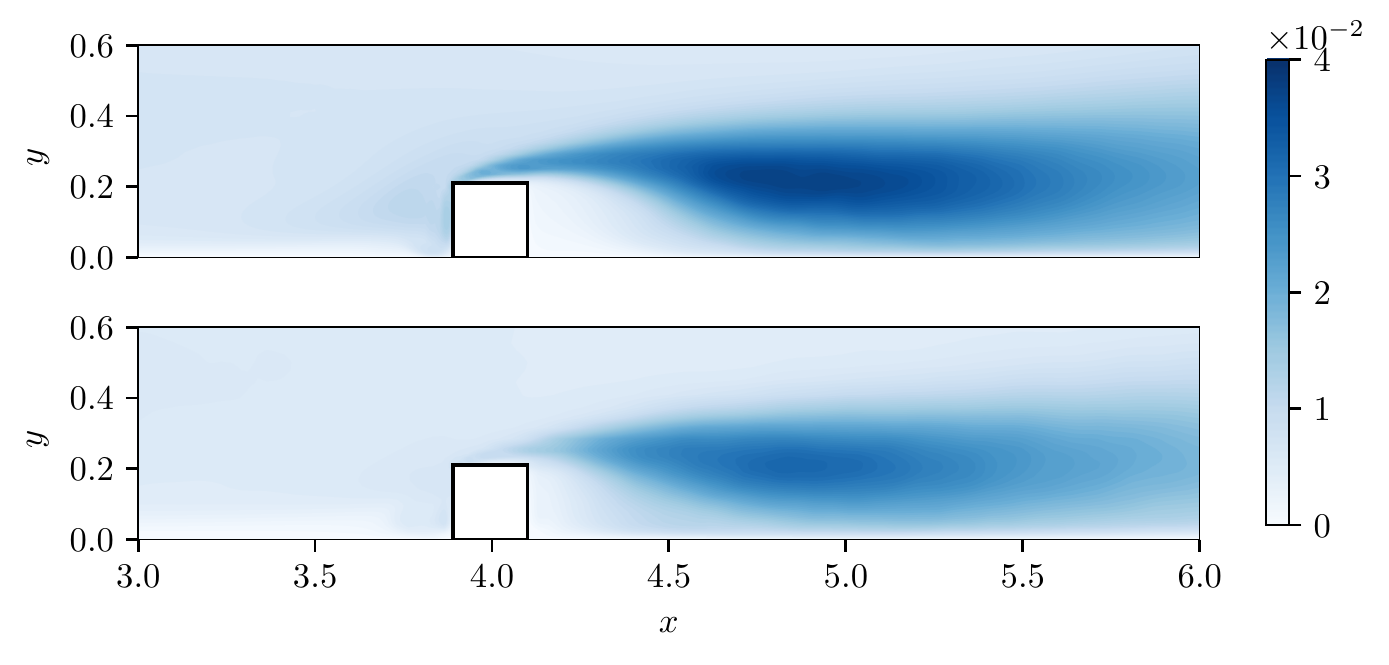}
\caption{Contour plots of turbulent kinetic energy in the region around the bar for uncontrolled (top) and controlled (bottom) case.}
\label{fig:tke}
\end{figure}
\begin{figure}
\centering
\includegraphics[width=0.9\textwidth]{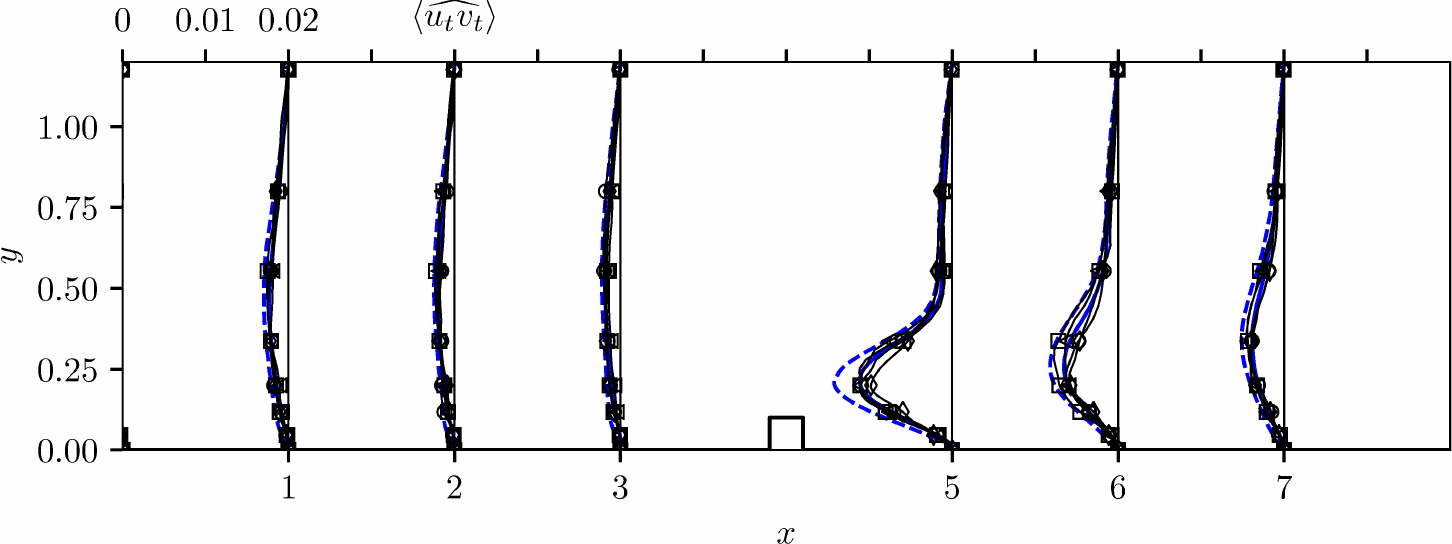}\\
\includegraphics[width=0.9\textwidth]{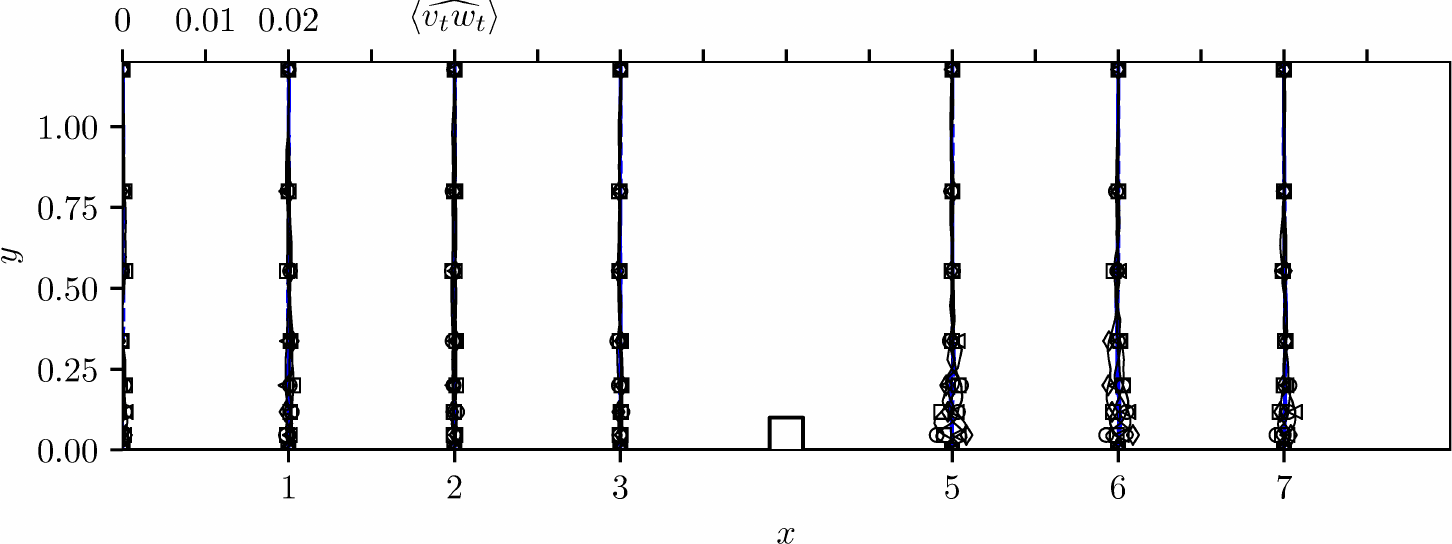}\\
\includegraphics[width=0.9\textwidth]{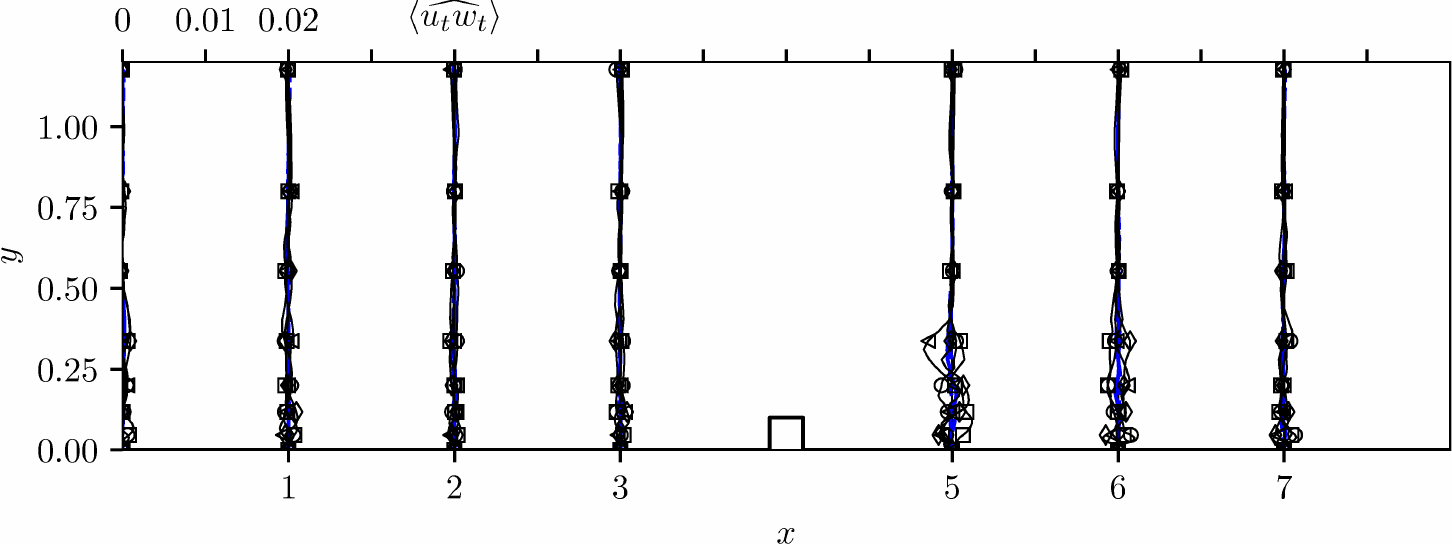}\\
\caption{Profiles of $\lela \wh{u_t v_t}\rira$ , $\lela \wh{v_t w_t}\rira$, $\lela \wh{w_t u_t}\rira$ at different $x$ stations and $\tau$ (thin solid lines with symbols). Lines and symbols are the same as in figure~\ref{fig:uuphase}. The bars are not drawn to scale.}
\label{fig:uvphase}
\end{figure}

Figure \ref{fig:wplus} depicts the profiles of $\lela \wh{w} \rira$ and $w_l$ for $\tau=0, T/4, T/2,3/4 T$ at different $x$ stations. The laminar solution $w_l$ agrees well with $\lela \wh{w} \rira$ outside the separation area ($x=2$ and $x=7$) for $\tau=T/4, 3/4 T$. At $\tau=0, T/2$, a large discrepancy occurs between the laminar and the turbulent profiles. The laminar $w_l$ tends to be null in outer region of $y^+>10$, contrarily to finite values of the turbulent $\lela \wh{w} \rira$. The differences between the laminar and turbulent flows are due to the transverse Reynolds stresses $-\p \avtez{u_t w_t}/\p x$ and $-\p \avtez{v_t w_t}/\p y$ in \eqref{eq:5}. 
Overall, the turbulent velocity $\lela \wh{w} \rira$ shows a satisfactory agreement with the laminar $w_l$, proving that the analysis of \S\ref{sec:laminar} gives a relevant physical model for the transverse oscillating mean flow obtained from the direct numerical simulations. This result is confirmed by the excellent laminar prediction of the spanwise power $\mc{P}_z$, shown in figure \ref{fig:turbpower}.  

The turbulence intensities, $\lela \wh{u_t u_t}\rira$, $\lela \wh{v_t v_t}\rira$, $\lela \wh{w_t w_t}\rira$ and $\lela \ov{u_t u_t}\rira$, $\lela \ov{v_t v_t}\rira$, $\lela \ov{w_t w_t}\rira$, are shown in figure \ref{fig:uuphase}. They are highest at the $x$ locations corresponding to the bar sides where peaks of $\lela \wh{u_t u_t}\rira$ and $\lela \ov{u_t u_t}\rira$ reach values two and four times higher than the values far from the separation regions ($x=3$ and $7$). The profiles of $\lela \wh{u_t u_t}\rira$ and $\lela \ov{u_t u_t}\rira$ show two local maxima far from the bar with the global maxima occurring at $y=0.1$ for $x=3$ and $y=0.5$ for $x=7$.    
Similar to the mean flow quantities, the phase-averaged quantities fluctuate more downstream ($4 < x < 7$) than upstream of the bar. The streamwise and spanwise correlations are more dominant than the wall-normal correlation in both the controlled and uncontrolled cases. Both $\lela \wh{w_t w_t}\rira$ and $\lela \ov{w_t w_t}\rira$ show a peak around the bar height $y=0.2$ for every $\tau$. The quantities are all reduced for any $\tau$ and at any location along the channel. Figure~\ref{fig:tke} confirms that the turbulent kinetic energy $0.5\left(\lela \ov{u_t u_t}\rira+\lela \ov{v_t v_t}\rira+\lela \ov{w_t w_t}\rira\right)$ decreases in the detached region downstream of the bar.

Figure \ref{fig:uvphase} compares the Reynolds stresses of the controlled and uncontrolled flows. The Reynolds stress $\lela \wh{u_t v_t}\rira$ is dominant over $\lela \wh{u_t w_t}\rira$ and $\lela \wh{v_t w_t}\rira$ and decreases with respect to the uncontrolled case for almost every $y$ at any $x$ and $\tau$, similar to \cite{jung-mangiavacchi-akhavan-1992}. The attenuating effect on $\lela \wh{u_t v_t}\rira$ is most intense downstream of the bar. 
The stresses $\lela \ov{v_t w_t}\rira$ and $\lela \ov{u_t w_t}\rira$ remain smaller than $\lela \wh{u_t v_t}\rira$ in the controlled case. (Note that $\lela \ov{u_t w_t}\rira$ vanishes in the averaged $z-$momentum equation of a flat-wall turbulent channel flow because of streamwise averaging, while in the channel with bars the streamwise derivative of this term is present in the balance, as shown in equation \eqref{eq:5}). The small amplitudes of these Reynolds stresses gives support to the assumption adopted in the laminar analysis of \eqref{eq:5}, where these quantities are neglected. 

\subsection{Role of the Reynolds stresses between bars} 
\label{sec:reynolds-bars}

In order to gain further insight into the role of the Reynolds stresses in the controlled channel flow, we consider a momentum balance relating the total drag to the Reynolds stresses $-\avtez{u_tv_t}$ along the horizontal line $l_m$ connecting the crests of two consecutive bars. The sum of the drag coefficients $C_p(\tau)+C_f(\tau)$ for the controlled flow is related to the product $\avtez{u}\avtez{v}$, the Reynolds stress $-\avtez{u_tv_t}$, and the friction term $Re_p^{-1}\, \p \avtez{u}/\p y|_h$ at $y=h$ via  

\begin{equation}
C_p(\tau)+C_f(\tau) \approx -\frac{L_y}{L_x}\int_{l_m} \avtez{u}\avtez{v} \,\dx-\frac{L_y}{L_x}\int_{l_m} \avtez{u_tv_t} \,\dx
+\frac{L_y}{L_x Re_p}\int_{l_m} \frac{\p \avtez{u}}{\p y} \dx,                                     
\label{eq:dragmech}
\end{equation}
where the phase dependence of the quantities is kept herein.

\begin{figure}
\centering
\includegraphics[width=0.47\textwidth]{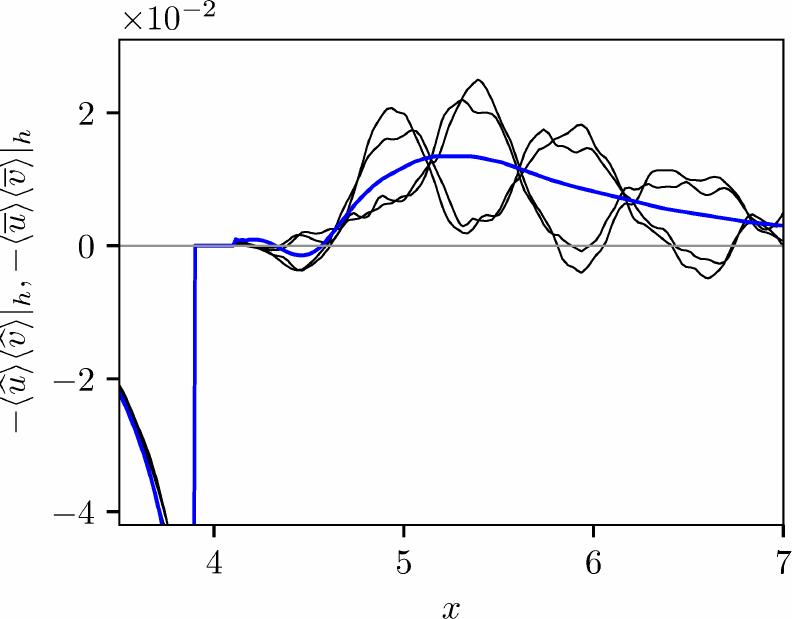}
\vspace{-3mm}
\includegraphics[width=0.47\textwidth]{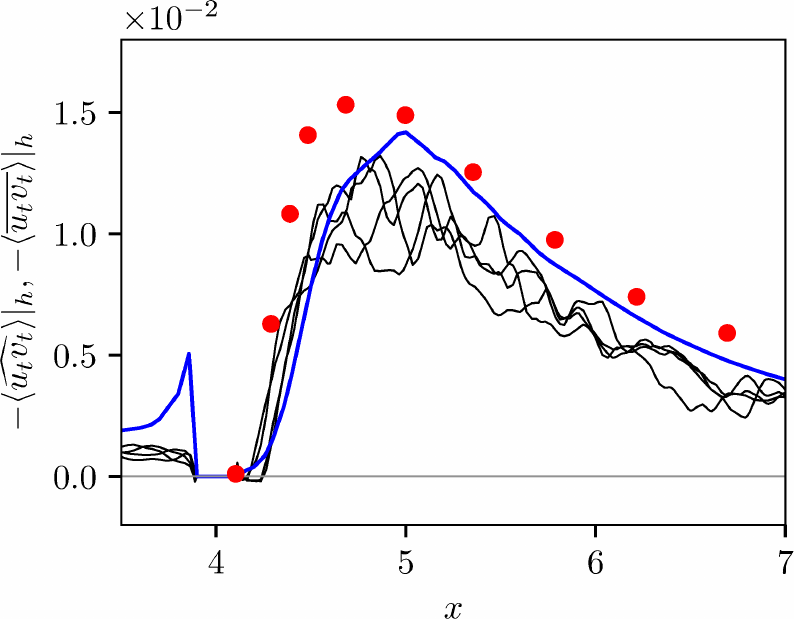}
\vspace{-3mm}
\includegraphics[width=0.47\textwidth]{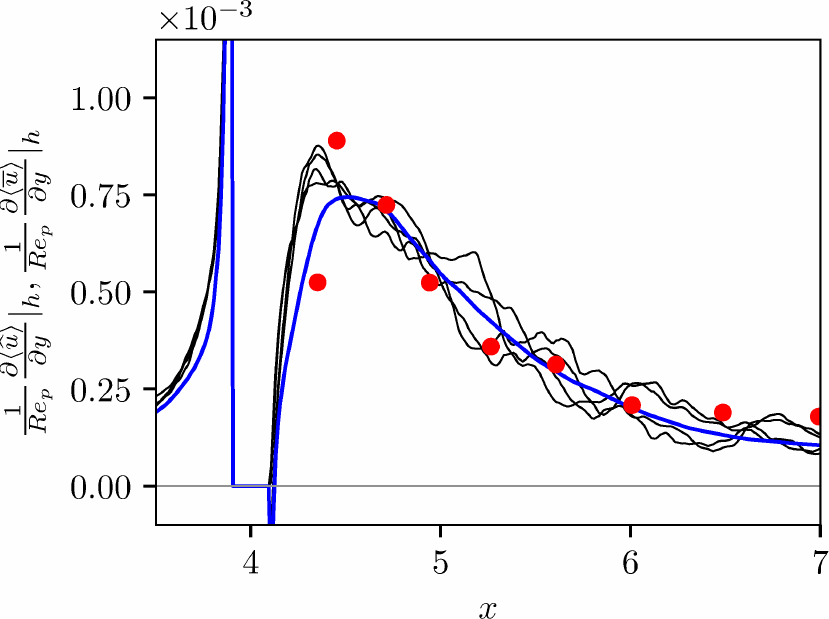}
\caption{Comparisons between the profiles of $-\avtez{u}\avtez{v}$, $-\avtez{u_tv_t}$, $Re_p^{-1} \p \avtez{u}/\p y\Bigr |_h$ of the controlled flow at phases $\tau=0, T/4, T/2, 3/4 T$ (thin solid lines) and the uncontrolled terms $-\lela \ov{u} \rira \lela \ov{v} \rira$, $-\lela{\ov{u_tv_t}}\rira$, $Re_p^{-1} \p \lela \ov{u} \rira /\p y\Bigr |_h$ (thick solid lines). The red circles are data from the direct numerical simulations of \cite{leonardi-etal-2003}.}
\label{fig:mechanism}
\end{figure}

Appendix \ref{sec:dragrey} presents the derivation of equation \eqref{eq:dragmech}. 
Note that in \eqref{eq:dragmech} the total drag $C_p(\tau)+C_f(\tau)$ is only approximated by the terms on the right hand side because $C_{cr}$, which is much smaller than $C_{ca}$, is neglected. Equation \eqref{eq:dragmech} differs from equation (5.3) in \cite{leonardi-etal-2003} as they did not include the term due to $\avtez{u}\avtez{v}$ and we use the time-ensemble average instead of the time average. 
Figure \ref{fig:mechanism} depicts the integrands of the terms in \eqref{eq:dragmech} for the controlled and uncontrolled flows. The Reynolds stresses $-\avtez{u_tv_t}$ and shear stresses $\p \avtez{u}/\p y$ are positive at every $x$ along the cavity mouth, giving a positive contribution to the overall drag for both flows. The term involving $\p \avtez{u}/\p y$ is however negligible compared to the other terms.
The term $-\avtez{u}\avtez{v}$ is positive and streamwise oscillatory downstream of the bar in the controlled case, while it is negative upstream of the bar ($2<x<4$) because the wall-normal velocity $\avtez{v}$ is large and positive. Due to conservation of mass, this upward velocity balances the decreasing streamwise mean velocity caused by the presence of the bar. 
Figure \ref{fig:mechanism} shows that the reduction of the total drag mainly occurs because the control attenuates the Reynolds stresses $-\avtez{u_tv_t}$, while the control effect of the time-average of $-\avtez{u}\avtez{v}$ and $\p \avtez{u}/\p y$ is marginal. The attenuation of $-\avtez{u_tv_t}$ occurs not only in the region where the flow is attached ($x<3.5$ and $x>5.5$), but also along $l_m$ above the vortex A ($3.5<x<4$). 
Good agreement is obtained behind the bar between our $\p \avtez{u}/\p y$ data along the channel mouth and the data from the direct numerical simulations of \cite{leonardi-etal-2003} (red circles), while the comparison between the $-\avtez{u_tv_t}$ data is only satisfactory downstream of the peak. The mismatch of the maximum values could be due to the large uncertainty caused by the large velocity fluctuations in the recirculation region.

We can interpret the reduction of the Reynolds stresses as the effect of the oscillation on the bulk turbulent flow that in turn manifests itself at the wall surfaces as the attenuation of the pressure difference on the vertical surfaces of the bars and of the wall-shear stress over the cavity wall. Further research work is needed to establish the link between the bulk Reynolds-stress reduction and the pressure changes at the bar sides.          

\section{Conclusions}
\label{sec:conclusions}

A drag reduction study on a channel flow with square bars located on each wall has been conducted using direct numerical simulations. The flow is controlled via a spanwise oscillating pressure gradient and a maximum total drag-reduction margin of $25\%$ is achieved. We have shown that a reduction of both pressure and skin-friction drag is possible via an active method, contrariwise to vortex generators and jets that may energize the fluid and increase the skin-friction drag to reduce the form drag. The drag reduction has been linked to a reduction of the Reynolds stress along the dividing streamline connecting the bar crest and the reattachment point along the cavity. The skin-friction drag is mostly reduced because the wall-shear stress of the attached flow along the cavity wall is decreased under the effect of the control, whereas the control has no significant effect on the friction drag of the separated flow and of the flow over the bar crest because the wall-shear stress is low or negative there. The pressure drag decreases because the pressure decreases in front of the bar and increases behind the bar. The lower pressure on the front of the bar is caused by the lower dynamical pressure of the mean flow approaching the bar, which is due to the reduced wall-shear stress along the cavity upstream of the bar.

The power to control the flow has been quantified and compared with the power saved thanks to drag reduction to compute the net energy saved. For the amplitudes and periods considered, the solution of the related spanwise laminar flow gives an excellent prediction of the turbulent power spent. No net power saved has been computed for the cases that feature the largest drag-reduction margins. The small net power savings, obtained for small drag-reduction margins, are within the uncertainty error of the computations, as discussed in Appendix \ref{sec:dragconv}.

Further studies are necessary to study the mechanism causing drag reduction by, for example, an analysis analogous to that of \cite{ricco-etal-2012} based on the turbulent enstrophy and dissipation or by extending the useful identity of \cite{fukagata-iwamoto-kasagi-2002} to the channel with bars to investigate the role of the Reynolds stresses on the drag-reduction effect. 
It is also interesting to determine the effect of the Reynolds number on the drag-reduction performance and whether some combinations of parameters and of the forcing actuation can further reduce drag. This technique is promising because it is shown that reduction of both pressure and skin-friction drag is possible, paving the way for research into more effective types of separation control. 

Future work should certainly be directed at improving the power balance to obtain a measurable net power saved.
It is known that a travelling wave forcing is more efficient than spanwise oscillation \citep{quadrio-ricco-viotti-2009} in reducing the skin-friction drag. A possibility to control separated flows could thus be a spanwise pressure gradient of the form $\Pi_z(x,t)=A\sin(k_x x-\omega t)$ and giving an effect similar to the one of \cite{quadrio-ricco-viotti-2009} in the attached-flow region. Such an active technique could be combined with a wall-based passive control method along the cavity where the flow separates. An idea could be the method proposed by \cite{heenan-morrison-1998}, who were able to reduce the separation drag by $9\%$ in a forward-facing step by making permeable the cavity wall in front of the step and obtaining a decrease of the root mean square of the pressure fluctuations.    

\vspace{0.25cm}
\section*{Acknowledgements}
We would like to thank the Department of Mechanical Engineering at the University of Sheffield and the Agency for Science, Technology and Research (A*STAR) in Singapore for funding this research. We would like to express our gratitude towards the A*STAR Computational Resource Centre for their kind assistance during the use of their high performance computing resource. 
GP chose the references, conceived and wrote \S\ref{sec:numerics}, \S\ref{sec:laminar}, \S\ref{sec:reynolds-bars}, and the Appendixes, modified Nek5000 for this problem, and generated half of the data at low $T$ in tables 1-3 and figures 1-5. Part of this work was presented by GP at the European Drag Reduction and Flow Control Meeting, Bad Herrenalb, Germany in March 2019. We would like to thank Professors S. Chernyshenko, S. He, and P. Luchini for the useful discussions. We are also indebted to Professor Paul Fischer for the advices about the mesh creation and for the useful information on running Nek5000. 

\vspace{0.25cm}
\section*{Declaration of Interests}
The authors report no conflict of interest.

\appendix

\section{Uncertainty of the drag components for the uncontrolled flow}
\label{sec:dragconv}

The sensitivity of the drag coefficients on the mesh resolution and the time step, for the uncontrolled case, is examined in table \ref{tab:tab1app}. Time accuracy is checked by comparing the two results obtained using the same grid with $l_x=6$, but with two different time s.pdf, $\Delta t=1.69 \cdot 10^{-4}$ and $\Delta t=8.45\cdot 10^{-5}$. A percentage variation of $0.4\%$ in the total drag $C_p+C_f$ is obtained and variations of all the other coefficients are less than $1\%$, indicating that the time step used is adequate. The accuracy of the drag with space resolution is evaluated comparing results with three different grids. Each grid has the same number of spectral elements, but a different number $l_x$ of GLL points is employed along each direction because the SEM order of convergence is spectral with $l_x$. The computation with $l_x=8$ gives variations of less than $1\%$ if compared to the one with $l_x=6$ except for $C_{cr}$ which, however, represents a small contribution to the total drag. The most resolved case ($l_x=10$) gives the highest percentage variation of the total drag ($1\%$) with respect to the less resolved run ($l_x=6$). These tests prove that the computation with $l_x=6$ and a time step $\Delta t=1.69 \cdot 10^{-4}$ guarantees accurate values of the drag coefficients and therefore this configuration is adopted for all the simulations with control.   
      
\begin{table}
\begin{center}
\begin{tabular}{c|c|c|c|c|c|c|c}
$l_x$  &    $\Delta t (10^{-4})$        & $t_f-t_i$   &  $C_p+C_f (10^{-3})$        &   $C_{cr} (10^{-4})$           &  $C_{ca} (10^{-3})$            &  $C_p (10^{-3})$                    \\
6     & 1.69  & 29.99  & 10.00  &  -1.10  &    2.01 &     8.11  \\

6     & 0.84  & 20.99  & 9.97   &  -1.11  &    2.01 &     8.06  \\                                                         

8     & 1.69  & 24.12  & 9.96   &  -1.12  &    1.99 &     8.07  \\                                                         

10    & 1.06  & 20.18  & 10.10  &  -1.11  &    2.00 &     8.15  \\                                                       
\end{tabular}
\end{center}
\caption{Space and time resolution checks for the drag components.}
\label{tab:tab1app}
\end{table}

\section{Drag balance and power for driving the flow along $x$}
\label{sec:dragbalpow}

\begin{figure}
\centering
\psfrag{x}{$x$}
\psfrag{y}{$y$}
\psfrag{Lx}{$L_x$}
\psfrag{Ly}{$L_y$}
\psfrag{Sxi}{${S_x}_i$}
\psfrag{Sxf}{${S_x}_f$}
\psfrag{S}{$S_w$}
\psfrag{i}{$\mb{i}$}
\psfrag{-i}{-$\mb{i}$}
\psfrag{j}{$\mb{j}$}

\psfrag{-j}{-$\mb{j}$}
\psfrag{Sl}{$S_l$}
\psfrag{Sr}{$S_r$}
\includegraphics[width=0.7\textwidth]{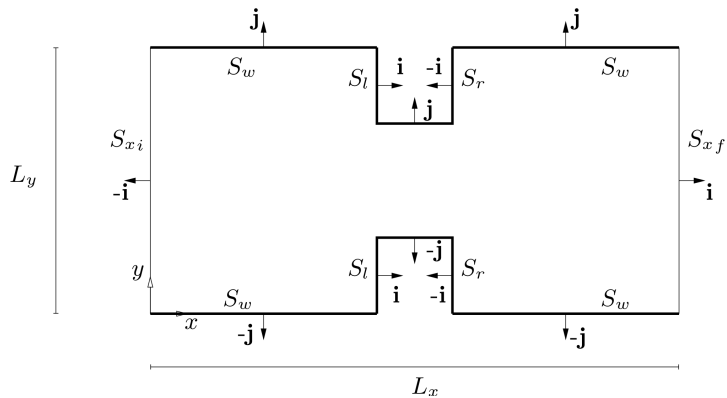}
\caption{Schematic of the fluid volume $V$. The unit normal vectors to each surface are represented by arrows with full tip.  The bars are not drawn to scale.}
\label{fig:dragbal}
\end{figure}

In this Appendix, we find the expression for the power spent to drive the flow along the streamwise direction. We first consider the volume integral of the streamwise momentum equation

\begin{equation}
\frac{\p }{\p t}\intdv{u}+\flux{u \mb{u}}=-\flux{p\, \mb{i}}+\frac{1}{Re_p}\flux {\nabla u}, 
\label{eq:dragbal1}
\end{equation}

where $S$ is the boundary surface of $V$, $\mb{u}$ is the velocity vector and $\mb{n}$ the unit vector pointing out of the control volume. The first integral in \eqref{eq:dragbal1} is null because the simulations are at constant mass flow rate. 
The boundary surface $S$ is composed of the opposite wall surfaces $S_w$, the plane surfaces ${S_x}_i$ and ${S_x}_f$ of normal $\mb{i}$ and $-\mb{i}$ and the plane surfaces ${S_z}_i$ and ${S_z}_f$ of the units vectors $\mb{k}$ and $-\mb{k}$, as shown in figure \ref{fig:dragbal}. The second integral in \eqref{eq:dragbal1} is null because $\mb{u} \cdot \mb{n}=0$ on $S_w$ and because the sum of the integrals over surfaces of opposite unit vectors is zero due to the periodic boundary conditions.

The pressure term in \eqref{eq:dragbal1} is obtained by defining the vector $\mb{p}=[p;0;0]$ and applying the divergence theorem, i.e.,

\begin{equation}
-\intdv{\frac{\p p}{\p x}}=-\intdv{\nabla \cdot \mb{p}}=-\flux{\mb{p}}=-\flux{p \mb{i}}.
\end{equation}

The surfaces $S_l$, $S_r$, ${S_x}_f$, and ${S_x}_i$ are the only ones with $\mb{i}\cdot \mb{n}\ne 0$ and the pressure integral in \eqref{eq:dragbal1} is 
 
\begin{equation}
\begin{aligned}
-\flux{p\, \mb{i}}&=-\ler \intsgn{p}{S_l}-\intsgn{p}{S_r}+\intsgn{p}{{S_x}_f}-\intsgn{p}{{S_x}_i} \rir=\\
                  &=-\ler C_pL_x L_z+\intyz{0}{L_y}{0}{L_z}{p|_{L_x}}-\intyz{0}{L_y}{0}{L_z}{p|_0}\rir=\\
                  &=-C_p(t)L_x L_z-\Pi_x(t)L_x L_y L_z,
\end{aligned}
\label{eq:dragbal2}
\end{equation} 

where the definitions $p|_{L_x}=\phi|_{L_x}+\Pi_z z$, $p|_0=\phi|_0+\Pi_x L_x+\Pi_z z$ and $\phi|_{L_x}=\phi|_0$ have been used.
The contributions of ${S_x}_i$, ${S_x}_f$, ${S_z}_i$, ${S_z}_f$ to the last term in \eqref{eq:dragbal1} is zero, similar to the term $\flux{u \mb{u}}$ and 

\begin{equation}
\begin{aligned}
\frac{1}{Re_p}\flux{\nabla u}=&\frac{1}{Re_p}\intsgn{ \nabla u \cdot \mb{n}}{S_{ca}}+\frac{1}{Re_p}\intsgn{ \nabla u \cdot \mb{n}}{S_{cr}}+\\
           &+\frac{1}{Re_p}\intsgn{ \nabla u \cdot \mb{n}}{S_l}+\frac{1}{Re_p}\intsgn{ \nabla u \cdot \mb{n}}{S_r}=\\
           =&\frac{1}{Re_p}\intsgn{ \nabla u \cdot \mb{n}}{S_{ca}}+\frac{1}{Re_p}\intsgn{ \nabla u \cdot \mb{n}}{S_{cr}}=\\
           =&-C_{ca}(t)L_x L_z-C_{cr}(t)L_x L_z=-C_f(t) L_x L_z,
\end{aligned}
\label{eq:dragbal3}
\end{equation}

where the minus sign is due to the orientation of $\mb{n}$, $S_{ca}$ and $S_{cr}$ are the cavity and crest surfaces, respectively. The integrals over $S_l$ and $S_r$ in \eqref{eq:dragbal3} are null because of continuity and the no-slip condition for $v$ and $w$, i.e., over the surface $S_l$:

\begin{equation}
\frac{1}{Re_p}\intsgn{ \nabla u \cdot \mb{n}}{S_l}=\frac{1}{Re_p}\intyz{0}{h}{0}{L_z}{\frac{\p u}{\p x}}=-\frac{1}{Re_p}\intyz{0}{h}{0}{L_z}{ \ler \frac{\p v}{\p y}+\frac{\p w}{\p z} \rir}=0.
\end{equation}   

By substituting \eqref{eq:dragbal3} and \eqref{eq:dragbal2} into \eqref{eq:dragbal1} we find

\begin{equation}
C_p(t)+C_f(t)=-\Pi_x(t) L_y
\label{eq:dragbal4}
\end{equation} 

or its time-averaged version $C_p+C_f=-\ov{\Pi}_x L_y$. 

We consider the first term of equation 1-108 in \cite{hinze-1975} averaged over $V$ 

\begin{equation}
-\frac{1}{V}\intdv{\frac{\p up}{\p x}}=-\frac{1}{V} \flux{up\, \mb{i}}.  
\label{eq:powx1}
\end{equation}
 
As the surfaces $S_l$, $S_r$, ${S_x}_f$ and ${S_x}_i$ are the only ones with $\mb{i}\cdot \mb{n}\ne 0$, we obtain

\begin{equation}
\begin{aligned}
\flux{up \mb{i}}&=\intsgn{up}{S_l}-\intsgn{up}{S_r}+\intsgn{up}{{S_x}_f}-\intsgn{up}{{S_x}_i}=\\
&=\intsgn{up}{{S_x}_f}-\intsgn{up}{{S_x}_i}=\Pi_x(t) L_x \intyz{0}{L_y}{0}{L_z}{u|_{L_x}},   
\label{eq:powx2}
\end{aligned}
\end{equation}

because of the no-penetration condition on $S_l$ and $S_r$, $p|_{L_x}=\phi|_{L_x}+\Pi_z z$, $p|_0=\phi|_0+\Pi_x L_x + \Pi_z z$ and $\phi|_{L_x}=\phi|_0$. The continuity equation in integral form states that the mass flow rate is the same at any cross section $S_x(x)$ of normal $\pm \mb{i}$ and independent of $x$  

\begin{equation}
\intsgn{u}{{S_x}(x)}=\intyz{0}{L_y}{0}{L_z}{u|_{L_x}}
\label{eq:powx3}
\end{equation}

The volume-averaged streamwise velocity $U_b(t)$ can be written using \eqref{eq:powx3} as

\begin{equation}
\begin{aligned}
U_b(t)&=\frac{1}{V} \intdv{u}=\frac{1}{V} \int_0^{L_x} \mr{d} x \intsgn{u}{{S_x}(x)}=\\
&=\frac{1}{V} \int_0^{L_x} \mr{d} x \intyz{0}{L_y}{0}{L_z}{u|_{L_x}}=\frac{L_x}{V} \intyz{0}{L_y}{0}{L_z}{u|_{L_x}},
\label{eq:powx4}
\end{aligned}
\end{equation}

where $U_b$ does not depend on $t$ because the mass flow rate is constant.
 
By using \eqref{eq:powx4} and \eqref{eq:powx2}, the power $\mr{P}_x$ per unit of volume used to drive the flow along $x$ is the time average of \eqref{eq:powx1}, i.e.,

\begin{equation}
\mr{P}_x=-\frac{1}{t_f-t_i}\int_{t_i}^{t_f} \frac{1}{V} \flux{up\, \mb{i}}\;\mr{d} t =-\ov{\Pi}_x U_b.
\label{eq:powx5}
\end{equation}

\section{Numerical solution of the laminar-flow equation}
\label{sec:numhelm}

Equation \eqref{eq:helmolts2} is discretized by using a centered finite-difference scheme

\begin{equation}
\wtil{w}_{a,b}=C_{a,b}\,\wtil{w}_{a,b-1}+D_{a,b}\,\wtil{w}_{a-1,b}+F_{a,b}\,\wtil{w}_{a+1,b}+G_{a,b}\,\wtil{w}_{a,b+1},
\label{eq:jacobi_stokes}
\end{equation}

where $a$ and $b$ are the indices to the grid points and $C_{a,b}$, $D_{a,b}$, $F_{a,b}$ and $G_{a,b}$ become 

\begin{subequations}
\begin{equation}
C_{a,b}=\frac{\frac{1}{\Delta y^2}+\frac{Re_p}{2\Delta y}\, \lela \wh{v}\rira_{a,b} }{i\,\omega\,Re_p+\frac{2}{\Delta x^2}+\frac{2}{\Delta y^2}}, \hspace{1cm} D_{a,b}=\frac{\frac{1}{\Delta x^2}+\frac{Re_p}{2\Delta x}\, \lela \wh{u}\rira_{a,b}   }{i\,\omega\,Re_p+\frac{2}{\Delta x^2}+\frac{2}{\Delta y^2}},  
\end{equation}
\begin{equation}
F_{a,b}=\frac{\frac{1}{\Delta x^2}-\frac{Re_p}{2\Delta x}\, \lela \wh{u}\rira_{a,b}   }{i\,\omega\,Re_p+\frac{2}{\Delta x^2}+\frac{2}{\Delta y^2}}, \hspace{1cm} G_{a,b}=\frac{\frac{1}{\Delta y^2}-\frac{Re_p}{2\Delta y}\, \lela \wh{v}\rira_{a,b} }{i\,\omega\,Re_p+\frac{2}{\Delta x^2}+\frac{2}{\Delta y^2}}.
\end{equation}
\end{subequations}

The boundary conditions \eqref{eq:bc_stokes_1}-\eqref{eq:bc_stokes_5} read:

\begin{subequations}
\begin{equation}
\wtil{w}_{1,b}=\wtil{w}_{l,b}=-\frac{A i}{\omega}\, \hspace{1cm}   1\leq b\leq m,
\label{eq:bc_helmolts_1}
\end{equation}
\begin{equation}
\wtil{w}_{a,m}=\wtil{w}_{a,1}=-\frac{A i}{\omega}\,   \hspace{1cm}   l\leq a\leq n_x,
\end{equation}
\begin{equation}
\wtil{w}_{l,b}=-\frac{A i}{\omega}\, \hspace{1cm}   1\leq b\leq m,
\end{equation}
\begin{equation}
\wtil{w}_{a,n_y+1}=\wtil{w}_{a,n_y-1},  \hspace{1cm} 1\le a\le n_x,
\end{equation}
\begin{equation}
\wtil{w}_{0,b}=\wtil{w}_{n_x-1,b},    \hspace{1cm} 1\le b\le n_y,  \hspace{1cm} \wtil{w}_{n_x,b}=\wtil{w}_{1,b},    \hspace{1cm} 1\le b\le n_y,
\label{eq:bc_helmolts_5}
\end{equation}
\end{subequations}

where $n_x$ and $n_y$ are the indices of the last points along $x$ and $y$. The scheme \eqref{eq:jacobi_stokes} can be directly applied to the interior points. The equation \eqref{eq:jacobi_stokes} is modified near the boundaries to take  boundary conditions \eqref{eq:bc_helmolts_1}-\eqref{eq:bc_helmolts_5} into account:
 
\begin{itemize}
\item ``square" scheme ($\wtil{w}_{0,b}=\wtil{w}_{n_x-1,b}$)
\end{itemize}

\begin{subequations}
\begin{equation}
\wtil{w}_{1,b}=C_{1,b}\,\wtil{w}_{1,b-1}+D_{1,b}\,\wtil{w}_{n_x-1,b}+F_{1,b}\,\wtil{w}_{2,b}+G_{1,b}\,\wtil{w}_{1,b+1}, 
\end{equation}
\begin{equation}
m+1\le\ b \le n_y-1,
\end{equation}
\end{subequations}

\begin{itemize}
\item ``triangle" scheme ($\wtil{w}_{a,n_y+1}=\wtil{w}_{a,ny-1}$)
\end{itemize}

\begin{subequations}
\begin{equation}
\wtil{w}_{a,n_y}=C_{a,n_y}\,\wtil{w}_{a,n_y-1}+D_{a,n_y}\,\wtil{w}_{a-1,n_y}+F_{a,n_y}\,\wtil{w}_{a+1,n_y}+G_{a,n_y}\,\wtil{w}_{a,ny-1},    
\end{equation}
\begin{equation}
2 \le a\le n_x-2,
\end{equation}
\end{subequations}

\begin{itemize}
\item ``empty circle" scheme ($\wtil{w}_{n_x,b}=\wtil{w}_{1,b}$)
\end{itemize}

\begin{subequations}
\begin{equation}
\wtil{w}_{n_x-1,b}=C_{n_x-1,b}\,\wtil{w}_{n_x-1,b-1}+D_{n_x-1,b}\,\wtil{w}_{n_x-2,b}+F_{n_x-1,b}\,\wtil{w}_{1,b}+G_{n_x-1,b}\,\wtil{w}_{n_x-1,b+1}, 
\end{equation}
\begin{equation}
3\le\ b \le n_y-1.
\end{equation}
\end{subequations}

Points with indices $1,n_y$, $n_x-1,2$ and $n_x-1,n_y$ are not marked with any symbol because they satisfy these two boundary conditions

\begin{itemize}
\item point $1,n_y$,   \hspace{0.5cm} ($\wtil{w}_{1,n_y+1}=\wtil{w}_{1,n_y-1}$, $\wtil{w}_{0,n_y}=\wtil{w}_{n_x-1,n_y}$),
\end{itemize}
 
\begin{equation}
\wtil{w}_{1,n_y}=C_{1,n_y}\,\wtil{w}_{1,n_y-1}+D_{1,n_y}\,\wtil{w}_{n_x-1,n_y}+F_{1,n_y}\,\wtil{w}_{2,n_y}+G_{1,n_y}\,\wtil{w}_{1,n_y-1},
\end{equation}

\begin{itemize}
\item point $n_x-1,2$,   \hspace{0.5cm} $\wtil{w}_{n_x-1,1}=-\frac{A}{\omega}\,i$,$\wtil{w}_{n_x,2}=\wtil{w}_{1,2}$,
\end{itemize}

\begin{equation}
\wtil{w}_{n_x-1,2}=-C_{n_x-1,2}\,\frac{A}{\omega}+D_{n_x-1,2}\,\wtil{w}_{n_x-2,2}+F_{n_x-1,2}\,\wtil{w}_{1,2}+G_{n_x-1,2}\,\wtil{w}_{n_x-1,3}
\end{equation}

\begin{itemize}
\item point $n_x-1,n_y$,   \hspace{0.5cm} $\wtil{w}_{n_x-1,n_y+1}=\wtil{w}_{n_x-1,n_y-1}$, $\wtil{w}_{n_x,n_y}=\wtil{w}_{1,n_y}$,
\end{itemize}

\begin{equation}
\begin{aligned}
\wtil{w}_{n_x-1,n_y}=&C_{n_x-1,n_y}\,\wtil{w}_{n_x-1,n_y-1}+D_{n_x-1,n_y}\,\wtil{w}_{n_x-2,n_y}+F_{n_x-1,n_y}\,\wtil{w}_{1,n_y}\\
&+G_{n_x-1,n_y}\,\wtil{w}_{n_x-1,n_y-1}.
\end{aligned}
\end{equation}

The equations are solved via a successive over-relaxation method 

\begin{equation}
\wtil{w}^{new}_{a,b}=\wtil{w}_{a,b}+\alpha\, r_{a,b},
\end{equation}

where the residue is defined as

\begin{equation}
r_{a,b}=C_{a,b}\,\wtil{w}_{a,b-1}+D_{a,b}\,\wtil{w}_{a-1,b}+F_{a,b}\,\wtil{w}_{a+1,b}+G_{a,b}\,\wtil{w}_{a,b+1}-\wtil{w}_{a,b}.
\end{equation}

\begin{sloppypar}
The convergence of the algorithm is guaranteed by monitoring $\left \| r \right \|=\sqrt{\sum_{a=1}^{n_x-1} \sum_{b=1}^{n_y} r_{a,b}\,r^{cc}_{a,b}}$, where $cc$ indicates the complex conjugate. 
\end{sloppypar}

\section{Relation with the laminar power of the oscillating wall}
\label{app:powoscwall}

We consider the smooth laminar channel with spanwise oscillation of walls 

\begin{equation}
\begin{cases}
\frac{\p w_s}{\p \tau}=\frac{1}{Re_p} \frac{\p^2 w_s}{\p y^2} \\
w_s(0,\tau)=w_s(2,\tau)=
A \sin{(\omega \tau)}=\omega \int_0^{\tau} A \cos{(\omega \wh{\tau})} \mr{d} \wh{\tau}=\omega \int_0^{\tau} \Pi_z \mr{d} \wh{\tau},
\end{cases}
\label{eq:wallosc1}
\end{equation} 

with the laminar power 

\begin{equation}
\mc{P}_s=\frac{1}{T Re_p}\int_0^T w_s \frac{\p w_s}{\p y} \bigg \vert_{y=0} \dta.
\label{eq:wallosc2}
\end{equation}
 
The system \eqref{eq:wallosc1} is transformed into the problem

\begin{equation}
\begin{cases}
\frac{\p w_l}{\p \tau}=\frac{1}{Re_p} \frac{\p^2 w_l}{\p y^2}-\Pi_z, \\
w_l(0,\tau)=w_l(2,\tau)=0, 
\end{cases}
\label{eq:wallosc3}
\end{equation} 

by using the equation 

\begin{equation}
w_s=\omega \ler w_l+\int_0^{\tau} \Pi_z\, \mr{d} \wh{\tau} \rir.
\label{eq:wallosc4}
\end{equation}

An expression for the spanwise volume-averaged velocity $W_{bs}=0.5\int_0^2 w_s \mr{d} y$ is obtained by averaging along the height of the channel ($0<y<2$),

\begin{equation}
\frac{\mr{d} W_{bs}}{\mr{d} \tau}=\frac{1}{2Re_p} \ler \frac{\p w_s}{\p y} \bigg \vert_{y=2}-\frac{\p w_s}{\p y} \bigg \vert_{y=0}  \rir=-\frac{1}{Re_p}\frac{\p w_s}{\p y} \bigg \vert_{y=0}.
\label{eq:wallosc6}
\end{equation}

The relationship between $W_{bs}$ and $W_{bl}$, defined in \S\ref{sec:laminar}, is obtained by averaging \eqref{eq:wallosc4} in the range $0<y<2$,

\begin{equation}
W_{bs}=\omega \ler W_{bl}+\int_0^{\tau} \Pi_z\, \mr{d} \wh{\tau} \rir.
\label{eq:wallosc7}
\end{equation} 

By substituting \eqref{eq:wallosc7} into \eqref{eq:wallosc6} one find

\begin{equation}
\frac{1}{Re_p}\frac{\p w_s}{\p y} \bigg \vert_{y=0}=-\omega\ler \frac{\mr{d} W_{bl}}{\mr{d} \tau}+ \Pi_z\rir.
\label{eq:wallosc8}
\end{equation}

By using \eqref{eq:wallosc8}, \eqref{eq:wallosc2} reads   

\begin{equation}
\begin{aligned}
\mc{P}_s
&=
-\frac{\omega^2}{T}\int_0^T \ler  \int_0^{\wh{\tau}} \Pi_z \dta \rir \ler \frac{\mr{d} W_{bl}}{\mr{d} \wh{\tau}}+ \Pi_z\rir \mr{d} \wh{\tau}=\\
&
=
-
\frac{\omega^2}{T} 
\left( 
\int_0^T \frac{\mr{d} W_{bl}}{\mr{d} \wh{\tau}} \; 
\int_0^{\wh{\tau}} \Pi_z \dta \;\mr{d} \wh{\tau}
+
\int_0^T \Pi_z \; \int_0^{\wh{\tau}} \Pi_z \dta \;\mr{d} \wh{\tau} 
\right).
\end{aligned}
\label{eq:wallosc9}
\end{equation}

By integrating by parts the first integral in the square bracket of \eqref{eq:wallosc9} one finds

\begin{equation}
\int_0^T \frac{\p W_{bl}}{\p \wh{\tau}} \; \int_0^{\wh{\tau}} \Pi_z \dta \mr{d} \wh{\tau}=\les W_{bl} \int_0^{\wh{\tau}} \Pi_z \dta \ris_0^T-\int_0^T W_{bl} \;\Pi_z \mr{d} \wh{\tau}=-\int_0^T W_{bl} \;\Pi_z \dta,
\label{eq:wallosc10}
\end{equation}

because $\int_0^T \Pi_z \dta=0$. The second integral in \eqref{eq:wallosc9} is null because 

\begin{equation}
\int_0^T \Pi_z \; \int_0^{\wh{\tau}} \Pi_z \dta\;\mr{d}\wh{\tau} = \les  \ler \int_0^{\wh{\tau}} \Pi_z \dta \rir^2 \ris_0^T-\int_0^T \Pi_z \; \int_0^{\wh{\tau}} \Pi_z \dta \;\mr{d}\wh{\tau},
\label{eq:wallosc11}
\end{equation}

and 

\begin{equation}
\int_0^T \Pi_z \; \int_0^{\wh{\tau}} \Pi_z \dta \;\mr{d}\wh{\tau} 
= 
\frac{1}{2}\les  \ler \int_0^{\wh{\tau}} \Pi_z \dta \rir^2 \ris_0^T=0.
\label{eq:wallosc12}
\end{equation}

The substitution of \eqref{eq:wallosc10} and \eqref{eq:wallosc12} into \eqref{eq:wallosc9} gives 

\begin{equation}
\mc{P}_s=-\omega^2\ler-\frac{1}{T}\int_0^T W_{bl} \;\Pi_z \dta \rir=-\omega^2 \mc{P}_{z,l}=-4\pi^2 T^{-2}\; \mc{P}_{z,l}.
\label{eq:wallosc13}
\end{equation}

Equation \eqref{eq:wallosc13} shows that the laminar power spent of a smooth oscillating wall $\mc{P}_s$ can be obtained by multiplying the laminar power spent of the oscillating pressure gradient $\mc{P}_{z,l}$ by a term proportional to $T^{-2}$.

\section{Drag dependence on the Reynolds stresses}
\label{sec:dragrey}

\begin{figure}
\centering
\includegraphics[width=1.0\textwidth]{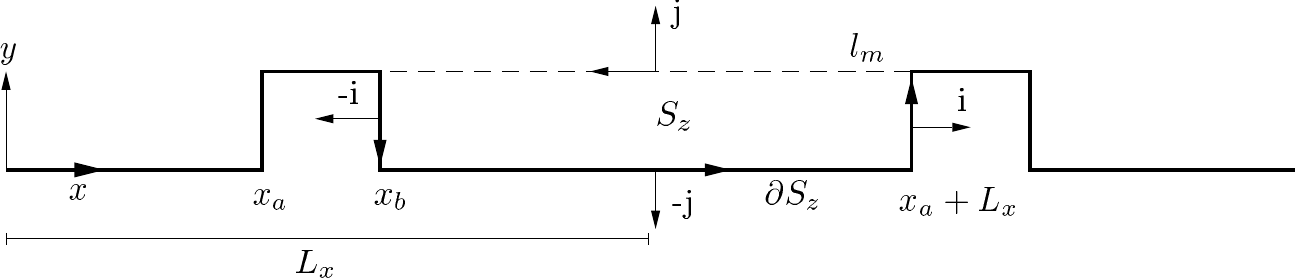}
\caption{Sketch of the rectangle $S_z$ of boundary $\p S_z$ used for \eqref{eq:tezavx4}. The tangent and normal unit vectors to $\p S_z$ are depicted as arrows with full tip. Only the normal unit vector is indicated. The locations $x_a$ and $x_b$ are the abscissae of the left and right vertical edge of the bar. The dashed line indicates the mouth of the cavity located at $y=h$. Only the bottom wall is shown. The distance between bars is not drawn to scale.}
\label{fig:Sz}
\end{figure}

In this Appendix, we study the mechanism by which the control decreases the time-ensemble averaged total drag $C_p(\tau)+C_f(\tau)$ at different phases of the oscillation. The dependence of the quantities on the phase is kept herein. We consider a momentum balance relating the total drag to the Reynolds stress $-\avtez{u_tv_t}$ along the horizontal line $l_m$ connecting the crests of two consecutive bars, i.e., the mouth of the cavity, as shown in figure \ref{fig:Sz}. Inspired by the works of \cite{gharib-roshko-1987} and \cite{leonardi-etal-2003}, we consider the time-ensemble and $z$-averaged streamwise momentum equation:

\begin{equation}
\frac{\p \avtez{u u}}{\p x}+ \frac{\p \avtez{u v}}{\p y}=-\frac{\p \lela \wh{p} \rira}{\p x}+\frac{1}{Re_p}\ler \frac{\p^2 \avtez{u}}{\p x^2}+\frac{\p^2 \avtez{u}}{\p y^2} \rir.
\label{eq:tezavx1}
\end{equation}
Equation \eqref{eq:tezavx1} is steady because $\p \avtez{u}/\p \tau=0$, as shown by figure \ref{fig:uphasemean}.
By defining the vector $\avtez{uu}\mb{i}+\avtez{uv}\mb{j}$ and noticing that $\p \lela \wh{p} \rira/\p x=\dive{(\lela \wh{p} \rira \mb{i})}$, we write \eqref{eq:tezavx1} as  

\begin{equation}
\dive{(\avtez{uu}\mb{i}+\avtez{uv}\mb{j})}
=
\dive{\ler-\lela \wh{p} \rira \mb{i}+\frac{1}{Re_p}\nabla {\avtez{u}}\rir}. 
\label{eq:tezavx2}
\end{equation}  

Integrating \eqref{eq:tezavx2} on a generic surface $S_z$ with boundary $\p S_z$ lying in the plane $x-y$ and applying the divergence theorem leads to
  
\begin{equation}
\intl{(\avtez{uu}\mb{i}+\avtez{uv}\mb{j})\cdot \mb{n}}{\p S_z}=-\intl{\lela \wh{p} \rira \mb{i}\cdot \mb{n}}{\p S_z}+\frac{1}{Re_p}\intl{\nabla \avtez{u}\cdot \mb{n}}{\p S_z}.
\label{eq:tezavx3}
\end{equation}

As the flow is periodic along $x$, it is possible to define $S_z$ as the rectangular area between two bars depicted in figure \ref{fig:Sz}. Equation \eqref{eq:tezavx3} becomes

\begin{equation}
\begin{aligned}
\int_{x_b}^{x_a+L_x} \avtez{uv}\Bigr|_{h} \,\dx&=\int_0^{h}\ler \avtez{p}\Bigr|_{x_b}-\avtez{p}\Bigr|_{x_a+L_x}\rir \,\dy+ 
\\
&
-\frac{1}{Re_p}\int_0^{h} \frac{\p \avtez{u}}{\p x}\Bigr|_{x_b} \dy-\frac{1}{Re_p}\int_{x_b}^{x_a+L_x} \frac{\p \avtez{u}}{\p y} \Bigr|_0 \dx+ 
\\
&
+\frac{1}{Re_p}\int_0^{h} \frac{\p \avtez{u}}{\p x}\Bigr|_{x_a+L_x} \dy+\frac{1}{Re_p}\int_{x_b}^{x_a+L_x} \frac{\p \avtez{u}}{\p y} \Bigr|_{h} \dx.                                          
\end{aligned}
\label{eq:tezavx4}
\end{equation}

The terms in \eqref{eq:tezavx4} involving $\p \avtez{u}/\p x$ are null because of continuity and the no-slip condition. Equation \eqref{eq:tezavx4} simplifies to

\begin{equation}
\begin{aligned}
\int_{x_b}^{x_a+L_x} \avtez{uv}\Bigr|_{h} \,\dx&
=\int_0^{h}\left(\avtez{p}\Bigr|_{x_b}-\avtez{p}\Bigr|_{x_a+L_x}\right) \,\dy-\frac{1}{Re_p} 
\int_{x_b}^{x_a+L_x}\frac{\p \avtez{u}}{\p y} \Bigr|_0 \dx+\\
&+\frac{1}{Re_p}\int_{x_b}^{x_a+L_x} \frac{\p \avtez{u}}{\p y} \Bigr|_{h} \dx.                                          
\end{aligned}
\label{eq:tezavx5}
\end{equation}

As the time-ensemble and spanwise-averaged pressure is $\avtez{p}=\avtez{\phi}+\wh{\Pi}_x (x-L_x)+0.5\wh{\Pi}_z L_z$ and $\avtez{\phi}=0$, we obtain 

\begin{equation}
\avtez{p}\Bigr|_{x_b}-\avtez{p}\Bigr|_{x_a+L_x}=\avtez{p}\Bigr|_{x_b}-\avtez{p}\Bigr|_{x_a}-\wh{\Pi}_x L_x.
\label{eq:tezavx6}
\end{equation}

Substituting \eqref{eq:tezavx6} into \eqref{eq:tezavx5} and considering the Reynolds decomposition $\avtez{uv}=\avtez{u}\avtez{v}+\avtez{u_tv_t}$ lead to 

\begin{equation}
\begin{aligned}
&\frac{1}{L_x}\int_0^{h}\left(\avtez{p}\Bigr|_{x_a}-\avtez{p}\Bigr|_{x_b}\right) \,\dy+\frac{1}{L_x Re_p}\int_{x_b}^{x_a+L_x} \frac{\p \avtez{u}}{\p y} \Bigr|_0 \dx+\wh{\Pi}_x h=\\
&-\frac{1}{L_x}\int_{x_b}^{x_a+L_x} \avtez{u}\avtez{v}\Bigr|_{h} \,\dx-\frac{1}{L_x}\int_{x_b}^{x_a+L_x} \avtez{u_tv_t}\Bigr|_{h} \,\dx+\frac{1}{L_x Re_p}\int_{x_b}^{x_a+L_x} \frac{\p \avtez{u}}{\p y} \Bigr|_{h} \dx.                                        
\end{aligned}
\label{eq:tezavx7}
\end{equation}

The first two terms of \eqref{eq:tezavx7} are half of the ensemble time-averaged coefficients $C_p(\tau)$ and $C_{ca}(\tau)$ as we only consider the bottom channel wall. Therefore \eqref{eq:tezavx7} reads

\begin{equation}
\begin{aligned}
\frac{C_p(\tau)+C_{ca}(\tau)}{2}+\wh{\Pi}_x h=&-\frac{1}{L_x}\int_{x_b}^{x_a+L_x} \avtez{u}\avtez{v}\Bigr|_{h} \,\dx-\frac{1}{L_x}\int_{x_b}^{x_a+L_x} \avtez{u_tv_t}\Bigr|_{h} \,\dx \\ 
&+\frac{1}{L_x Re_p}\int_{x_b}^{x_a+L_x} \frac{\p \avtez{u}}{\p y} \Bigr|_{h} \dx.                                        
\end{aligned}
\label{eq:tezavx8}
\end{equation}  

As in \cite{gharib-roshko-1987}, the crest skin-friction drag is two order of magnitude smaller than the total drag $-\Pi_x(\tau) L_y=C_p(\tau)+C_{f,ca}(\tau)+C_{f,cr}(\tau)\approx C_p(\tau)+C_{f,ca}(\tau)$. Equation \eqref{eq:tezavx8} becomes
 
\begin{equation}
C_p(\tau)+C_f(\tau) \approx -\frac{L_y}{L_x}\int_{l_m} \avtez{u}\avtez{v} \,\dx-\frac{L_y}{L_x}\int_{l_m} \avtez{u_tv_t} \,\dx
+\frac{L_y}{L_x Re_p}\int_{l_m} \frac{\p \avtez{u}}{\p y} \dx.                                        
\label{eq:tezavx9}
\end{equation}  

Equation \eqref{eq:tezavx9} shows that the time-ensemble average of the total drag is due to the product $\avtez{u}\avtez{v}$, the Reynolds stresses $\avtez{u_tv_t}$, and the friction term $Re_p^{-1}\p \avtez{u}/\p y$ along the mouth of the cavity. 

\bibliographystyle{jfm}
\bibliography{pr}

\begin{thebibliography}{49}
\expandafter\ifx\csname natexlab\endcsname\relax\def\natexlab#1{#1}\fi
\def\au#1{#1} \def\ed#1{#1} \def\yr#1{#1}\def\at#1{#1}\def\jt#1{\textit{#1}}
  \def\bt#1{#1}\def\bvol#1{\textbf{#1}} \def\vol#1{#1} \def\pg#1{#1}
  \def\publ#1{#1}\def\arxiv#1{#1}\def\org#1{#1}\def\st#1{\textit{#1}}

\bibitem[Banchetti {\em et~al.\/}(2020)Banchetti, Luchini \&
  Quadrio]{banchetti-etal-2020}
{\sc \au{Banchetti, J.}, \au{Luchini, P.} \& \au{Quadrio, M.}} \yr{2020}
  \at{Turbulent drag reduction over curved walls}.  \jt{J. Fluid Mech.}
  \bvol{896}~(A10).

\bibitem[Brackston {\em et~al.\/}(2016{\natexlab{{\em a\/}}})Brackston,
  Garc{\'\i}a De La~Cruz, Wynn, Rigas \& Morrison]{brackston-etal-2016}
{\sc \au{Brackston, R.D.}, \au{Garc{\'\i}a De La~Cruz, J.M.}, \au{Wynn, A.},
  \au{Rigas, G.} \& \au{Morrison, J.F.}} \yr{2016{\natexlab{{\em a\/}}}}
  \at{Stochastic modelling and feedback control of bistability in a turbulent
  bluff body wake}.  \jt{J. Fluid Mech.}  \bvol{802},  \pg{726--749}.

\bibitem[Brackston {\em et~al.\/}(2016{\natexlab{{\em b\/}}})Brackston, Wynn \&
  Morrison]{brackston-etal-2016-eif}
{\sc \au{Brackston, R.D.}, \au{Wynn, A.} \& \au{Morrison, J.F.}}
  \yr{2016{\natexlab{{\em b\/}}}}  \at{Extremum seeking to control the
  amplitude and frequency of a pulsed jet for bluff body drag reduction}.
  \jt{Exp. Fluids}  \bvol{57}~(10),  \pg{159}.

\bibitem[Brackston {\em et~al.\/}(2018)Brackston, Wynn \&
  Morrison]{brackston-etal-2018}
{\sc \au{Brackston, R.D.}, \au{Wynn, A.} \& \au{Morrison, J.F.}} \yr{2018}
  \at{Modelling and feedback control of vortex shedding for drag reduction of a
  turbulent bluff body wake}.  \jt{Int. J. Heat and Fluid Flow}  \bvol{71},
  \pg{127--136}.

\bibitem[Bradley \& Wray(1974)]{bradley-wray-1974}
{\sc \au{Bradley, R.} \& \au{Wray, W.}} \yr{1974}  \at{A conceptual study of
  leading-edge-vortex enhancement by blowing}.  \jt{J. Aircraft}
  \bvol{11}~(1),  \pg{34--38}.

\bibitem[Bragg \& Gregorek(1987)]{bragg-gregorek-1987}
{\sc \au{Bragg, M.B.} \& \au{Gregorek, G.M.}} \yr{1987}  \at{Experimental study
  of airfoil performance with vortex generators}.  \jt{J. Aircraft}
  \bvol{24}~(5),  \pg{305--309}.

\bibitem[Calarese \& Crisler(1985)]{calarese-crisler-1985}
{\sc \au{Calarese, W.} \& \au{Crisler, W.P.}} \yr{1985}  \at{Aim-85-0354
  afterbody drag reduction by vortex generators} .

\bibitem[Chang(2014)]{chang-2014}
{\sc \au{Chang, P.K.}} \yr{2014} {\em Separation of flow\/}.  \publ{Elsevier}.

\bibitem[Chng {\em et~al.\/}(2009)Chng, Rachman, Tsai \& Zha]{chng-etal-2009}
{\sc \au{Chng, T.L.}, \au{Rachman, A.}, \au{Tsai, H.M.} \& \au{Zha, G.}}
  \yr{2009}  \at{Flow control of an airfoil via injection and suction}.  \jt{J.
  Aircraft}  \bvol{46}~(1),  \pg{291--300}.

\bibitem[Cho {\em et~al.\/}(2016)Cho, Choi \& Choi]{cho-etal-2016}
{\sc \au{Cho, M.}, \au{Choi, S.} \& \au{Choi, H.}} \yr{2016}  \at{Control of
  flow separation in a turbulent boundary layer using time-periodic forcing}.
  \jt{J. Fluids Eng.}  \bvol{138}~(10).

\bibitem[Choi {\em et~al.\/}(2014)Choi, Lee \& Park]{choi-etal-2014}
{\sc \au{Choi, H.}, \au{Lee, J.} \& \au{Park, H.}} \yr{2014}  \at{Aerodynamics
  of heavy vehicles}.  \jt{Ann. Rev. Fluid Mech.}  \bvol{46},  \pg{441--468}.

\bibitem[Chun {\em et~al.\/}(1999)Chun, Lee \& Sung]{chun-lee-sung-1999}
{\sc \au{Chun, S.}, \au{Lee, I.} \& \au{Sung, H.J.}} \yr{1999}  \at{Effect of
  spanwise-varying local forcing on turbulent separated flow over a
  backward-facing step}.  \jt{Exp. Fluids}  \bvol{26}~(5),  \pg{437--440}.

\bibitem[Corke {\em et~al.\/}(2002)Corke, Jumper, Post, Orlov \&
  McLaughlin]{corke-etal-2002}
{\sc \au{Corke, T.~C.}, \au{Jumper, E.~J.}, \au{Post, M.~L}, \au{Orlov, D.} \&
  \au{McLaughlin, T.~E.}} \yr{2002}  \at{Application of weakly-ionized plasmas
  as wing flow-control devices}.  \jt{AIAA Paper}  \bvol{350},  \pg{2002}.

\bibitem[Fischer(2017)]{fischer-2017}
{\sc \au{Fischer, P.}} \yr{2017} {NEK5000} {V}ersion 17.0.4. {D}ec. 17, 2017.
  {ANL}, {I}llinois. available: https://nek5000.mcs.anl.gov.

\bibitem[Fukagata {\em et~al.\/}(2002)Fukagata, Iwamoto \&
  Kasagi]{fukagata-iwamoto-kasagi-2002}
{\sc \au{Fukagata, K.}, \au{Iwamoto, K.} \& \au{Kasagi, N.}} \yr{2002}
  \at{Contribution of {R}eynolds stress distribution to the skin friction in
  wall-bounded flows}.  \jt{Phys. Fluids}  \bvol{14}~(11),  \pg{73--76}.

\bibitem[Gharib \& Roshko(1987)]{gharib-roshko-1987}
{\sc \au{Gharib, M} \& \au{Roshko, A}} \yr{1987}  \at{The effect of flow
  oscillations on cavity drag}.  \jt{J. Fluid Mech.}  \bvol{177},
  \pg{501--530}.

\bibitem[Greenblatt {\em et~al.\/}(2006)Greenblatt, Paschal, Yao, Harris,
  Schaeffler \& Washburn]{greenblatt-etal-2006a}
{\sc \au{Greenblatt, D.}, \au{Paschal, K.B.}, \au{Yao, C.S.}, \au{Harris, J.},
  \au{Schaeffler, N.W.} \& \au{Washburn, A.~E.}} \yr{2006}  \at{Experimental
  investigation of separation control. {P}art 1: Baseline and steady suction}.
  \jt{AIAA J.}  \bvol{44}~(12),  \pg{2820--2830}.

\bibitem[Gad-el Hak {\em et~al.\/}(1998)Gad-el Hak, Pollard \&
  Bonnet]{gadelhak-pollard-bonnet-1998}
{\sc \au{Gad-el Hak, M.}, \au{Pollard, A.} \& \au{Bonnet, J.P.}} \yr{1998} {\em
  Flow control: fundamentals and practices\/}.  \publ{Springer}.

\bibitem[Heenan \& Morrison(1998)]{heenan-morrison-1998}
{\sc \au{Heenan, A.~F.} \& \au{Morrison, J.F.}} \yr{1998}  \at{Passive control
  of backstep flow}.  \jt{Exp. Th. Fluid Sc.}  \bvol{16},  \pg{122--132}.

\bibitem[Hinze(1975)]{hinze-1975}
{\sc \au{Hinze, J.O.}} \yr{1975} {\em Turbulence\/}.  \publ{McGraw Hill, Inc.
  -- Second Edition}.

\bibitem[Huang {\em et~al.\/}(2006)Huang, Corke \&
  Thomas]{huang-corke-thomas-2006}
{\sc \au{Huang, J.}, \au{Corke, T.~C.} \& \au{Thomas, F.~O.}} \yr{2006}
  \at{Plasma actuators for separation control of low-pressure turbine blades}.
  \jt{AIAA J.}  \bvol{44}~(1),  \pg{51--57}.

\bibitem[Ikeda \& Durbin(2007)]{ikeda-durbin-2007}
{\sc \au{Ikeda, T.} \& \au{Durbin, P.A.}} \yr{2007}  \at{Direct simulations of
  a rough-wall channel flow}.  \jt{J. Fluid Mech.}  \bvol{571},  \pg{235}.

\bibitem[Jeong \& Hussain(1995)]{jeong-hussain-1995}
{\sc \au{Jeong, J.} \& \au{Hussain, F.}} \yr{1995}  \at{On the identification
  of a vortex}.  \jt{J. Fluid Mech.}  \bvol{285},  \pg{69--94}.

\bibitem[Jukes \& Choi(2012)]{jukes-choi-2012}
{\sc \au{Jukes, T.N.} \& \au{Choi, K.S.}} \yr{2012}
  \at{Dielectric-barrier-discharge vortex generators: characterisation and
  optimisation for flow separation control}.  \jt{Exp. Fluids}  \bvol{52}~(2),
  \pg{329--345}.

\bibitem[Jung {\em et~al.\/}(1992)Jung, Mangiavacchi \&
  Akhavan]{jung-mangiavacchi-akhavan-1992}
{\sc \au{Jung, W.J.}, \au{Mangiavacchi, N.} \& \au{Akhavan, R.}} \yr{1992}
  \at{Suppression of turbulence in wall-bounded flows by high-frequency
  spanwise oscillations}.  \jt{Phys. Fluids A}  \bvol{4}~(8),  \pg{1605--1607}.

\bibitem[Leonardi {\em et~al.\/}(2003)Leonardi, Orlandi, Smalley, Djenidi \&
  Antonia]{leonardi-etal-2003}
{\sc \au{Leonardi, S.}, \au{Orlandi, P.}, \au{Smalley, R.J.}, \au{Djenidi, L.}
  \& \au{Antonia, R.A.}} \yr{2003}  \at{Direct numerical simulations of
  turbulent channel flow with transverse square bars on one wall}.  \jt{J.
  Fluid Mech.}  \bvol{491},  \pg{229--238}.

\bibitem[Lin {\em et~al.\/}(1994)Lin, Robinson, McGhee \&
  Valarezo]{lin-etal-1994}
{\sc \au{Lin, J.C.}, \au{Robinson, S.K.}, \au{McGhee, R.J.} \& \au{Valarezo,
  W.O.}} \yr{1994}  \at{Separation control on high-lift airfoils via
  micro-vortex generators}.  \jt{J. Aircraft}  \bvol{31}~(6),  \pg{1317--1323}.

\bibitem[Lin {\em et~al.\/}(1991)Lin, Selby \& Howard]{lin-selby-howard-1991}
{\sc \au{Lin, J.C.}, \au{Selby, G.V.} \& \au{Howard, F.G.}} \yr{1991}
  \at{Exploratory study of vortex-generating devices for turbulent flow
  separation control}.  \jt{AIAA Paper}  \bvol{42}.

\bibitem[Maday {\em et~al.\/}(1990)Maday, Patera \&
  R{\o}nquist]{maday-patera-ronquist-1990}
{\sc \au{Maday, Y.}, \au{Patera, A.T.} \& \au{R{\o}nquist, E.M.}} \yr{1990}
  \at{An operator-integration-factor splitting method for time-dependent
  problems: application to incompressible fluid flow}.  \jt{J. Sci. Comp.}
  \bvol{5}~(4),  \pg{263--292}.

\bibitem[Meyer \& Seginer(1994)]{meyer-seginer-1994}
{\sc \au{Meyer, J.} \& \au{Seginer, A.}} \yr{1994}  \at{Effects of periodic
  spanwise blowing on delta-wing configuration characteristics}.  \jt{AIAA
  journal}  \bvol{32}~(4).

\bibitem[Minelli {\em et~al.\/}(2019)Minelli, Tokarev, Zhang, Liu, Chernoray,
  Basara \& Krajnovi{\'c}]{minelli-etal-2019}
{\sc \au{Minelli, G.}, \au{Tokarev, M.}, \au{Zhang, J.}, \au{Liu, T.},
  \au{Chernoray, V.}, \au{Basara, B.} \& \au{Krajnovi{\'c}, S.}} \yr{2019}
  \at{Active aerodynamic control of a separated flow using streamwise synthetic
  jets}.  \jt{Flow Turb. Comb.}  \bvol{103}~(4),  \pg{1039--1055}.

\bibitem[Neretti(2016)]{neretti-2016}
{\sc \au{Neretti, G.}} \yr{2016}  \at{Active flow control by using plasma
  actuators}.  \jt{Rec. Prog. Aircr. Tech.}  \pg{pp. 57--76}.

\bibitem[Nuber \& Needham(1948)]{nuber-needham-1948}
{\sc \au{Nuber, R.J.} \& \au{Needham, J.R.}} \yr{1948}  \at{Exploratory
  wind-tunnel investigation of the effectiveness of area suction in eliminating
  leading-edge separation over an {NACA} {641A212} airfoil}.  \jt{NASA Langley
  Research Center} .

\bibitem[Oliveira \& Younis(2000)]{oliveira-younis-2000}
{\sc \au{Oliveira, P.J.} \& \au{Younis, B.A.}} \yr{2000}  \at{On the prediction
  of turbulent flows around full-scale buildings}.  \jt{J. Wind Eng. Ind.
  Aero.}  \bvol{86}~(2-3),  \pg{203--220}.

\bibitem[Pope(2000)]{pope-2000}
{\sc \au{Pope, S.B.}} \yr{2000} {\em Turbulent {F}lows\/}.  \publ{Cambridge
  University Press}.

\bibitem[Post \& Corke(2004)]{post-corke-2004}
{\sc \au{Post, M.L.} \& \au{Corke, T.C.}} \yr{2004}  \at{Separation control on
  high angle of attack airfoil using plasma actuators}.  \jt{AIAA J.}
  \bvol{42}~(11).

\bibitem[Post \& Corke(2006)]{post-corke-2006}
{\sc \au{Post, M.L.} \& \au{Corke, T.C.}} \yr{2006}  \at{Separation control
  using plasma actuators: dynamic stall vortex control on oscillating airfoil}.
   \jt{AIAA J.}  \bvol{44}~(12),  \pg{3125--3135}.

\bibitem[Quadrio \& Ricco(2003)]{quadrio-ricco-2003}
{\sc \au{Quadrio, M.} \& \au{Ricco, P.}} \yr{2003}  \at{Initial response of a
  turbulent channel flow to spanwise oscillation of the walls}.  \jt{J.
  Turbul.}  \bvol{4 007}.

\bibitem[Quadrio \& Ricco(2004)]{quadrio-ricco-2004}
{\sc \au{Quadrio, M.} \& \au{Ricco, P.}} \yr{2004}  \at{Critical assessment of
  turbulent drag reduction through spanwise wall oscillations}.  \jt{J. Fluid
  Mech.}  \bvol{521},  \pg{251--271}.

\bibitem[Quadrio \& Ricco(2011)]{quadrio-ricco-2011}
{\sc \au{Quadrio, M.} \& \au{Ricco, P.}} \yr{2011}  \at{The laminar generalized
  {S}tokes layer and turbulent drag reduction}.  \jt{J. Fluid Mech.}
  \bvol{667},  \pg{135--157}.

\bibitem[Quadrio {\em et~al.\/}(2009)Quadrio, Ricco \&
  Viotti]{quadrio-ricco-viotti-2009}
{\sc \au{Quadrio, M.}, \au{Ricco, P.} \& \au{Viotti, C.}} \yr{2009}
  \at{Streamwise-travelling waves of spanwise wall velocity for turbulent drag
  reduction}.  \jt{J. Fluid Mech.}  \bvol{627},  \pg{161--178}.

\bibitem[Ricco \& Hahn(2013)]{ricco-hahn-2013}
{\sc \au{Ricco, P.} \& \au{Hahn, S.}} \yr{2013}  \at{Turbulent drag reduction
  through rotating discs}.  \jt{J. Fluid Mech.}  \bvol{722},  \pg{267--290}.

\bibitem[Ricco \& Hicks(2018)]{ricco-hicks-2018}
{\sc \au{Ricco, P.} \& \au{Hicks, P.D.}} \yr{2018}  \at{Streamwise-travelling
  viscous waves in channel flows}.  \jt{J. Eng. Math.}  \bvol{120},
  \pg{861--869}.

\bibitem[Ricco {\em et~al.\/}(2012)Ricco, Ottonelli, Hasegawa \&
  Quadrio]{ricco-etal-2012}
{\sc \au{Ricco, P.}, \au{Ottonelli, C.}, \au{Hasegawa, Y.} \& \au{Quadrio, M.}}
  \yr{2012}  \at{Changes in turbulent dissipation in a channel flow with
  oscillating walls}.  \jt{J. Fluid Mech.}  \bvol{700},  \pg{77--104}.

\bibitem[Ricco \& Quadrio(2008)]{ricco-quadrio-2008}
{\sc \au{Ricco, P.} \& \au{Quadrio, M.}} \yr{2008}  \at{Wall-oscillation
  conditions for drag reduction in turbulent channel flow}.  \jt{Int. J. Heat
  Fluid Flow}  \bvol{29},  \pg{601--612}.

\bibitem[Seifert \& Pack(2002)]{seifert-pack-2002}
{\sc \au{Seifert, A.} \& \au{Pack, L.~G.}} \yr{2002}  \at{Active flow
  separation control on wall-mounted hump at high {R}eynolds numbers}.
  \jt{AIAA J.}  \bvol{40}~(7),  \pg{1363--1372}.

\bibitem[Touber \& Leschziner(2012)]{touber-leschziner-2012}
{\sc \au{Touber, E.} \& \au{Leschziner, M.A.}} \yr{2012}  \at{Near-wall streak
  modification by spanwise oscillatory wall motion and drag-reduction
  mechanisms}.  \jt{J. Fluid Mech.}  \bvol{693},  \pg{150--200}.

\bibitem[Trujillo {\em et~al.\/}(1997)Trujillo, Bogard \&
  Ball]{trujillo-bogard-ball-1997}
{\sc \au{Trujillo, S.M.}, \au{Bogard, D.G.} \& \au{Ball, K.S.}} \yr{1997}
  \at{Turbulent boundary layer drag reduction using an oscillating wall}.
  \jt{AIAA Paper}  \bvol{97-1870}.

\bibitem[Wong \& Kontis(2007)]{wong-kontis-2007-a}
{\sc \au{Wong, C.} \& \au{Kontis, K.}} \yr{2007}  \at{Flow control by spanwise
  blowing on a {NACA} 0012}.  \jt{J. Aircraft}  \bvol{44}~(1),  \pg{337--340}.

\end{thebibliography}

\end{document}